\documentclass[twocolumn]{aa}
\usepackage{epsfig}
\usepackage{amsmath}
\usepackage{amssymb}
\usepackage{natbib}
\usepackage[caption=false]{subfig}
\usepackage{float}
\usepackage{longtable}
\usepackage{graphicx}
\usepackage{textcomp}
\usepackage{color}
\usepackage{url}
\usepackage{epstopdf}

\begin{document} %
\def\simlt{\mathrel{\rlap{\lower 3pt\hbox{$\sim$}}\raise 2.0pt\hbox{$<$}}}
\def\simgt{\mathrel{\rlap{\lower 3pt\hbox{$\sim$}} \raise
2.0pt\hbox{$>$}}}
\def\lsim{\,\lower2truept\hbox{${<\atop\hbox{\raise4truept\hbox{$\sim$}}}$}\,}
\def\gsim{\,\lower2truept\hbox{${> \atop\hbox{\raise4truept\hbox{$\sim$}}}$}\,}

\title{AGN torus emission for a homogeneous sample of bright FSRQs}
   \titlerunning{Torus emission in FSRQs}
\authorrunning{G. Castignani \& G. De Zotti}
   \author{G. Castignani
          \inst{1}\fnmsep\thanks{e-mail: castigna@sissa.it}
   \and
          G. De Zotti\inst{2,1}
          }

   \institute{SISSA, Via Bonomea 265, 34136, Trieste, Italy
         \and
          INAF-Osservatorio Astronomico di Padova, Vicolo dell'Osservatorio 5, I-35122 Padova, Italy
             }

   \date{July 23, 2014}


\abstract
{We have selected a complete sample of 80 flat-spectrum radio quasars (FSRQs) from the WMAP 7-yr catalog within
the SDSS area, all with measured redshift, and have inspected their SEDs looking for evidence of an AGN torus emission. {A  SED fitting algorithm has found such evidence for seven objects and an uncertain indication for one more. A further analysis has picked out four additional FSRQs whose observed SEDs may be consistent with the presence of a torus.} All these $8+4$ FSRQs belong to the sub-sample of 55 sources showing the optical--ultraviolet bump interpreted as thermal emission from a standard accretion disc. {Torus luminosities have been estimated for the eight objects whose torus was identified by the fitting algorithm. For the other 47 FSRQs in the sub-sample, including the four with indications of torus emission not spotted by the algorithm,} we have derived upper limits to the torus luminosity. Our analysis shows that the torus can show up clearly only under quite special conditions: low luminosity and preferentially low peak frequency of the beamed synchrotron emission from the jet; high torus luminosity, close to that of the accretion disc. This implies that the inferred ratios of torus to disc luminosity are biased high. The median value, considering upper limits as detections {(survival analysis techniques are found to be inapplicable because of the excessive fraction of upper limits)}, is $L_{\rm torus}/L_{\rm disc}\sim 1$ while studies of radio quiet quasars yield  average ratios $\langle L_{\rm torus}/L_{\rm disc}\rangle \simeq 1/3$--$1/2$. On the other hand, although our poor statistics does not allow us to draw firm conclusions, our results are compatible with the FSRQ tori having the same properties of those of radio quiet quasars. At variance with \citet{Plotkin2012}, who investigated a sample of optically selected BL Lacs, we find that the Wide-field Infrared Survey Explorer (WISE) infrared colors do not allow us to draw any conclusion on the presence or absence of tori associated with WMAP selected blazars. With the latter selection blazars of all types (FSRQs with and without evidence of torus, BL Lacs, blazars of unknown type) occupy the same region of the WISE color - color plane, and their region overlaps that of SDSS quasars with point-like morphology. }


\maketitle

\section{Introduction}\label{par:intro}

The circum-nuclear tori around Active Galactic Nuclei (AGNs) have a key role in determining both their physical properties and their observed spectral energy distributions. According to the unification scheme, a fraction of the radiation produced by accretion into the central super-massive black hole is absorbed and re-emitted mainly at mid-infrared (MIR) wavelengths by a dusty, massive, possibly clumpy \citep{Markowitz2014} medium commonly referred to as the torus \citep[see][for a recent review]{Hoenig2013}. While there is broad agreement about the nitty-gritty of the unified scheme concept, the detailed properties of the tori remain unsettled.

Clues to the morphology of the torus come from studies of obscuration of the nuclei and of the re-emission by heated dust. Their obscuration properties have been recently investigated using large samples of X-ray \citep{Hao2013,Lusso2013,Merloni2014} or optically \citep{Calderone2012,MaWang2013,Roseboom2013} selected AGNs. Two approaches have been used. One is `demographic': it statistically gauges the covering factor of the obscuring medium from the ratio of obscured to unobscured AGNs { \citep{ogle2006,lawrence_elvis2010}}. Alternatively, the covering factor is estimated from the spectral energy distribution (SED) and in particular from the infrared (IR) to bolometric flux ratio of each individual source
{ \citep[SED-based approach,][]{hatziminaoglou2008,hatziminaoglou2009,alonsoherrero2011}}; this ratio is a measure of the fraction of the nuclear emission that is absorbed by the dusty torus and re-emitted in the IR.

Little is known about the properties of tori associated with flat-spectrum radio quasars (FSRQs) although indications of their presence have been occasionally reported \citep[e.g.,][]{Sbarrato2013}. On the contrary, \citet{Plotkin2012} did not detect any observational signature of a dusty torus from a sample of $\sim 100$ BL Lacertae (BL Lac) objects which, together with FSRQs, constitute the AGN population called `blazars'\footnote{Blazars are thought to be radio sources whose relativistic jet is closely aligned with the line-of-sight \citep{UrryPadovani1995}. In the framework of unified models for radio-loud AGNs, BL Lacs and FSRQs are interpreted as relativistically beamed Fanaroff-Riley type I and II \citep{FanaroffRiley1974} sources, respectively.}.

In this paper we investigate the SEDs of a complete sample of 80 FSRQs drawn from the Wilkinson Microwave Anisotropy Probe (WMAP) 7-yr catalog, looking for evidence of emission from a torus. The aims of this study are to check whether indeed FSRQs show unambiguous evidence of torus emission and, if so, to characterize its properties also in comparison with those of tori associated with radio-quiet AGNs. With respect to the latter objects, FSRQs have the great advantage that tori are most likely seen almost perfectly face-on, as they are expected to be almost perpendicular to the jet direction, which is closely aligned with the line of sight. This removes the large uncertainty plaguing estimates of the torus luminosity of radio-quiet AGNs due to the fact that the inclination is generally unknown. On the other hand, in the case of blazars the synchrotron emission from the jet may swamp the emission from the torus.

The plan of the paper is the following. In Sect.~\ref{sec:sample} we present the adopted sample,  in Sect.~\ref{sec:SED_modelling} we model the SEDs, in Sect.~\ref{sec:color_plots} we discuss the inferred constraints on the torus properties. Finally, in Sect.~\ref{sec:conclusions} we summarize our conclusions.

We adopt a flat $\Lambda \rm CDM$ cosmology with matter density $\Omega_{\rm m} = 0.32$, dark energy density $\Omega_{\Lambda} = 0.68$ and Hubble constant $h=H_0/100\, \rm km\,s^{-1}\,Mpc^{-1} = 0.67$ \citep{PlanckCollaborationXVI2013}.

\section{The sample}\label{sec:sample}
The WMAP satellite has provided the first all-sky survey at high radio frequencies ($\ge 23\,$GHz). At these frequencies blazars are the dominant radio-source population. We have selected a complete blazar sample, flux-limited at 23 GHz (K band), drawn from the 7-year WMAP point source catalog \citep{Gold2011}.

The basic steps of our selection procedure are described in \citet[][hereafter C13]{castignani2013}. The initial sample consists of 255 blazars, 245 of which have redshift measurements (compared to C13 we have found in the literature the redshifts of two more blazars, WMAP7\#\,354 and 365).


Following C13, to ensure a coverage of the SED as thorough and homogeneous as possible, still preserving the completeness of the sample, we have restricted the analysis to blazars within the area covered by the Tenth Data Release\footnote{\url{https://www.sdss3.org/dr10/}} (DR10) of the Sloan Digital Sky Survey (SDSS), totalling over 14,000 square degrees and providing simultaneous five-band photometry in the $\textsf{u}$, $\textsf{g}$, $\textsf{r}$, $\textsf{i}$ and $\textsf{z}$ bands. Of the 105 blazars in this area (two more, WMAP7 \#\,64 and 220, than found by C13 in the Eight Data Release, DR8), 19 are classified, in the most recent version of the blazar catalog BZCAT \citep{Massaro2013BZCAT}\footnote{\url{www.asdc.asi.it/bzcat/}} as BL Lacs, 6, including WMAP7 \#\,64, as blazars of unknown type and 80, including WMAP7 \#\,220, as FSRQs. In the following we will deal only with the latter objects.

We have recovered and updated the photometric data collected by C13. In particular, we have replaced the SDSS/DR8 and the 2012 Wide-field Infrared Survey Explorer \citep[WISE;][]{Wright2010} All-Sky Data Release photometry with the more recent SDSS/DR10 and AllWISE\footnote{\url{http://wise2.ipac.caltech.edu/docs/release/allwise/}} source catalog, i.e. the latest version of the WISE catalog. In many cases the SDSS/DR10 catalog gives multi-epoch photometry of the sources. We have adopted the median values and associated with them an error equal to the rms dispersion, likely due to variability, which is generally much larger than the photometric errors. The adopted  magnitudes, denoted e.g. as \textsf{dered\_g}, are corrected for Galactic extinction. As suggested in the DR8 tutorial\footnote{\url{www.sdss3.org/dr10/algorithms/fluxcal.php\#SDSStoAB}} we have decreased the DR10 \textsf{u}-band magnitudes by 0.04 to bring them to the AB system. We have also corrected the magnitudes for the Inter Galactic Medium (IGM) absorption, as described in C13.



WISE data are especially important for the study of the torus emission.
{WISE photometric data are provided at 3.4, 4.6, 12, and 22 $\mu$m. Hereafter we denote as $W1$, $W2$, $W3$, and $W4$ the WISE Vega magnitudes at these wavelengths, respectively.}
Except for WMAP7\#\,153, which was not detected, all the FSRQs in the sample have unambiguous WISE counterparts with signal-to-noise ($S/N$) ratios larger than three in the first three channels, and 75 (out of 80) were detected with $S/N>3$ also in the fourth channel (W4). 

{ WISE Vega magnitudes have been converted into flux densities by adopting the photometric calibrations and the color corrections for flat-spectrum sources given by \citet{Wright2010} and \citet{Jarrett2011}. The color corrections are less than 1$\%$ for the $W1$, $W2$, and $W4$ filters and $\leq8\%$ for $W3$ filter.}

\begin{figure*} \centering
\subfloat{\includegraphics[width=0.32\textwidth,natwidth=610,natheight=642]{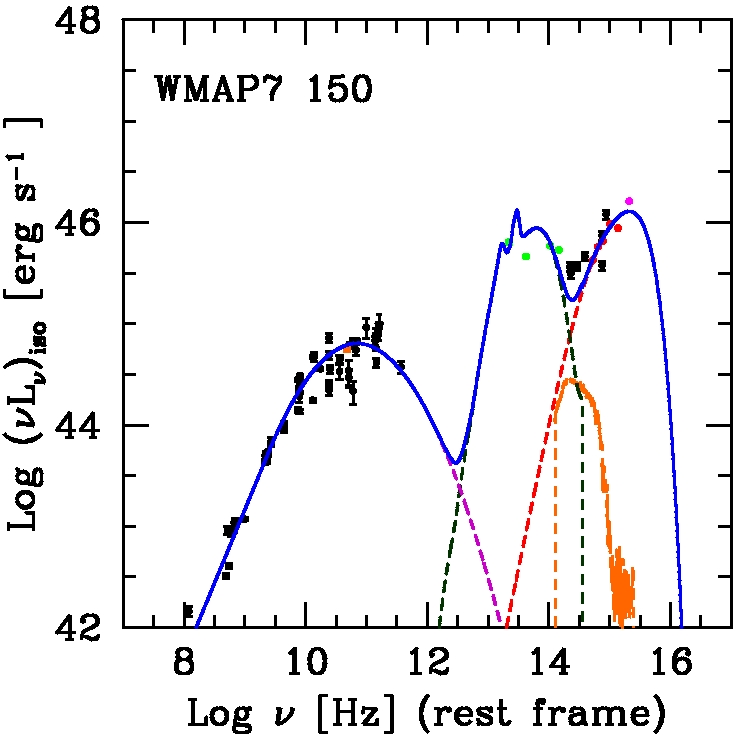}}
\subfloat{\includegraphics[width=0.32\textwidth,natwidth=610,natheight=642]{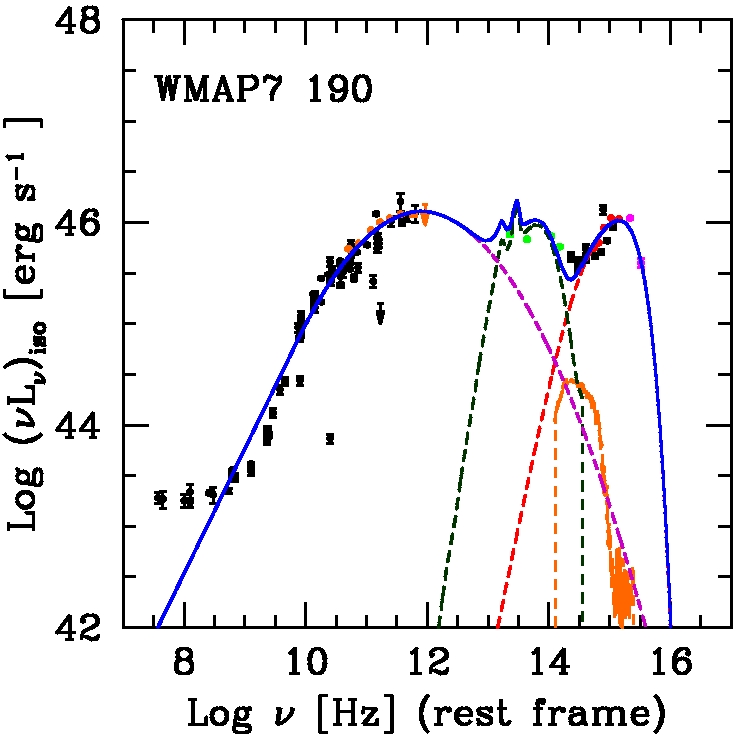}}
\subfloat{\includegraphics[width=0.32\textwidth,natwidth=610,natheight=642]{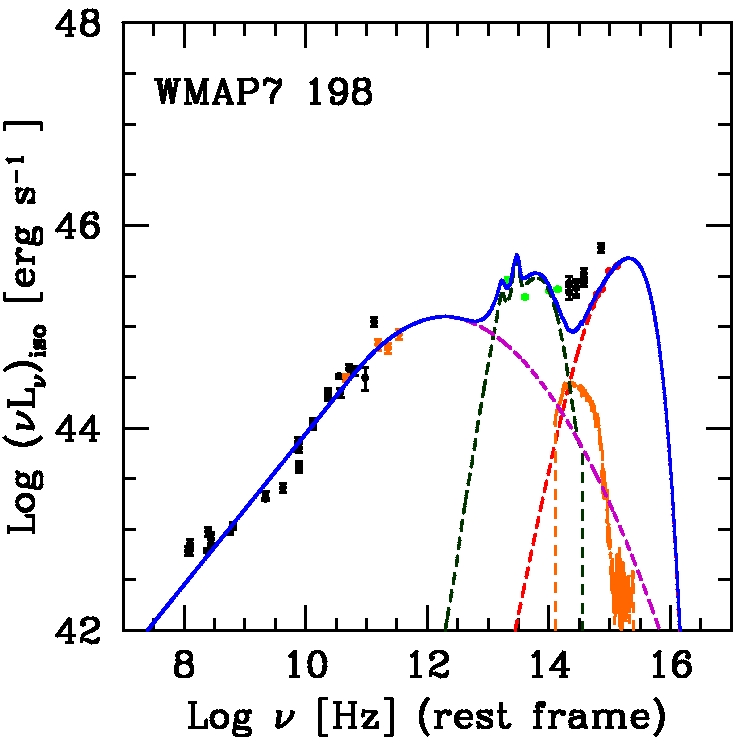}}\qquad
\subfloat{\includegraphics[width=0.32\textwidth,natwidth=610,natheight=642]{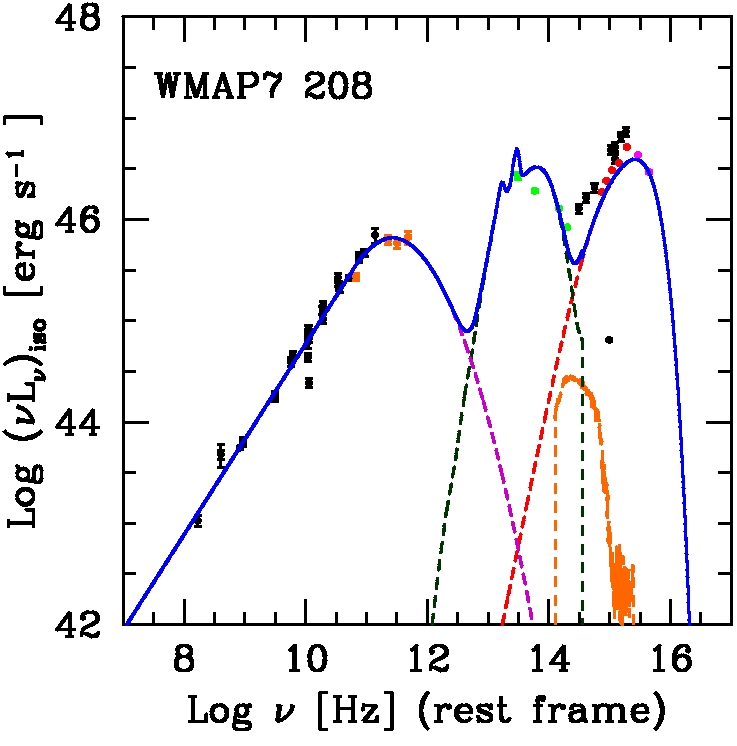}}
\subfloat{\includegraphics[width=0.32\textwidth,natwidth=610,natheight=642]{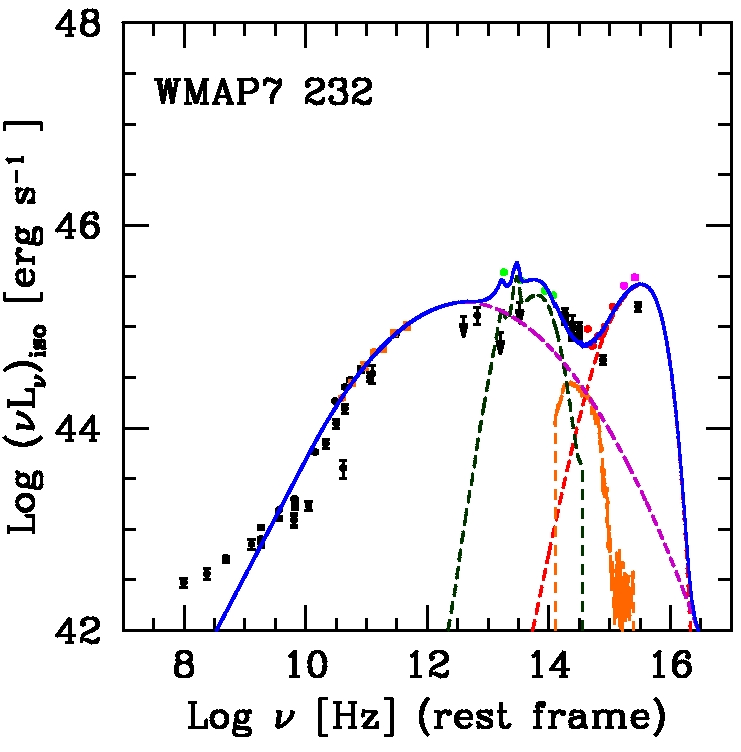}}
\subfloat{\includegraphics[width=0.32\textwidth,natwidth=610,natheight=642]{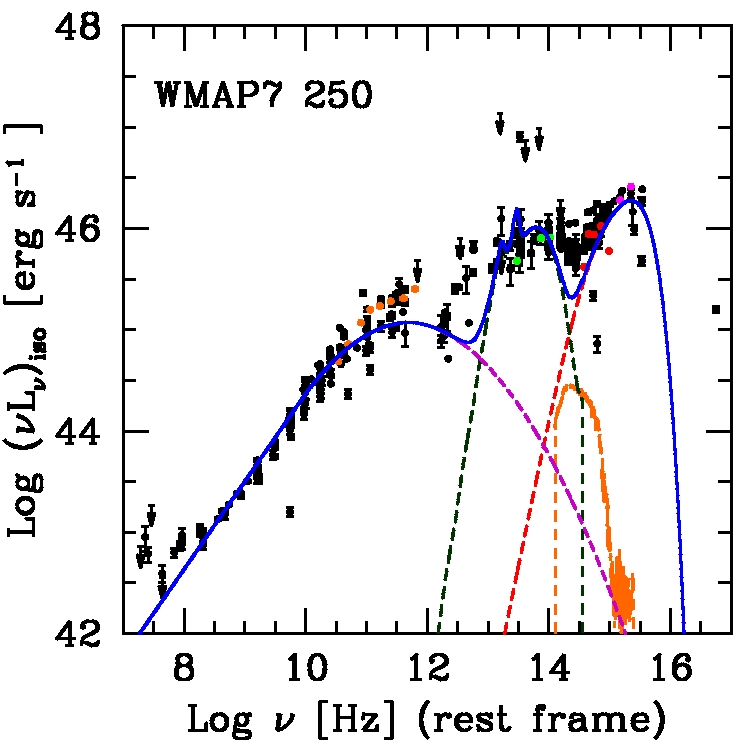}}\qquad
\subfloat{\includegraphics[width=0.32\textwidth,natwidth=610,natheight=642]{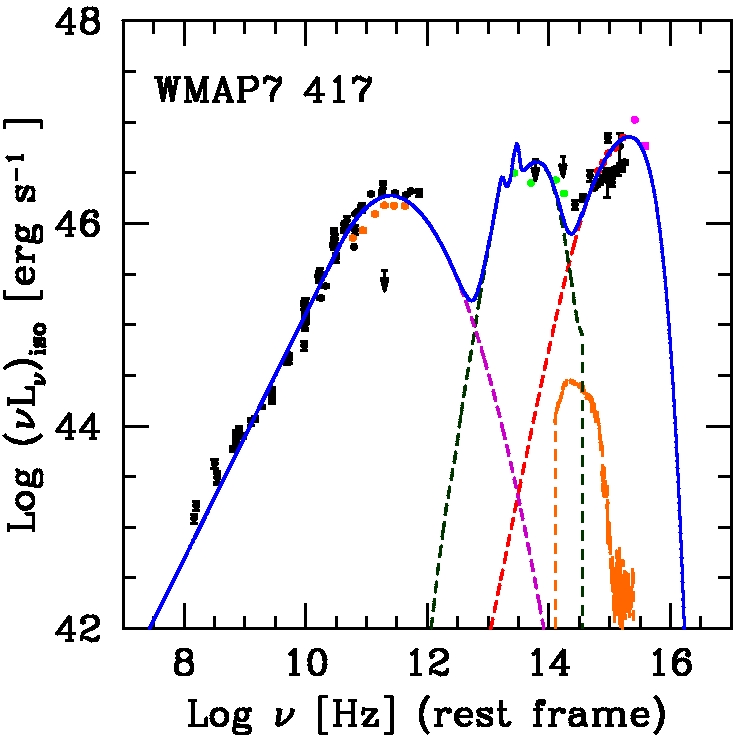}}
\subfloat{\includegraphics[width=0.32\textwidth,natwidth=610,natheight=642]{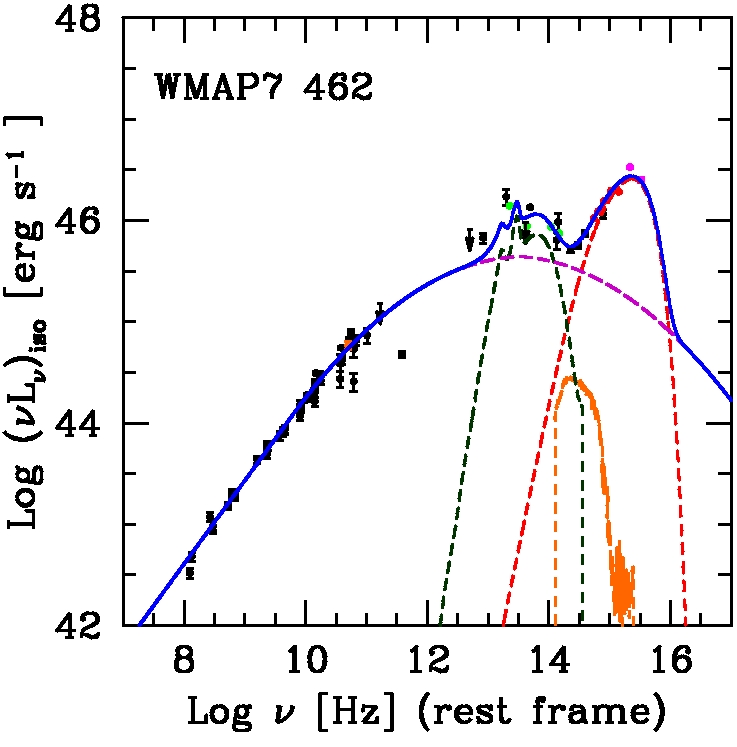}}\qquad
\caption{SEDs for the FSRQs with evidence (or indication) of torus emission {identified by the SED fitting algorithm}. Solid blue line: total SED, which includes synchrotron emission (dashed violet line), host galaxy (dashed orange line), disc (dashed red line) and torus (dark green dashed line) emissions. The host galaxy was taken to be a passive elliptical with $M_{R}=-23.7$. Data points: Planck (orange); WISE (green); SDSS (red); GALEX (magenta). Black points are data taken from the NASA/IPAC Extragalactic Database (NED). Note that, at variance with what was done to compute both $L_{\rm disc}$ and $L_{\rm torus}$ (see text), the luminosities shown here are computed assuming isotropic emission.}
\label{fig:SEDs_FSRQs_with_torus}
\end{figure*}

\begin{figure*} \centering
\subfloat{\includegraphics[width=0.32\textwidth,natwidth=610,natheight=642]{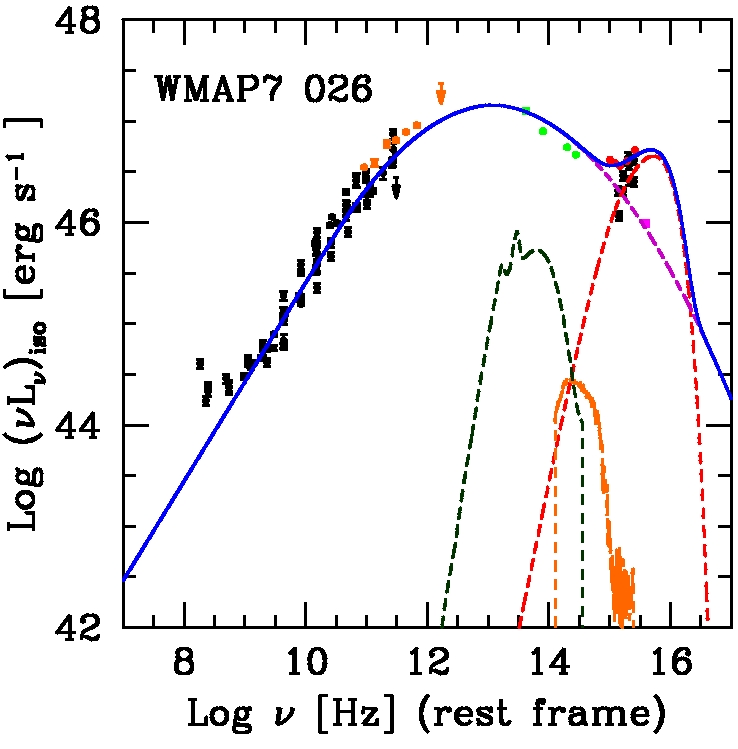}}
\subfloat{\includegraphics[width=0.32\textwidth,natwidth=610,natheight=642]{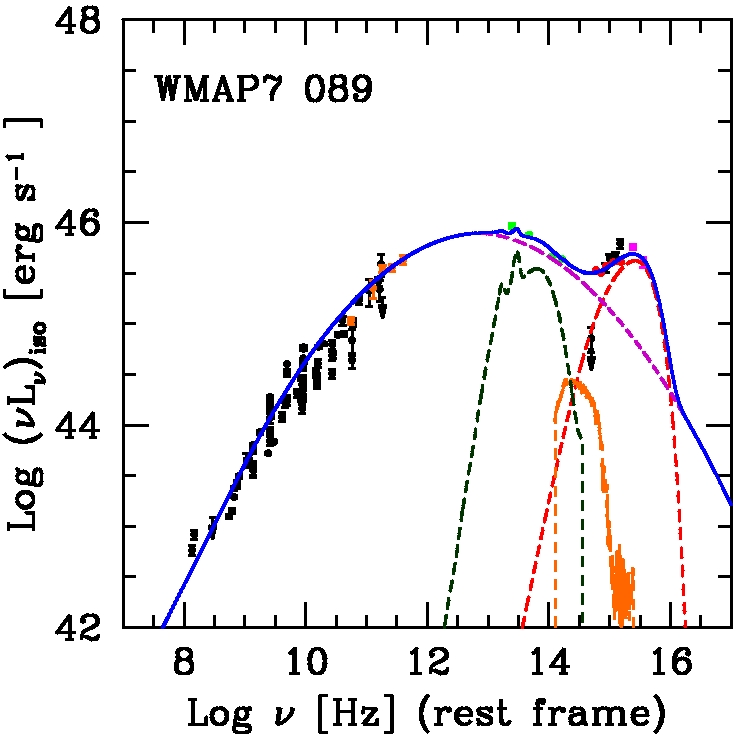}}
\subfloat{\includegraphics[width=0.32\textwidth,natwidth=610,natheight=642]{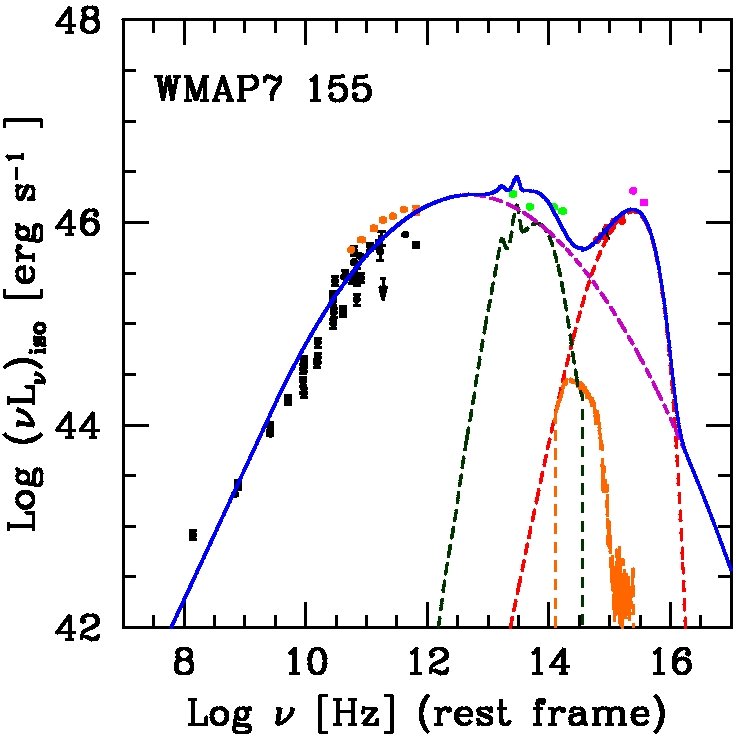}}\qquad
\caption{Examples of FSRQ SEDs with no evidence of torus. The same color code of Figure~\ref{fig:SEDs_FSRQs_with_torus} is adopted. The dashed dark green lines show the upper limits to the torus component determined with the procedure described in the text.}
\label{fig:SEDs_FSRQs_NO_torus}
\end{figure*}

\section{SED modeling}\label{sec:SED_modelling}

Of the 80 FSRQs in our sample 55 (i.e. 69\%) show clear evidence of the optical/ultraviolet (UV) bump, interpreted as the emission from a standard optically thick, geometrically thin accretion disc model \citep{ShakuraSunyaev1973}. In C13 we estimated the black hole mass for 54 of these sources by means of the SED fitting. The additional source is the FSRQ WMAP7 \#\,220, which did not have SDSS counterpart at the time of the C13 analysis. Using the C13 approach we estimate, for this source, a black hole mass M$_\bullet=1.6\times10^9$~M$_\odot$ and an accretion rate $\dot{M}=3.2$~M$_\odot$~yr$^{-1}$, corresponding to an Eddington ratio of 0.09. No other estimates of the black hole mass of this source were found in the literature.

None of the 25 FSRQs without a clear evidence of optical/UV bump shows signs of torus emission in their SEDs. The absence of the optical/UV bump deprives us of an important constraint on the intensity of the torus emission that, for almost face-on tori as those of FSRQs are expected to be, cannot exceed that of the disc. Because of that, we could not obtain meaningful constraints on the torus luminosity for these objects.

For the 55 FSRQs with evidence of the bump we attempted a SED modeling taking into account the Doppler boosted synchrotron continuum described as in \citet{Donato2001}, the passive elliptical host galaxy template by \citet{Mannucci2001}, the \citet{ShakuraSunyaev1973} accretion disc model and the clumpy AGN torus model by \citet{hoenig_kishimoto2010} provided in the CAT3D library\footnote{\url{http://cat3d.sungrazer.org/}}. For the latter we adopted the following parameter values: power law index of the radial dust cloud distribution $a=-1.5$; half opening angle of the torus $\theta_0=60\,$degrees; mean number of clouds along an equatorial line-of-sight $N_0=5$; optical depth of individual clouds $\tau_V=30$; outer radius of the torus $R_{\rm out}=150\,$pc; sublimation radius for a disc luminosity of $10^{46}\,\hbox{erg}\,\hbox{s}^{-1}$, $r_{\rm sub,0}=0.9\,$pc. The chosen values for the model parameters are consistent with those adopted in \citet{hoenig2011} to reproduce the mean SED of a complete sample of quasars and radio galaxies at $1.0\leq z\leq 1.4$.

{We caution that attempts to fit the full blazar SEDs built collecting data from the literature are fraught with difficulties coming from incomplete spectral coverage, variability (data are generally non simultaneous), and uncertainties on assumptions behind the models and on model parameters. However the main purpose of the present paper is only to look for the presence of the AGN torus. To  this end we take advantage of simultaneous WISE data. Our fits of the global SEDs do not pretend to characterize accurately the other components but are only aimed at checking whether they may contaminate the torus emission.}



The fits of the global SEDs were made using five free parameters. Four are those of the blazar sequence model for
the synchrotron emission \citep[the 5~GHz luminosity, the 5~GHz spectral index, the junction frequency between the low- and the high-frequency synchrotron template and the peak frequency of $\nu{\rm L}_\nu$;][]{Donato2001}. The remaining parameter is the normalization of the torus template. Since the routine used to compute the minimum $\chi^2$ fit did not converge in most cases when we attempted to use more than five parameters, the other components { (i.e. the host-galaxy and the optical/UV bump templates)} were kept fixed. An absolute magnitude of $M_{R}=-23.7$ was attributed to the host galaxy, as in \citet{Giommi2012}. The normalization and the peak frequency of the accretion disc template were fixed to the values found by C13.

{In the presence of strong variability, the choice of the error to be used in the calculation of the $\chi^2$ is not a trivial task. 
Based on blazar variability studies \citep[e.g.][]{Hovatta2008}  
we have adopted an error  $\sigma=\max\{0.4; \sigma_{\rm obs}\}$ whenever multiple observations at the same frequency, with dispersion $\sigma_{\rm obs}$ around the mean $\log(\nu{\rm L}_\nu)$,  are available, and
$\sigma=0.4$ for the other non-simultaneous data points.
The low-frequency ($\nu < 5\,$GHz) measurements, that likely refer to a different, frequently extended, emission component, not accounted for by the \citet{Donato2001} model, were excluded from the calculation of the $\chi^2$. }


The fitting algorithm found statistically significant (significance level $\geq3\sigma$, see below) signatures of a torus in seven FSRQs, i.e. in $\simeq 9\%$ of the 80 FSRQs in the sample or in $\simeq 13\%$ of the 55 FSRQs with evidence of the accretion disc bump. A $\sim2\sigma$ torus detection was found for one additional object, WMAP7\#\,250, i.e. 3C~273. Our best fit SEDs for these objects, with the contributions of the components mentioned above, are compared with the data in Figure~\ref{fig:SEDs_FSRQs_with_torus}.
Although the evidence of the torus is statistically significant, the estimate of the torus luminosity is endowed with a considerable uncertainty,  difficult to quantify.  
{ This is because the available data constrain only poorly the torus SED.
Our estimates therefore depend on the adopted model. As illustrated by the figure, contamination by other components is likely to be a minor issue. In particular the torus emission is far above that of our galaxy template. Since the latter corresponds to a giant elliptical, the contamination from the host galaxy is likely negligible.}


{According to the best fit produced by our algorithm, the jet emission at mid-IR frequencies of the remaining 47 FSRQs swamps that from the torus, if present.} This prevents an estimate of the torus luminosity for the majority of our sources. Upper limits to such luminosity were derived as follows. The minimum $\chi^2$ per degree of freedom ($\chi^2_\nu$) obtained from the fitting procedure corresponds to a negligible contribution from the torus. 
{If $\chi^2_\nu$ is $\simeq 1$ the $1\,\sigma$ upper limit to the amplitude of the torus template can be obtained gradually increasing such amplitude and redoing the fit for the other components until the minimum $\chi^2$ increases by $\delta\chi^2=1$ \citep{Cash1976}. In our case, the minimum $\chi^2_\nu$ is generally $\gg 1$ (the median value is 8.6). There are two main reasons for that. One is that variability frequently modifies also the SED \textit{shape}; thus a single SED cannot be simultaneously consistent with all the photometric data. The second one is that the adopted SED model is anyway too simple to provide an accurate description of the data; but, as mentioned above,  we cannot afford additional parameters.  To overcome this problem we have applied the \citet{Cash1976} method after having increased all the error bars by a constant factor equal to $(\chi^2_\nu)^{1/2}$.}
{ By the same token, the evidence of torus emission was considered to be significant if setting to zero the amplitude of the torus template increases the re-scaled minimum $\chi^2$ by $\delta\chi^2\geq9$, corresponding to a significance $\geq3\,\sigma$ for one interesting parameter (the normalization of the torus template).}

For WMAP7\#\,153, which is missing mid-IR measurements, we have adopted $3\,\sigma$ flux density upper limits of 0.6 and 3.6 mJy, from instrumental noise alone \citep{Wright2010}, for the WISE channels $W3$ and $W4$, respectively.
For channels $W1$ and $W2$ the $3\,\sigma$ limits are set by confusion noise and amount to 0.31 and 0.17~mJy, respectively \citep{Jarrett2011}. For the 4 objects detected in the $W4$ channel with $\hbox{S/N}<3$ we have adopted An upper limit equal to three times the error.

The procedure described above yielded upper limits to the torus luminosity for 25 out of the 47 FSRQs.  {For the remaining 22 objects either the minimization routine did not converge (13 cases) or the result was determined by the few data points with the lowest error bars and did not look credible at visual inspection (5 cases) or the upper limits exceeded the accretion disk luminosity (4 cases). For these objects the upper limits to the torus luminosity were obtained requiring that the torus template does not exceed both the mid-IR measurements and the disc luminosity.} Examples of SEDs which do not show evidence of torus emission are shown in Fig.~\ref{fig:SEDs_FSRQs_NO_torus}. The dashed green lines show the torus SED with luminosity equal to the adopted upper limit.\footnote{Figures showing the fits to the SEDs of all { 80} FSRQs { in our sample are
reported in Appendix~\ref{app:SEDs}.}}

{As mentioned above, our algorithm did not always come to a sensible solution. This has motivated a further analysis of the SEDs. A visual inspection of the fits yielded by the algorithm (see Appendix~\ref{app:SEDs}) indicates that  in several cases (WMAP7\#\;042, 137, 166, 191, 221, 224, 295, 327, 265, 278, 306, 407, 412, 428, and 434)  the synchrotron emission might sink down at lower frequencies than implied by those fits, leaving room for a torus contribution. However, except for the four cases discussed below, the shape of the SEDs of these sources in the frequency region where the torus is expected to show up are more akin to synchrotron than to the torus template; we have therefore kept the upper limits to the torus luminosity derived as described above.} 
{For sources WMAP7\#\;042, 166, 191, and 428, all at $z<1.5$,  the WISE photometry seems to be consistent with the torus template, although the fact that the fitting algorithm prefers a different solution indicates that it is hard to draw any firm conclusion. Thus we regard these four objects as showing tentative, certainly not unambiguous, evidence of torus emission. To better illustrate this point, the torus template has been fitted by hand to the WISE photometry, and the corresponding luminosities have been adopted as upper limits, in place of those estimated as described above. The new fits are shown in Appendix~\ref{app:SEDs}.   In in the case of WMAP7~\#~428 we also show the best fit solution for the torus yielded by the algorithm. For the other three objects the best fit torus emission is negligible. }

From the SED fits of all the 55 FSRQs we have thus derived synchrotron, disc, and torus luminosities (or upper limits). We have taken into account that both the disc and the torus emissions are anisotropic and the result depends on their inclination, $i$, with respect to the plane of the sky. As argued above, the FSRQs should have $i\simeq 0^\circ$. The jet synchrotron emission is obviously also highly anisotropic but we do not have enough information to  properly take the anisotropy into account; therefore we have computed the synchrotron luminosity assuming isotropic emission. This assumption has no impact on the conclusions of this paper.




The disc luminosity is estimated as in C13. The dependence of the observed torus flux density on the inclination angle $i$ has been derived from the CAT3D model templates reported for different values of $i$ in the CAT3D library to obtain:
\begin{equation}\label{eq:torus_lum}
F_\nu(i) = F_\nu(0) \frac{0.56+0.88|\cos i|}{0.56+0.88}\hspace{0.2cm}.
\end{equation}
The intrinsic luminosity, obtained integrating over all possible viewing angles ($0^\circ\leq i\leq 180^\circ$) is thus lower by a factor of $1/(0.56+0.88)\simeq 0.69$ than would be obtained from the observed flux density ($i=0^\circ$) assuming isotropic emission.

\begin{figure*} \centering
\subfloat[]{\includegraphics[width=0.5\textwidth,natwidth=610,natheight=642]{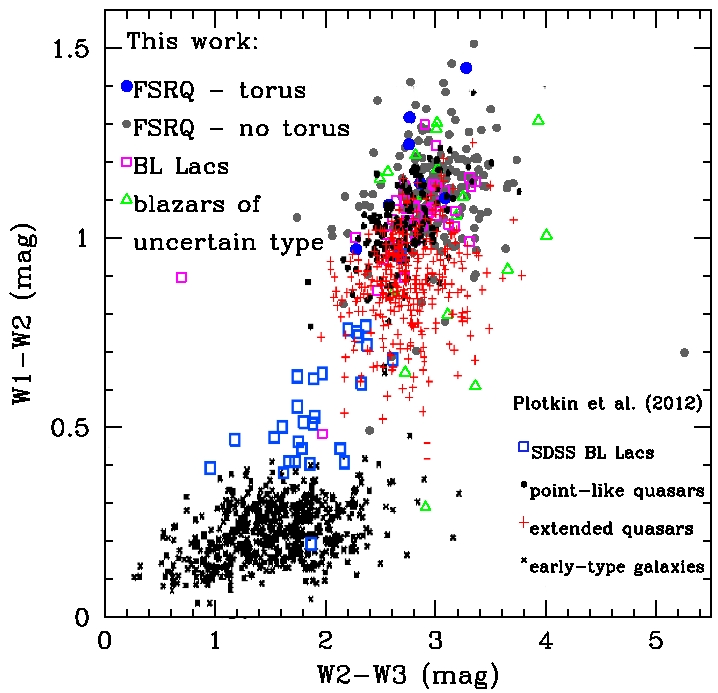}}
\subfloat[]{\includegraphics[width=0.45\textwidth,natwidth=610,natheight=642]{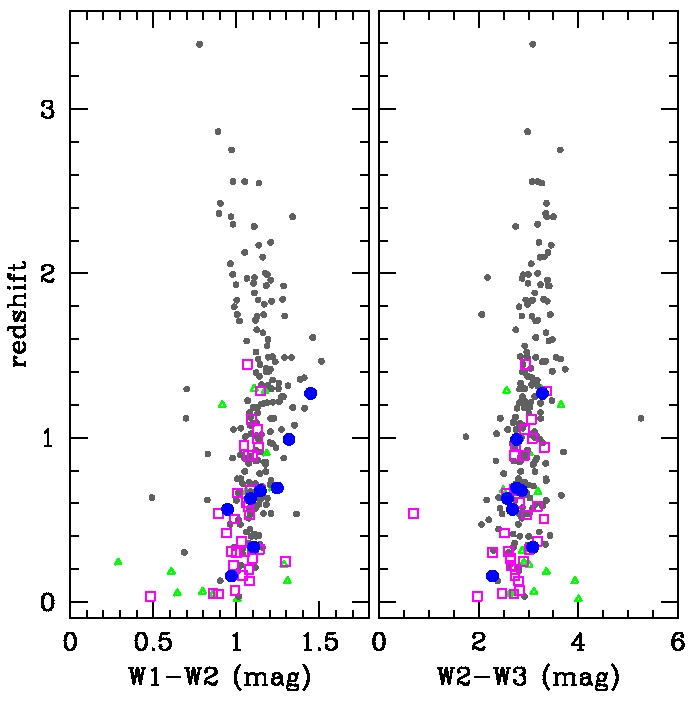}}\qquad
\caption{Left-hand panel: mid-IR (WISE) colors of WMAP-selected blazars of different kind (filled blue circles: FSRQs with evidence of torus, { including WMAP7\#\;250}; gray points: FSRQs without evidence of torus; open magenta squares: BL~Lacs; open green triangles: blazars of uncertain type) compared with those of sources shown in the top left panel of Fig. 3 of \citet{Plotkin2012}, namely SDSS BL Lacs (open blue squares), quasars with point-like (filled black circles) and extended (red $+$ signs) morphology in SDSS imaging and early-type galaxies (black $\times$ signs). Right-hand panels:  WISE colors vs. redshift for WMAP-selected blazars; points have the same meaning as in the left panel.}
\label{fig:color_plots_WISE}
\end{figure*}

\begin{figure*} \centering
\subfloat{\includegraphics[width=0.45\textwidth,natwidth=610,natheight=642]{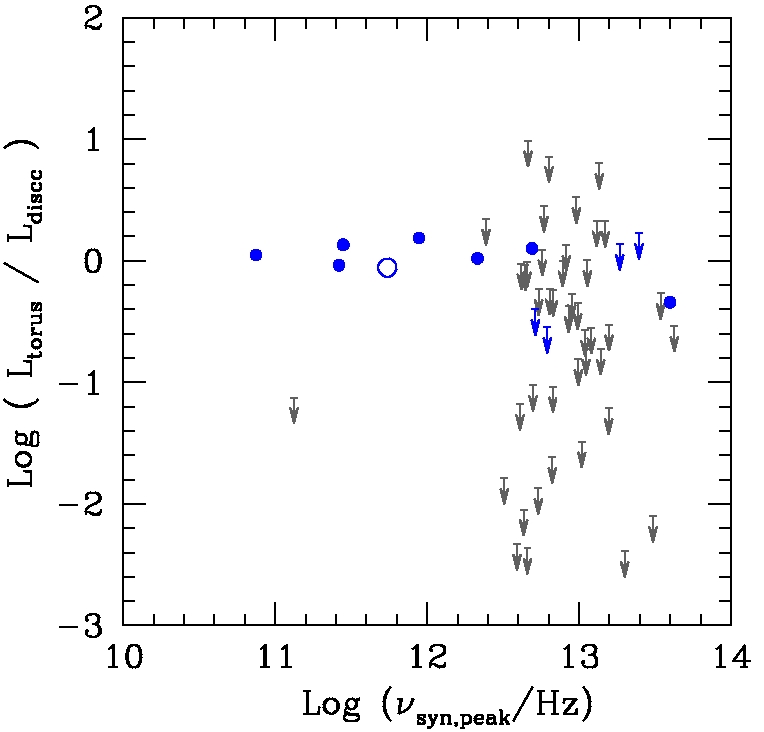}}
\subfloat{\includegraphics[width=0.45\textwidth,natwidth=610,natheight=642]{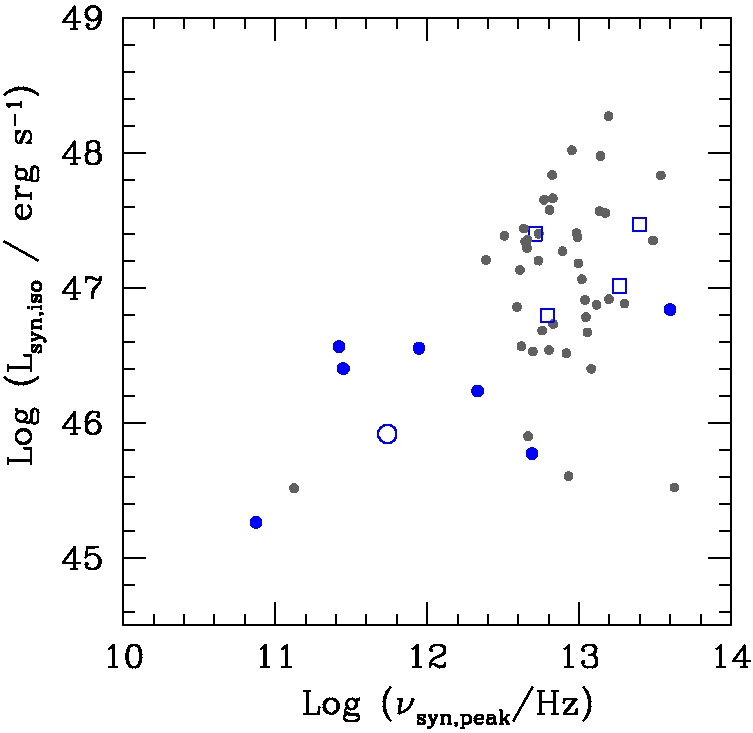}}
\caption{Torus to disc luminosity (left panel) and synchrotron luminosity versus synchrotron peak frequency (right panel) for the 55 FSRQs with evidence of blue bump. The filled blue points are the FSRQs with evidence of torus, the open blue point is WMAP7\#\,250, for which the presence of torus is less certain. { Blue upper limits (left panel) and open blue squares (right panel) correspond to sources WMAP7\#\;042, 166, 191, and 428 which show tentative evidence of torus.} Gray  symbols refer to the { remaining} FSRQs without evidence of torus.}
\label{fig:torus_correlations_plot4}
\end{figure*}

\begin{figure*} \centering
\subfloat{\includegraphics[width=0.45\textwidth,natwidth=610,natheight=642]{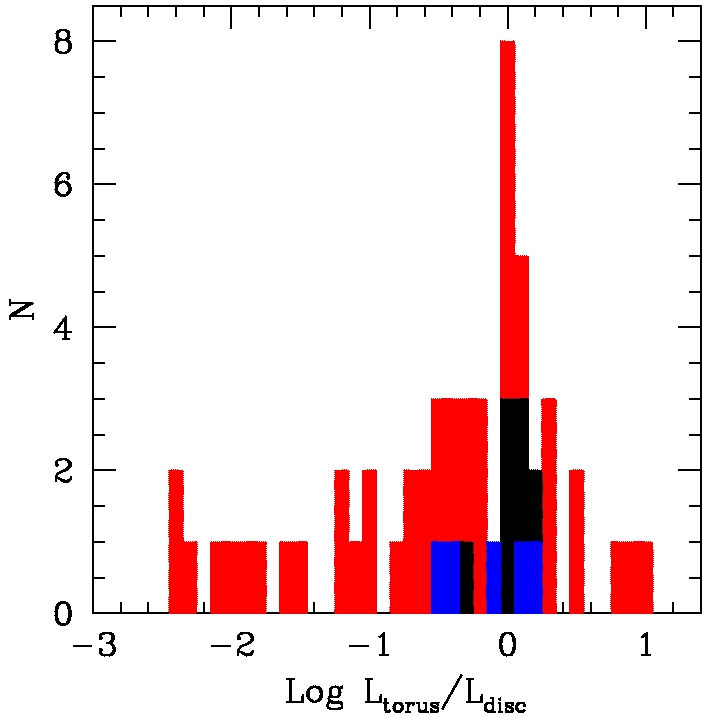}}
\subfloat{\includegraphics[width=0.45\textwidth,natwidth=610,natheight=642]{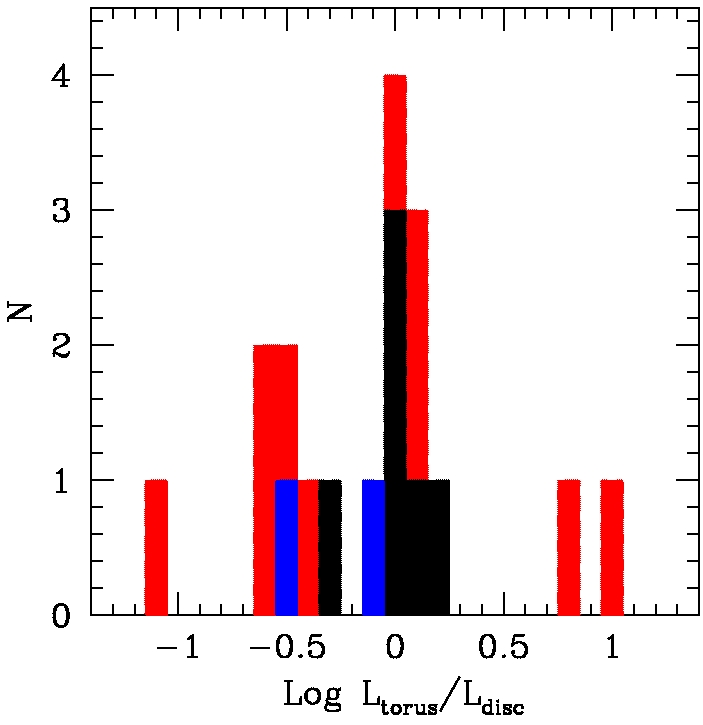}}\qquad
\caption{Left panel: distribution of the ratio of the torus to the disc luminosities for the 55 FSRQs with both disc and torus luminosity estimates (or upper limits). Right panel: distribution for the sub-sample of 18 sources located at $z\leq1$. In both panels, the black histogram show the sources with evidence of torus, the red histogram those with only upper limits to the torus luminosity. For the latter the luminosity ratios are upper limits. { The blue bars  correspond to WMAP7\#\;042, 166, 191, 250, and 428 for which the evidence for the torus is tentative; among them only WMAP7\#\;250 and 428 are at $z\leq1$ and therefore appear also in the right panel.}}
\label{fig:histo_Ltorus_Ldisc}
\end{figure*}

\section{Color plots and torus properties}\label{sec:color_plots}

\citet{Plotkin2012} argued that statistical evidence of torus emission from AGNs can be provided by mid-IR colors. Their investigation of the WISE data for $\sim 100$ BL Lacs selected from the SDSS showed that even the most weakly beamed ones, whose mid-IR emission should not be completely dominated by the jet emission, have IR colors too blue for a significant contribution from the torus be present.

In the left-hand panel of Fig.~\ref{fig:color_plots_WISE} we show the distribution of our WMAP selected blazars of different types in the same WISE color diagram used by \citet{Plotkin2012}. There is no clear separation in the $(W1-W2)$--$(W2-W3)$ plane between the FSRQs with evidence of torus emission and objects of other kinds: FSRQs without evidence of torus, WMAP selected BL Lacs, WMAP blazars of unknown type. All these objects occupy the same region as SDSS quasars with point-like morphology. Hence, at variance with the findings by \citet{Plotkin2012} for their optically selected BL Lac sample, this diagram does not provide any indication on the presence or absence of torus emission in blazar SEDs.

{The outlier with $W1-W2 = 0.7$ and $W2-W3 = 5.3$ is WMAP7\#\,376, i.e. PKS~B1908-201. Its anomalous WISE colors might be due to contamination by diffuse Galactic emission, visible in the IRAS maps, that affects mostly the $W3$ channel both because of the Galactic emission spectrum and because of the poorer angular resolution. }

The figure also shows that the 28 SDSS BL Lacs at $z<0.3$ in the \citet{Plotkin2012} sample (none of which is in common with our sample) occupy, in the color-color diagram, a different region than our WMAP-selected BL Lacs. This implies that the conclusions by \citet{Plotkin2012} do not apply to the general BL Lac population but are specific to their optical selection. The fact that the \citet{Plotkin2012} BL Lacs have colors intermediate between quasars and early-type galaxies may suggest substantial contamination from the host galaxy.

The right-hand panels of Fig.~\ref{fig:color_plots_WISE} show that the FSRQs with evidence of torus emission are all at relatively low redshift ($z<1.5$) and roughly span the same redshift range as WMAP selected BL Lacs. Since the sample is flux-limited at 23 GHz, where the emission is dominated by synchrotron, lower $z$ objects are those with lower synchrotron luminosity. The latter is thus less likely to swamp the torus emission. This is quantified in the right panel of Fig.~\ref{fig:torus_correlations_plot4} which also shows that tori are preferentially detected in FSRQs with lower synchrotron peak frequencies. The left panel of the same figure shows that the detected tori have luminosities close to the accretion disc luminosity, i.e. are at the upper limit of the physically plausible range for face-on objects. On the whole, it appears that the torus can show up only under quite special conditions: low luminosity and preferentially low peak frequency of the beamed synchrotron emission from the jet { as well as} high torus luminosity, close to that of the accretion disc. { These conditions must hold at once.} In other words, the lack of torus detection does not necessarily mean that blazars have weaker tori than radio quiet AGNs with similar accretion disc luminosity. In particular, the lack of detections at $z>1.5$ for our flux limited sample does not necessarily mean that tori are weaker at high $z$.

In the left panel of Fig.~\ref{fig:histo_Ltorus_Ldisc} we report the distribution of the ratios of the torus to the disc luminosities (or of upper limits to this ratio) for the 55 FSRQs with disc and torus luminosity estimates (or upper limits in the case of tori). In the right panel we report the same distribution only for the sub-sample of 18 sources located at $z\leq1$, which include six out of the seven FSRQs with evidence of torus and the uncertain cases WMAP7\#\,250 {and 428}. For this subsample the median logarithmic ratio, treating upper limits as detections, is $\langle\log(L_{\rm torus}/L_{\rm disc})\rangle = -0.02$. The median decreases to  $\langle\log(L_{\rm torus}/L_{\rm disc})\rangle = -0.24$ for the full sample.

{The fraction of FSRQs with measured torus emission is too low to allow a {proper} use of survival analysis techniques \citep{FeigelsonNelson1985,Schmitt1985} to reconstruct the distribution of luminosity ratios taking the upper limits into account. Our attempt to apply anyway the \citet{kaplan_meyer1958} estimator did not produce any valid estimate of the median $\langle\log(L_{\rm torus}/L_{\rm disc})\rangle$ for the full sample. In the case of the sub-sample of the 18 FSRQs at $z\leq1$ the estimator gave a  median $\langle\log(L_{\rm torus}/L_{\rm disc}) \rangle= 0.09$ with undetermined uncertainties. As mentioned above, torus to disc luminosity ratios $>1$ are unphysical. }

As mentioned above, the median torus to disc luminosity ratio is biased high because only tori with luminosity close to the physical upper limit can possibly be detected against the strong dilution by the beamed jet emission. In fact the median ratios quoted above are higher than found for radio quiet quasars. For example, based on a large sample drawn from the fifth edition of the SDSS quasar catalog \citet{Calderone2012} estimated that the torus reprocesses on average $\sim 1/3$ to $\sim 1/2$ of the accretion disc luminosity. \citet{Hao2013} did not find indications of evolution with redshift of the SEDs (hence also of the torus/disc luminosity ratio) of their 407 X-ray selected AGNs.
We have also looked for correlations of the torus luminosity with several physical quantities, including the black hole masses, the synchrotron properties (peak frequency, peak luminosity, total luminosity) and the disc luminosity. No significant correlation was found, not surprisingly given the poor statistics.

\section{Conclusions}\label{sec:conclusions}

We have investigated the SEDs of a complete sample of 80 FSRQs, flux limited at 23 GHz, drawn from the WMAP 7-yr catalog and located within the area covered by the SDSS DR10 catalog. We have found evidence of torus emission for 7 objects, all included in the sub-sample of 55 sources showing the optical--UV bump interpreted as thermal emission from a standard accretion disc. An uncertain indication of torus emission was found for one additional object. For the other 47 FSRQs in the accretion disc sub-sample we have derived upper limits to the torus luminosity. {The WISE data for four of the latter objects are compatible with torus emission, although our SED fitting algorithm favours an interpretation in terms of synchrotron emission.}

Our analysis has shown that the Doppler boosted synchrotron emission from the relativistic jet strongly hampers the detectability of the FSRQ torus emission. The detection was only possible for objects with torus luminosity close to the disc luminosity, which constitutes a physical upper limit to it. Even in this limiting case, the jet emission swamps that from the torus in the majority of objects unless it peaks at frequencies much lower than the mid-IR ones.
This implies that the inferred ratios of torus to disc luminosity are biased high. The median values, considering upper limits as detections, indicate $L_{\rm torus}/L_{\rm disc}\sim 1$ while studies of radio quiet quasars yield  average ratios $\langle L_{\rm torus}/L_{\rm disc}\rangle \simeq 1/3$--$1/2$ \citep{Calderone2012}. On the other hand, although our poor statistics does not allow us to draw firm conclusions, our results are compatible with the FSRQ tori having the same properties as those of radio quiet quasars, consistent with the unified scenario for AGNs.

At variance with \citet{Plotkin2012}, who investigated a sample of optically selected BL Lacs, we find that the WISE colors do not allow us to draw any conclusion on the presence or absence of tori associated with WMAP selected blazars. With the latter selection blazars of all types (FSRQs with and without evidence of torus, BL Lacs, blazars of unknown type) occupy the same region of the $(W1-W2)$--$(W2-W3)$ plane, and their region overlaps that of SDSS quasars with point-like morphology.

\begin{acknowledgements}
We thank the referee for helpful comments and Luigi Danese and Joaquin Gonzalez-Nuevo for fruitful discussion. Work supported in part by ASI/INAF agreement 2014-024-R.0 for the {\it Planck} LFI activity of Phase E2. This publications has made use of data products from the Wide-field Infrared Survey Explorer (WISE), the Sloan Digital Sky Survey (SDSS) and the NASA/IPAC Extragalactic Database (NED). The WISE is a joint project of the University of California, Los Angeles, and the Jet Propulsion Laboratory/California Institute of Technology, funded by the National Aeronautics and Space Administration. The NED  is operated by the Jet Propulsion Laboratory, California Institute of Technology, under contract with the National Aeronautics and Space Administration. Funding for the SDSS and SDSS-II has been provided by the Alfred P. Sloan Foundation, the Participating Institutions, the National Science Foundation, the U.S. Department of Energy, the National Aeronautics and Space Administration, the Japanese Monbukagakusho, the Max Planck Society, and the Higher Education Funding Council for England. The SDSS Web Site is http://www.sdss.org/. The SDSS is managed by the Astrophysical Research Consortium for the Participating Institutions. The Participating Institutions are the American Museum of Natural History, Astrophysical Institute Potsdam, University of Basel, University of Cambridge, Case Western Reserve University, University of Chicago, Drexel University, Fermilab, the Institute for Advanced Study, the Japan Participation Group, Johns Hopkins University, the Joint Institute for Nuclear Astrophysics, the Kavli Institute for Particle Astrophysics and Cosmology, the Korean Scientist Group, the Chinese Academy of Sciences (LAMOST), Los Alamos National Laboratory, the Max-Planck-Institute for Astronomy (MPIA), the Max-Planck-Institute for Astrophysics (MPA), New Mexico State University, Ohio State University, University of Pittsburgh, University of Portsmouth, Princeton University, the United States Naval Observatory, and the University of Washington.
\end{acknowledgements}

{}
\begin{appendix}
 \section{Spectral Energy Distributions}\label{app:SEDs}
{ We report the SEDs of all 80 FSRQs in our sample within the SDSS area. Those corresponding to the 55 FSRQs with evidence of optical/UV bump  are in Figure~\ref{fig:SEDs_FSRQs_SDSS_disc}. The SEDs of the remaining 25 FSRQs with no clear evidence of bump are shown in Figure~\ref{fig:SEDs_FSRQs_SDSS_NOdisc}.}

\begin{figure*} \centering
\subfloat{\includegraphics[width=0.32\textwidth,natwidth=610,natheight=642]{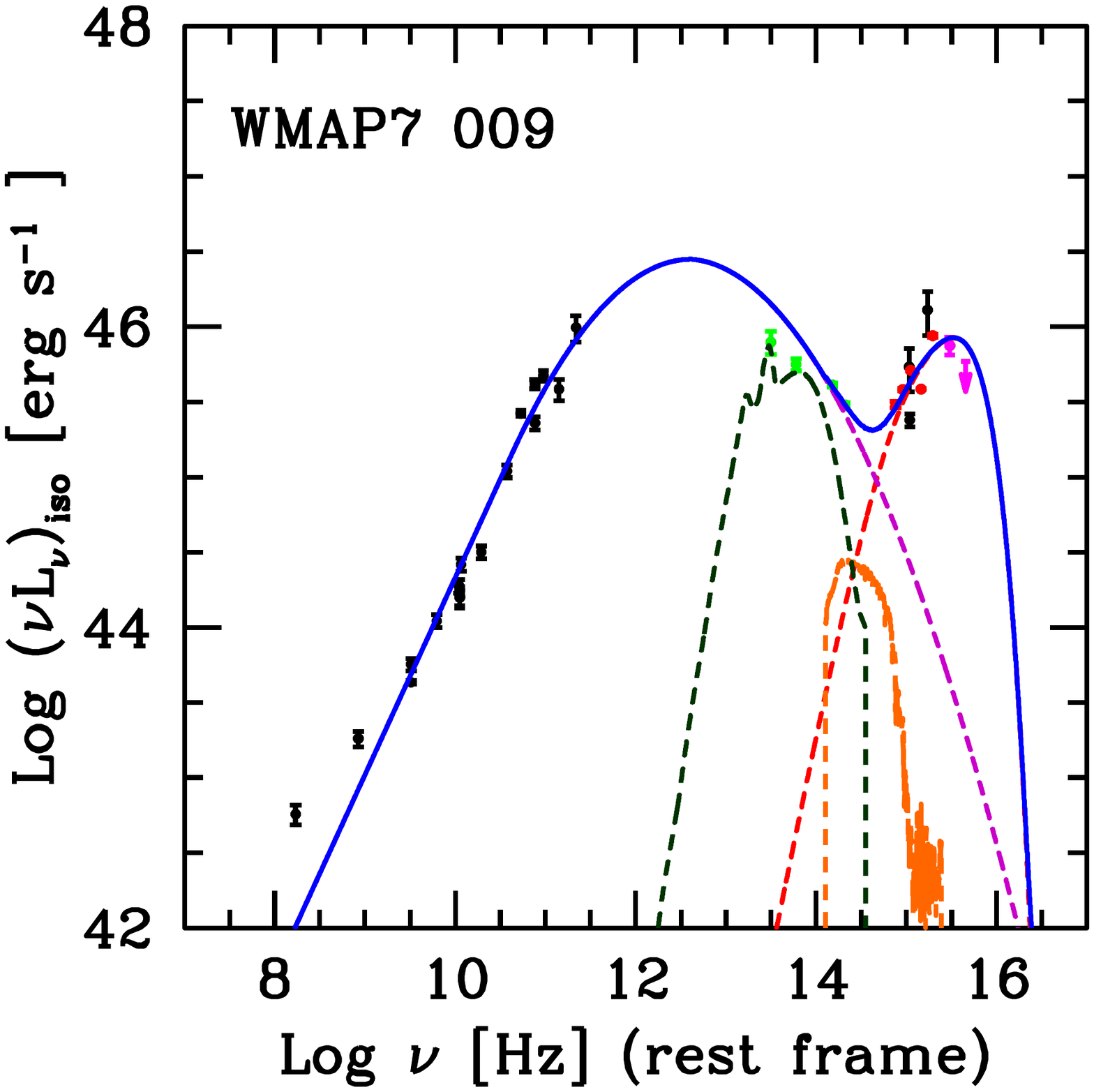}}
\subfloat{\includegraphics[width=0.32\textwidth,natwidth=610,natheight=642]{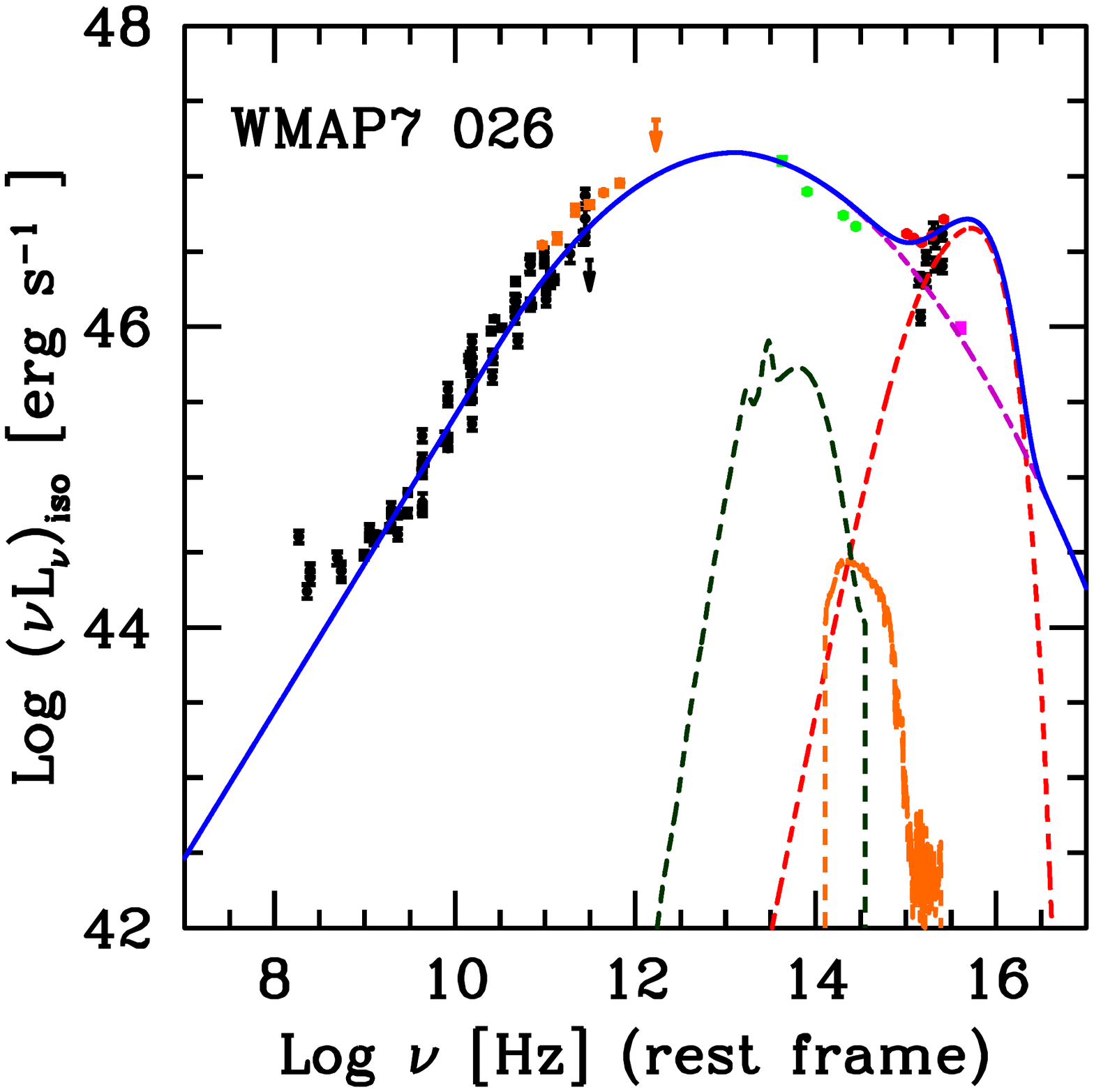}}
\subfloat{\includegraphics[width=0.32\textwidth,natwidth=610,natheight=642]{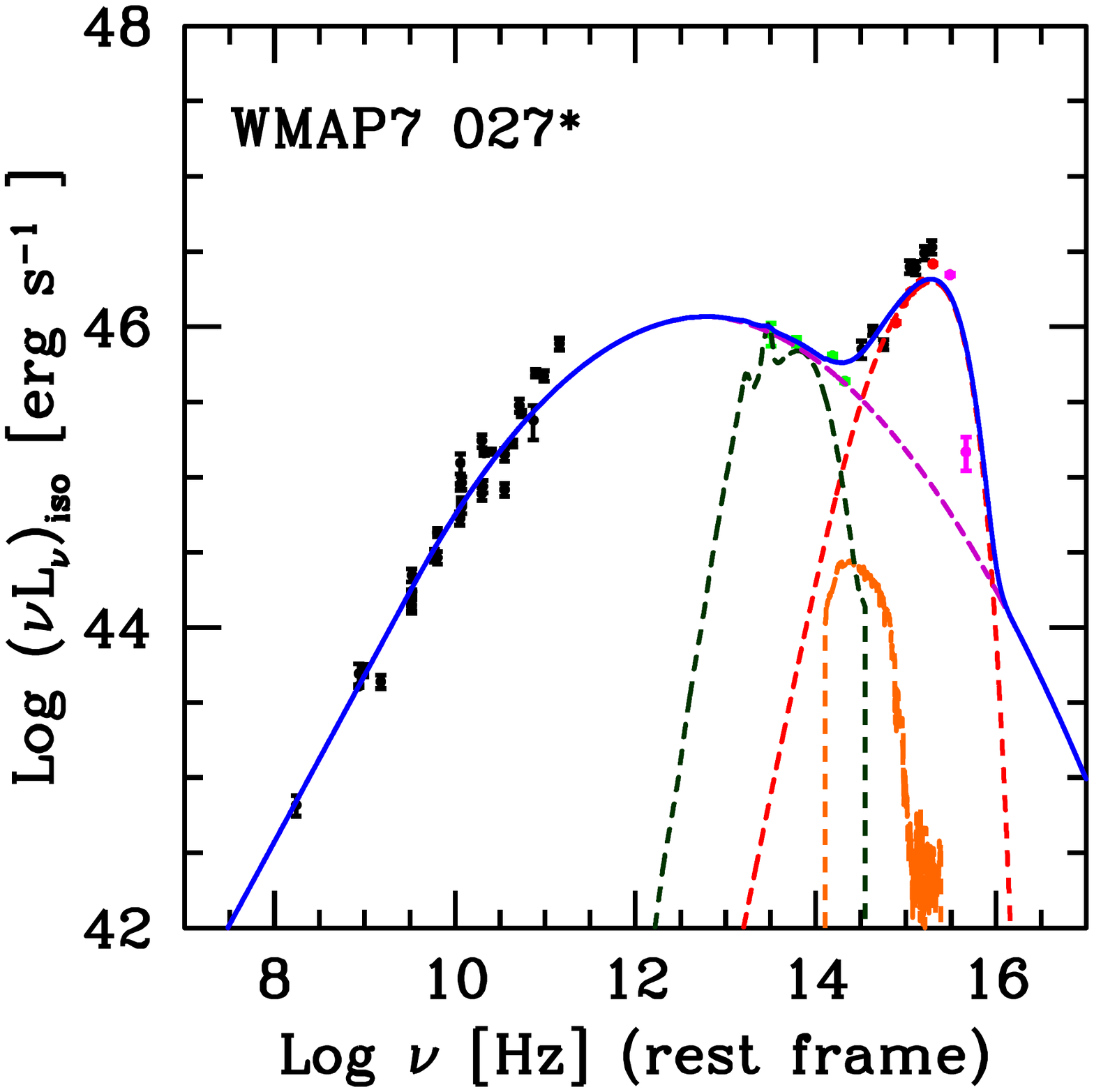}}\qquad
\subfloat{\includegraphics[width=0.32\textwidth,natwidth=610,natheight=642]{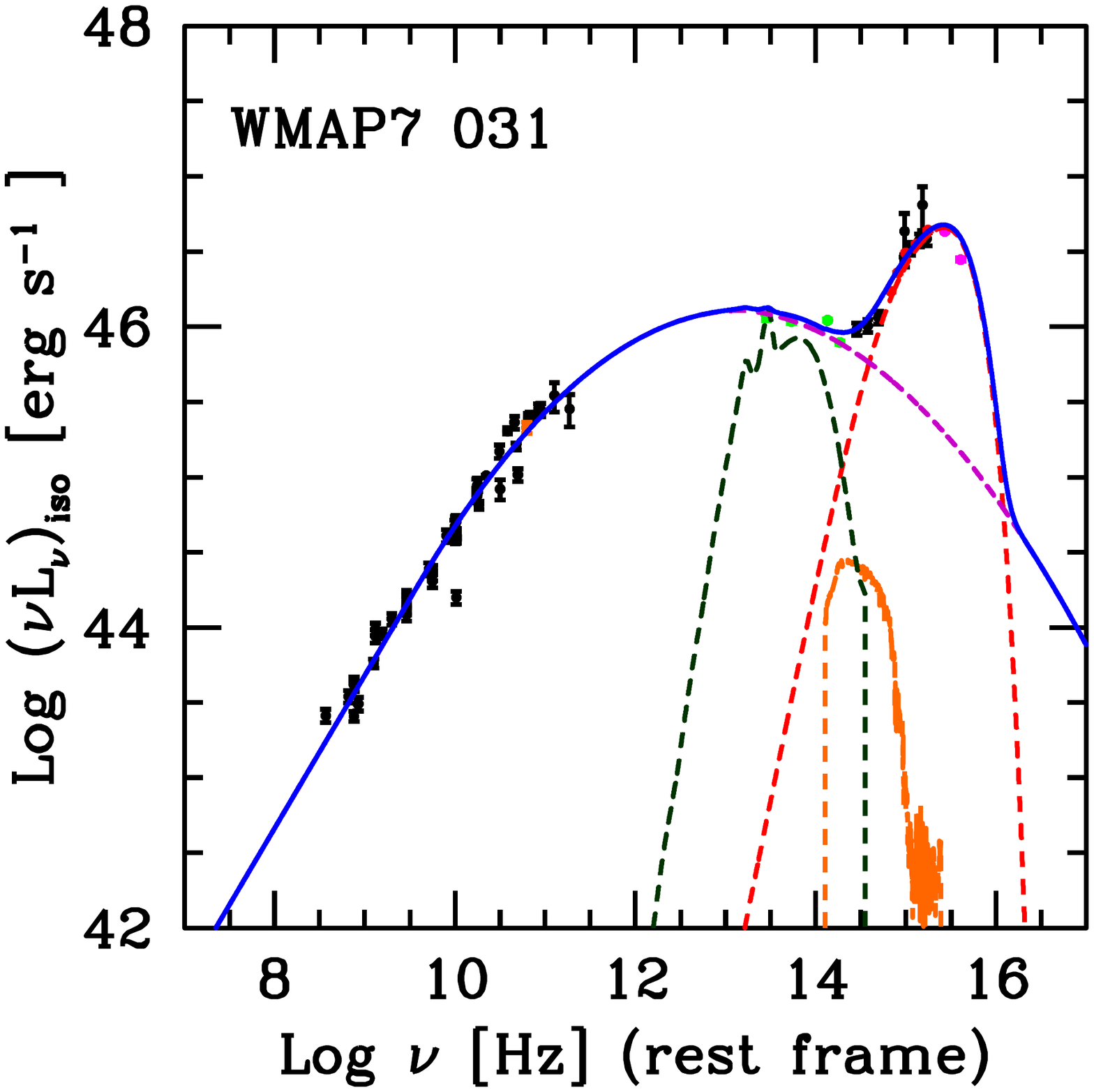}}
\subfloat{\includegraphics[width=0.32\textwidth,natwidth=610,natheight=642]{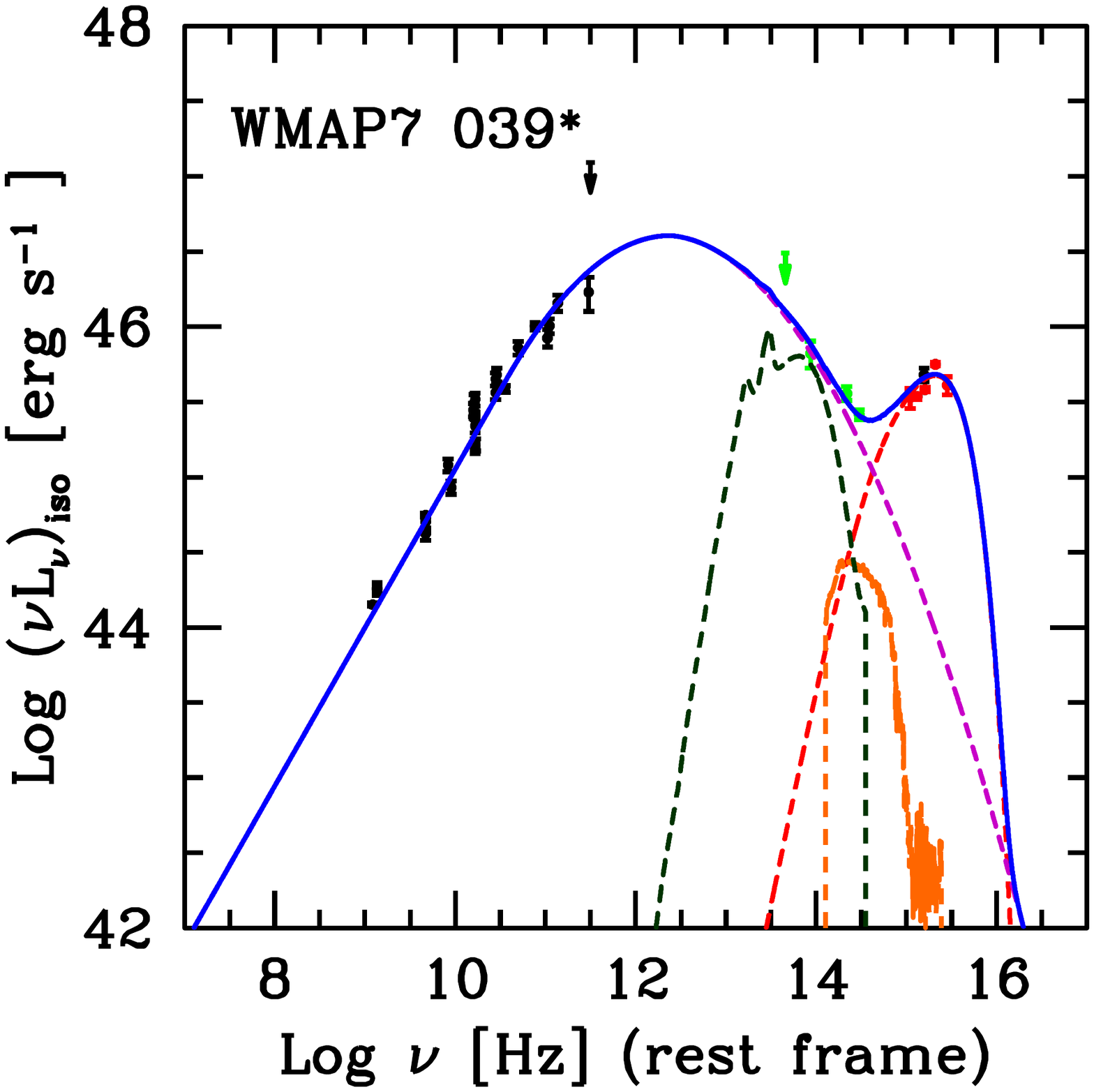}}
\subfloat{\includegraphics[width=0.32\textwidth,natwidth=610,natheight=642]{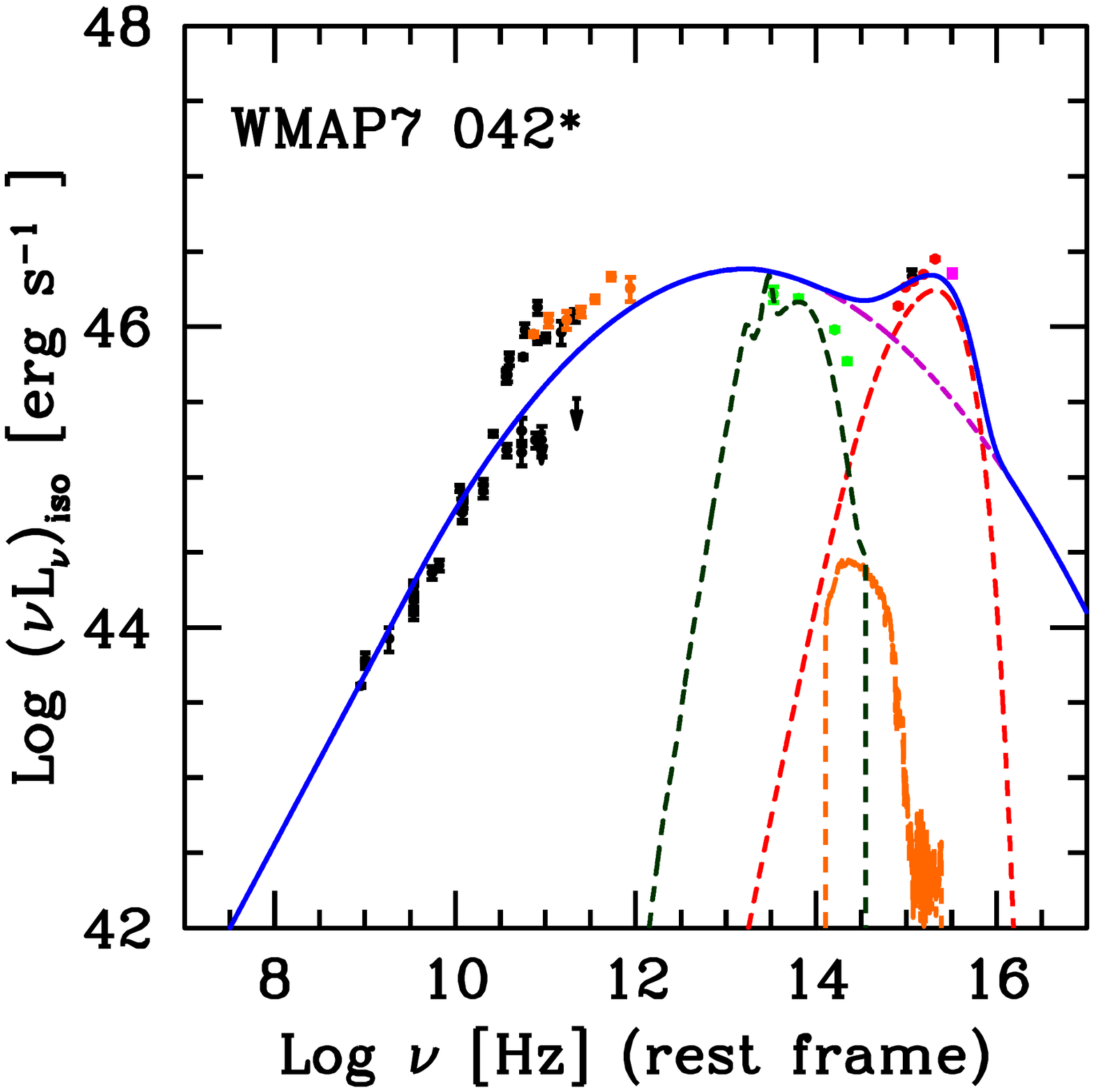}}\qquad
\caption{ SEDs of the 55 FSRQs in the sample with evidence of optical/UV bump. The meaning of the lines and of data points is the same as in Fig.~\ref{fig:SEDs_FSRQs_with_torus}. The 22 FSRQs for which the $\chi^2$ minimization procedure did not produce valid upper limits are flagged with an asterisk. For these sources upper limits to the torus luminosity were derived as described in the text and are reported in the SEDs. For WMAP7\#\;428, which shows tentative evidence of torus, we report both the best fit torus template and the upper limit such that the model does not exceed the mid-IR WISE measurements.}
\label{fig:SEDs_FSRQs_SDSS_disc}
\end{figure*}

\begin{figure*} \centering
\ContinuedFloat
\subfloat{\includegraphics[width=0.32\textwidth,natwidth=610,natheight=642]{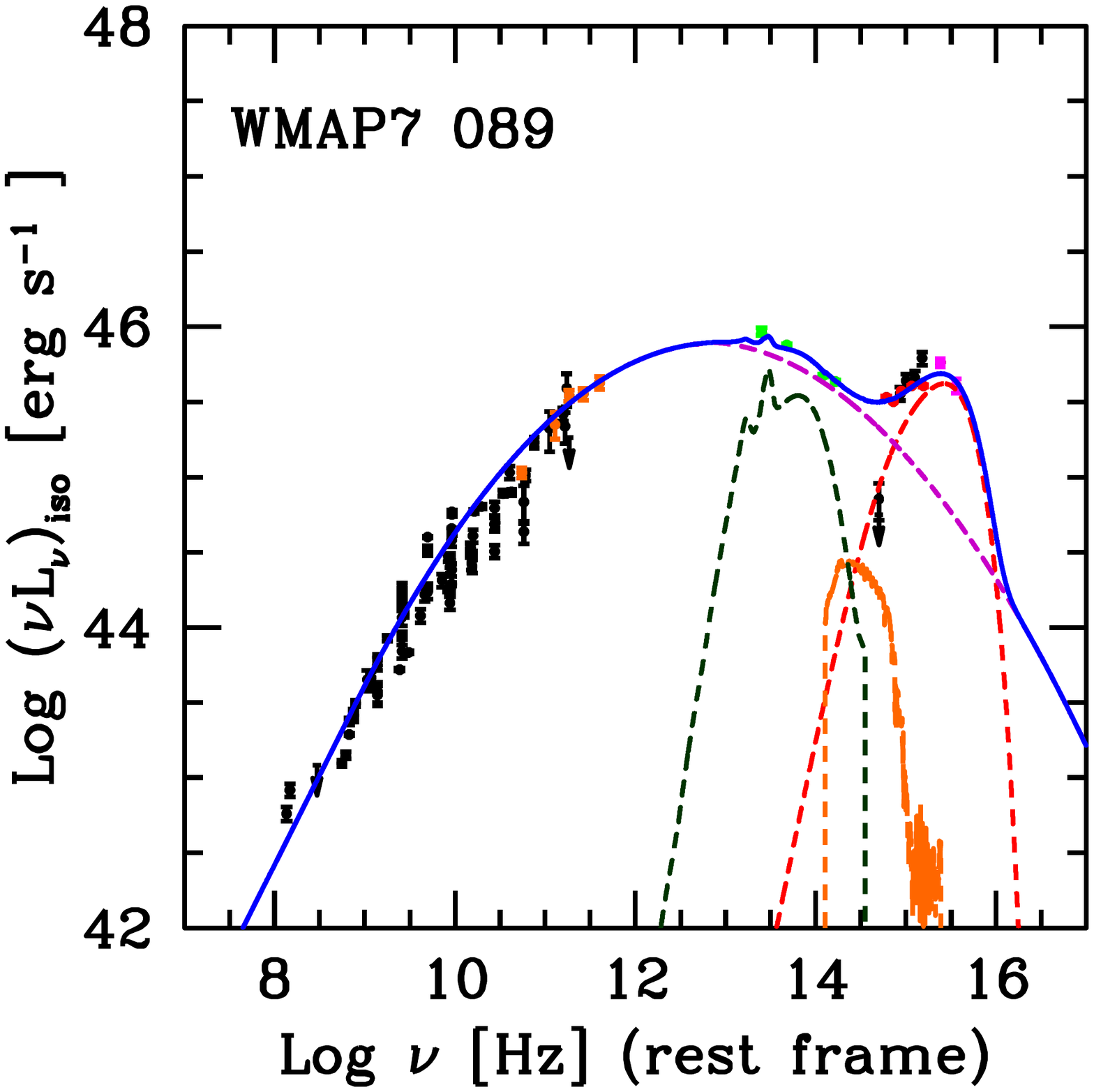}}
\subfloat{\includegraphics[width=0.32\textwidth,natwidth=610,natheight=642]{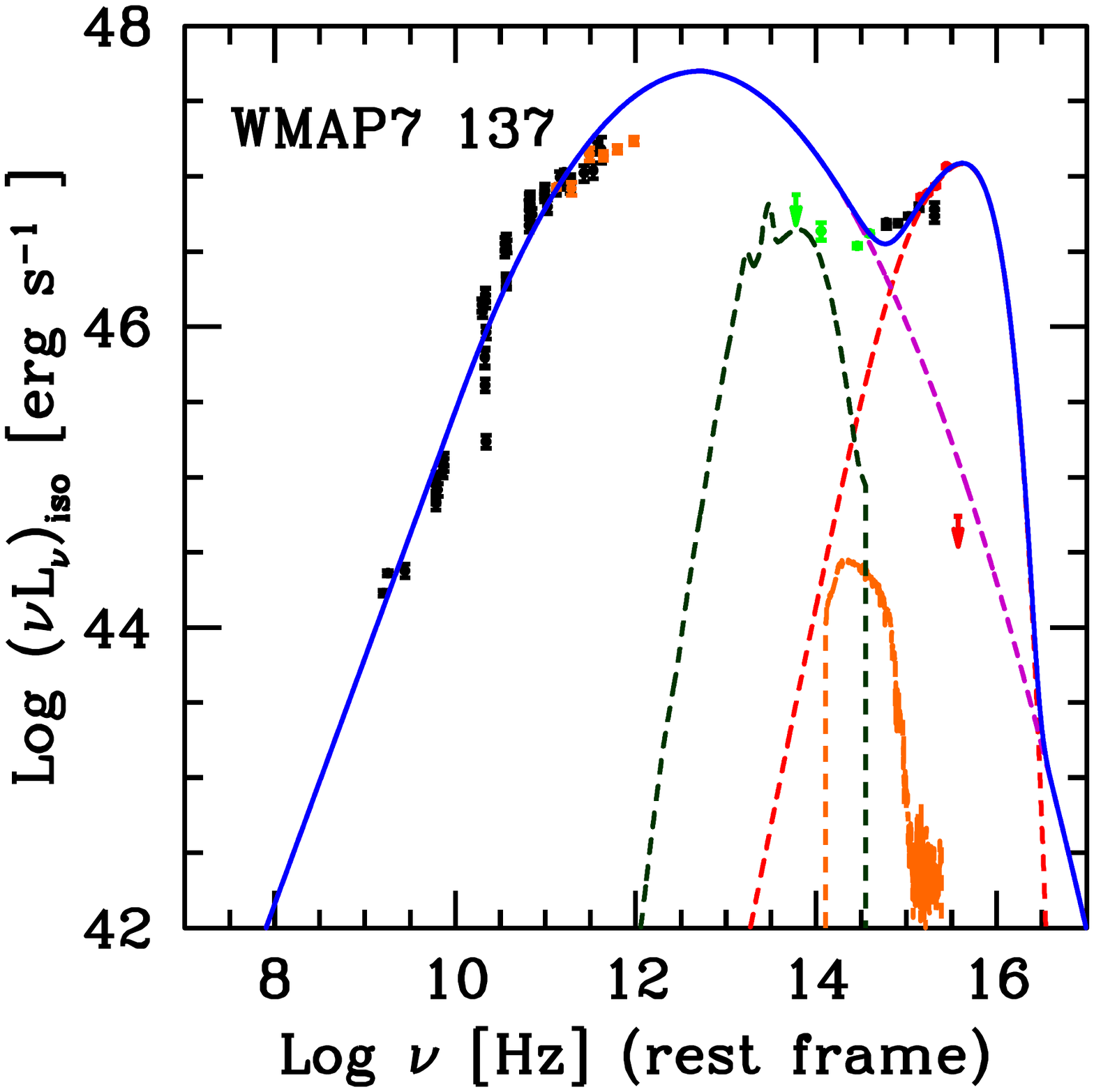}}
\subfloat{\includegraphics[width=0.32\textwidth,natwidth=610,natheight=642]{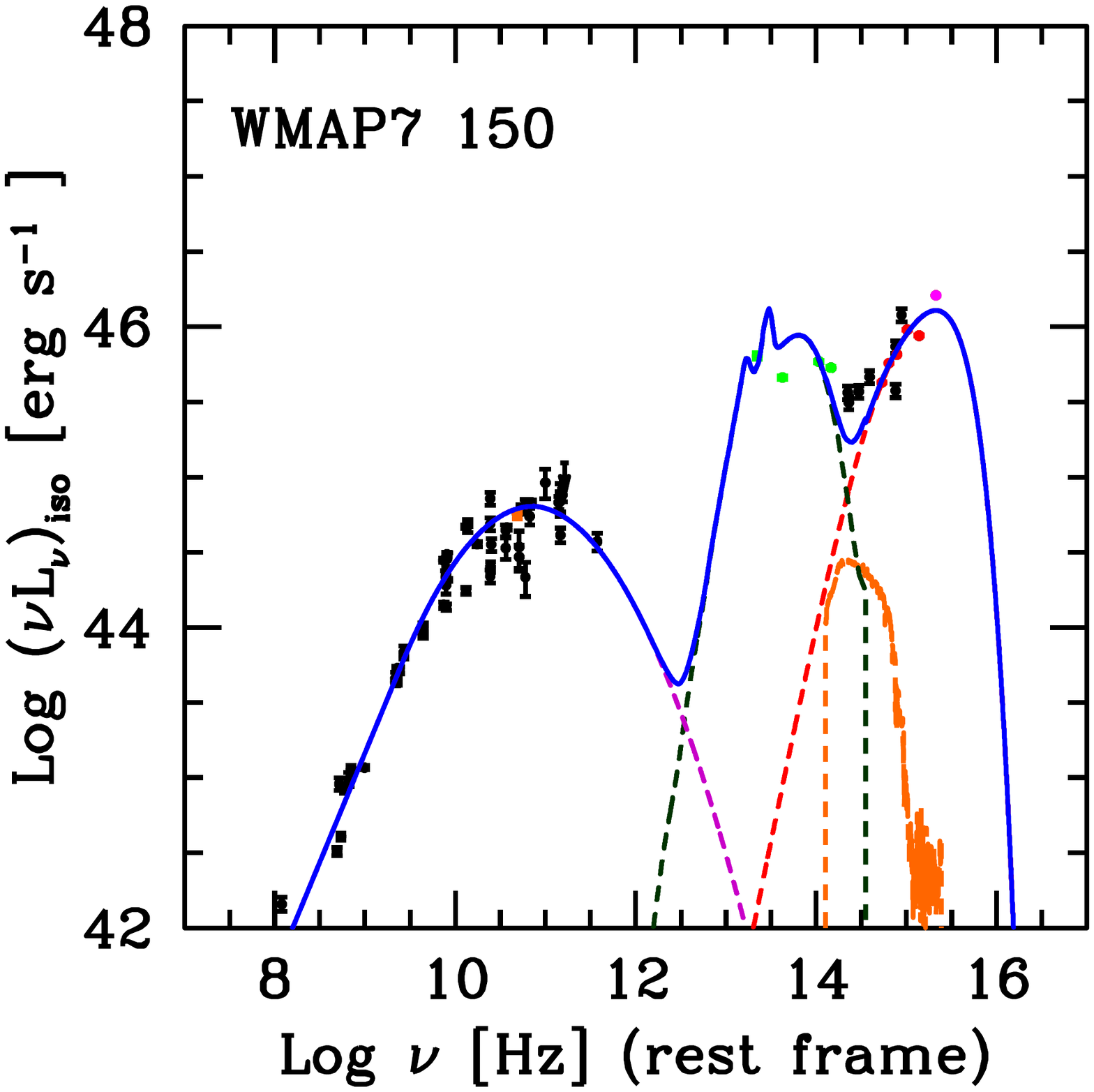}}\qquad
\subfloat{\includegraphics[width=0.32\textwidth,natwidth=610,natheight=642]{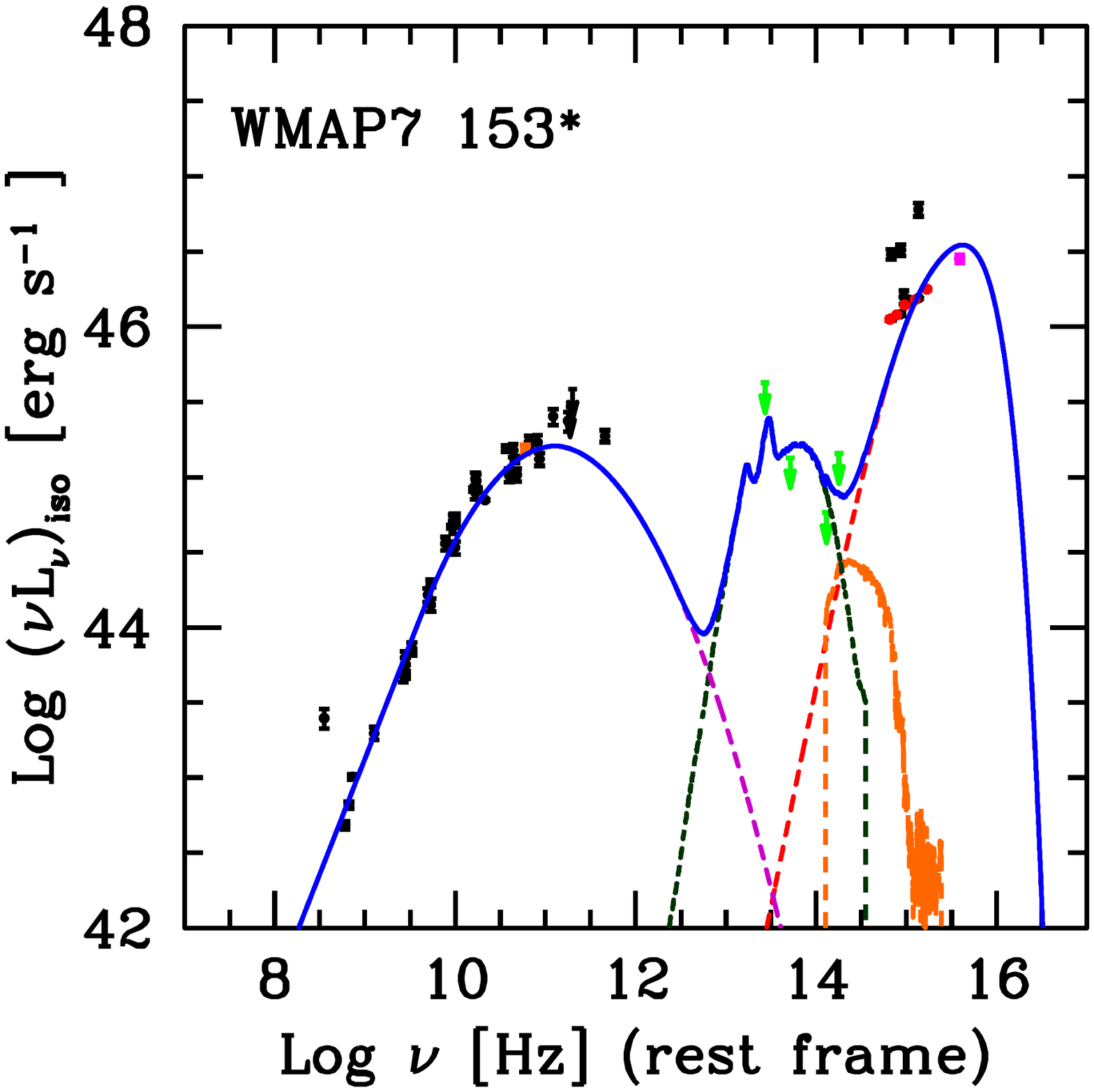}}
\subfloat{\includegraphics[width=0.32\textwidth,natwidth=610,natheight=642]{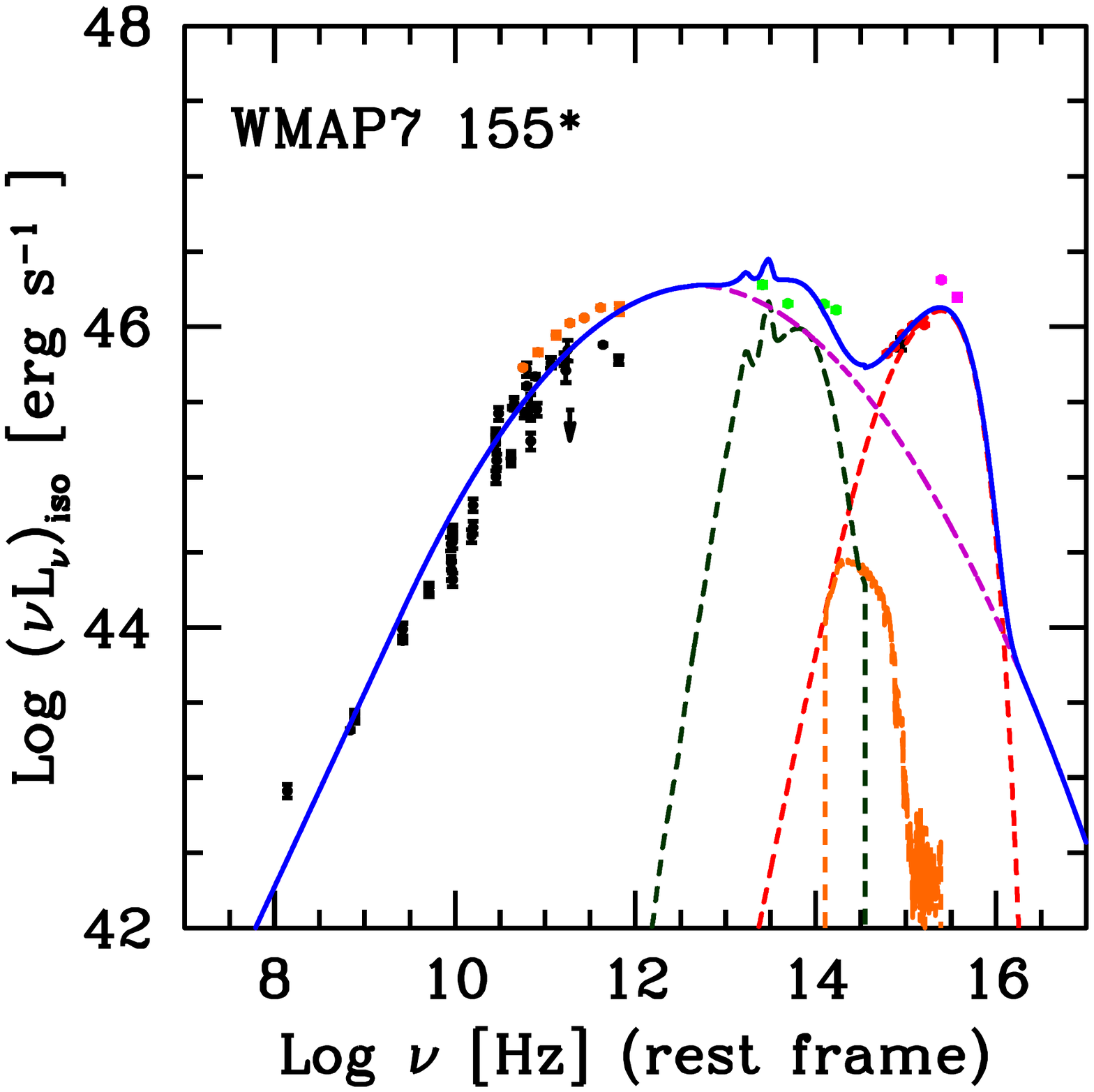}}
\subfloat{\includegraphics[width=0.32\textwidth,natwidth=610,natheight=642]{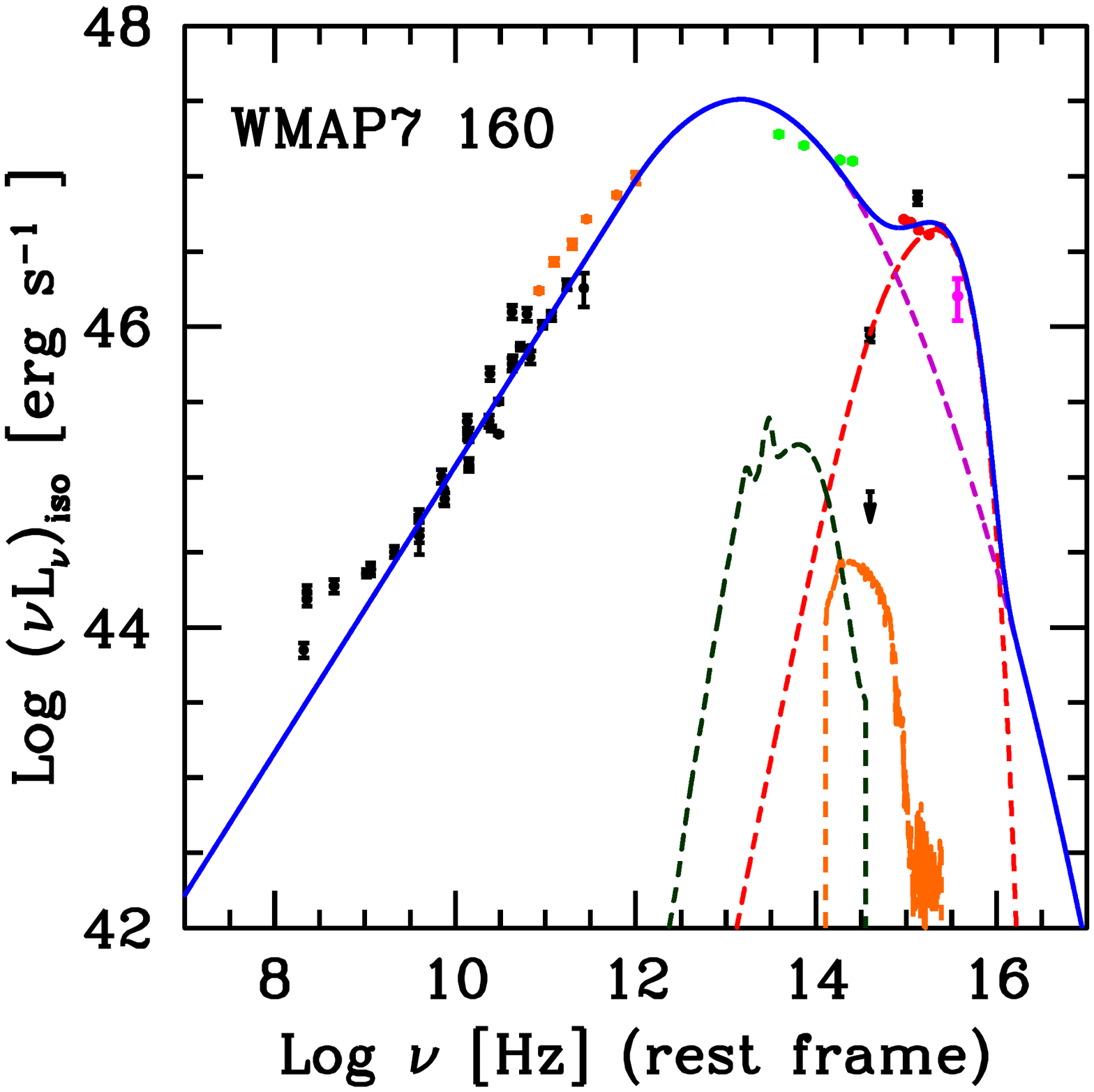}}\qquad
\subfloat{\includegraphics[width=0.32\textwidth,natwidth=610,natheight=642]{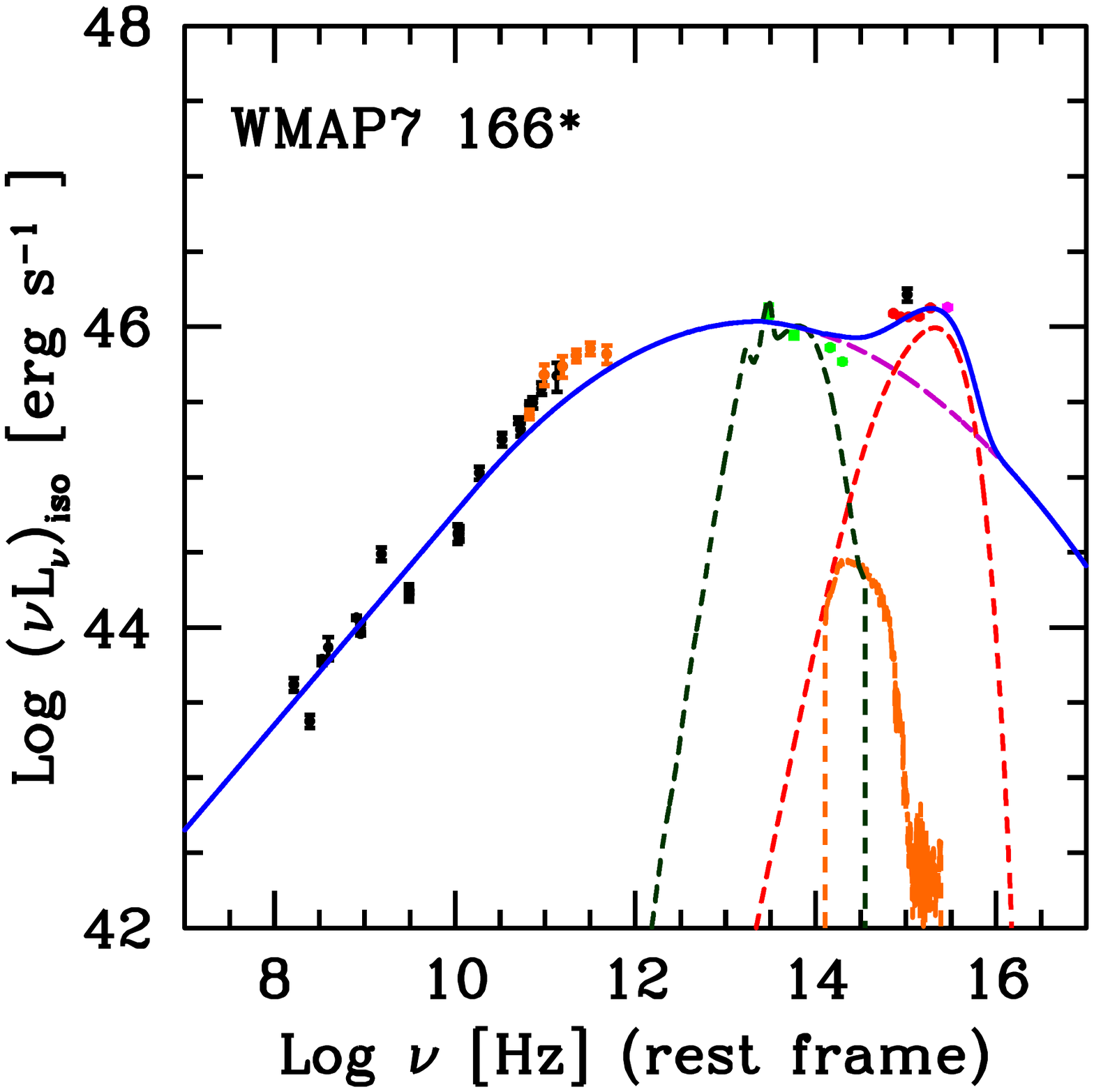}}
\subfloat{\includegraphics[width=0.32\textwidth,natwidth=610,natheight=642]{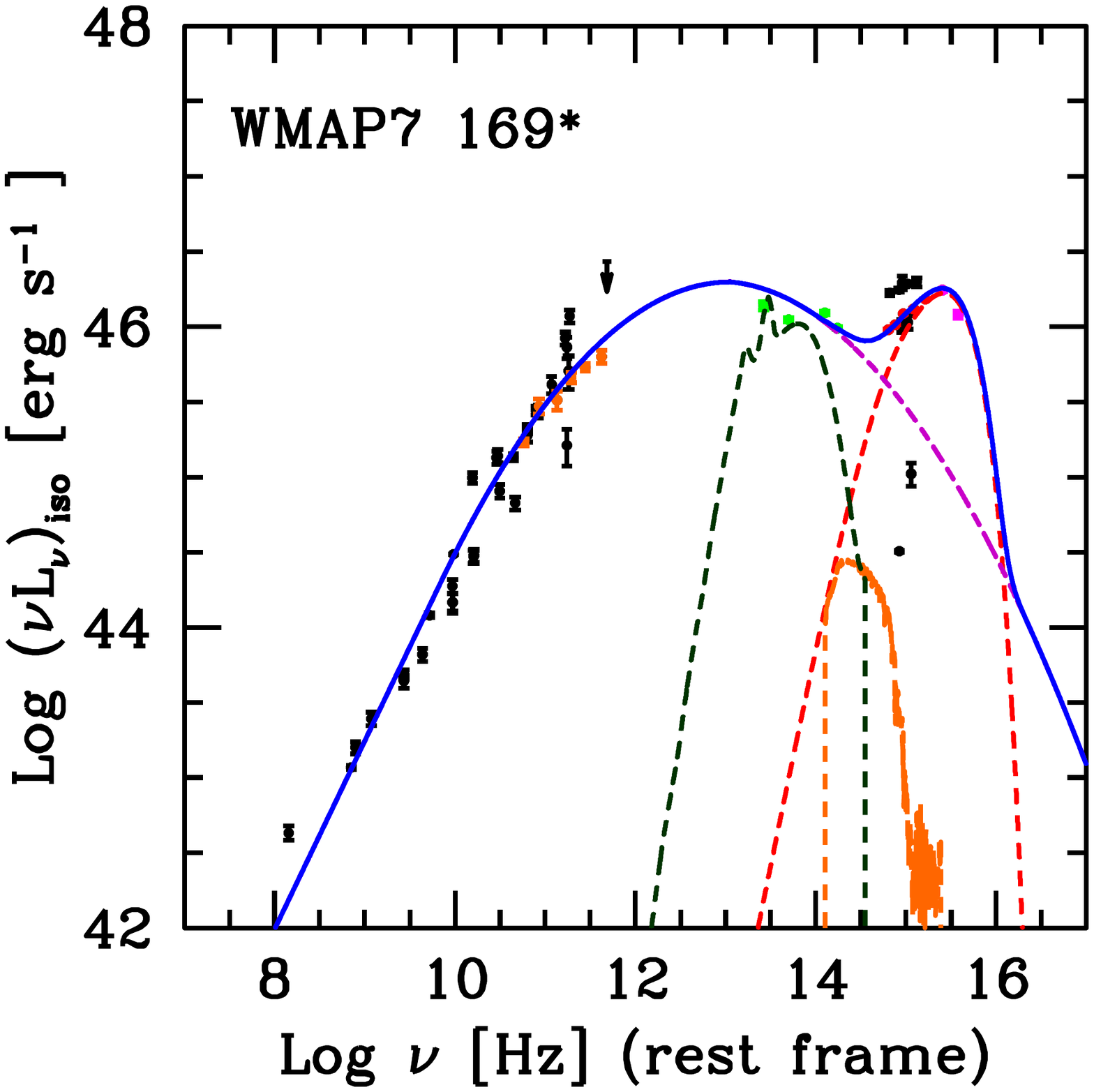}}
\subfloat{\includegraphics[width=0.32\textwidth,natwidth=610,natheight=642]{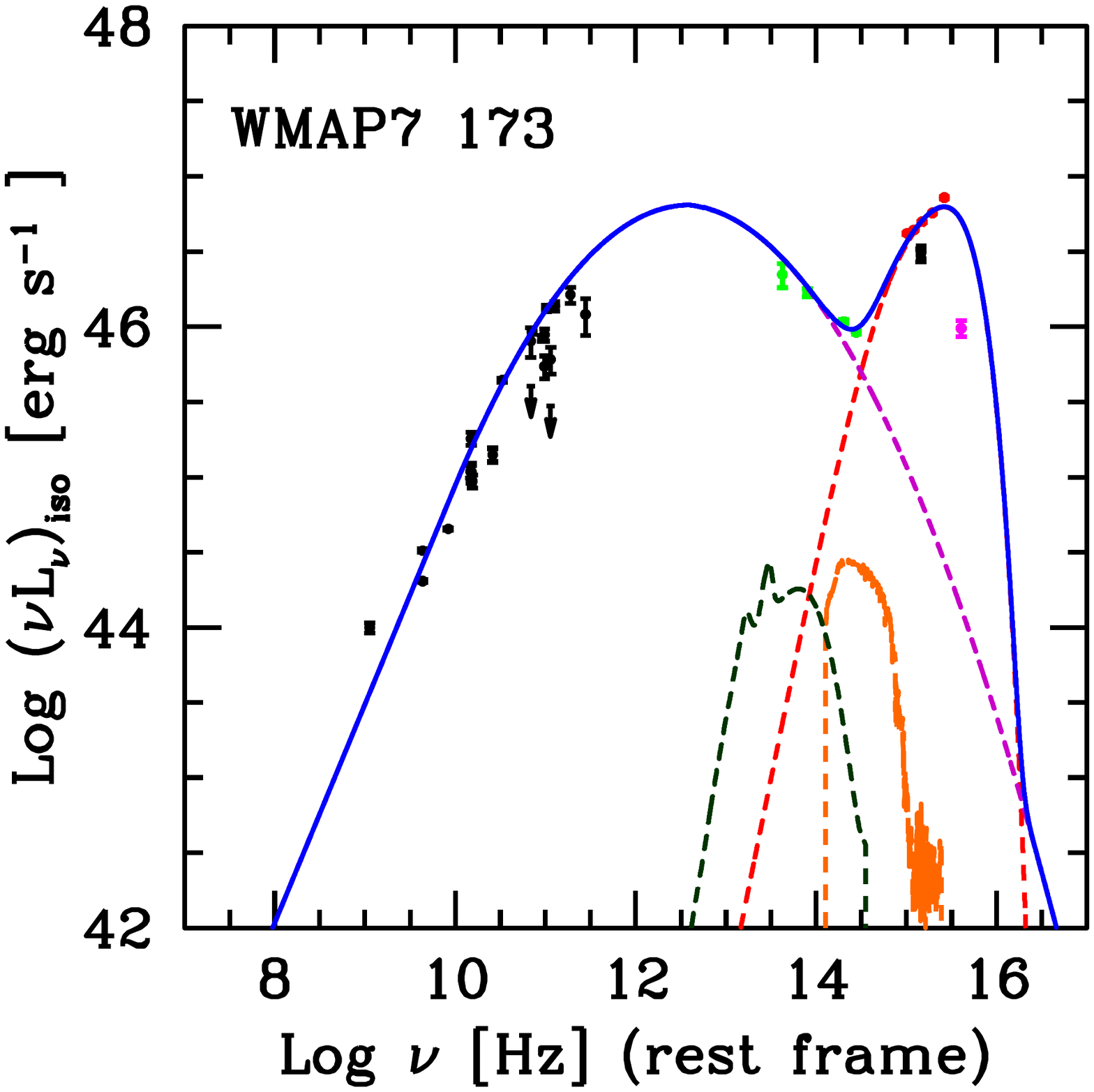}}\qquad
\caption{Continued.}
\end{figure*}

\begin{figure*} \centering
\ContinuedFloat
\subfloat{\includegraphics[width=0.32\textwidth,natwidth=610,natheight=642]{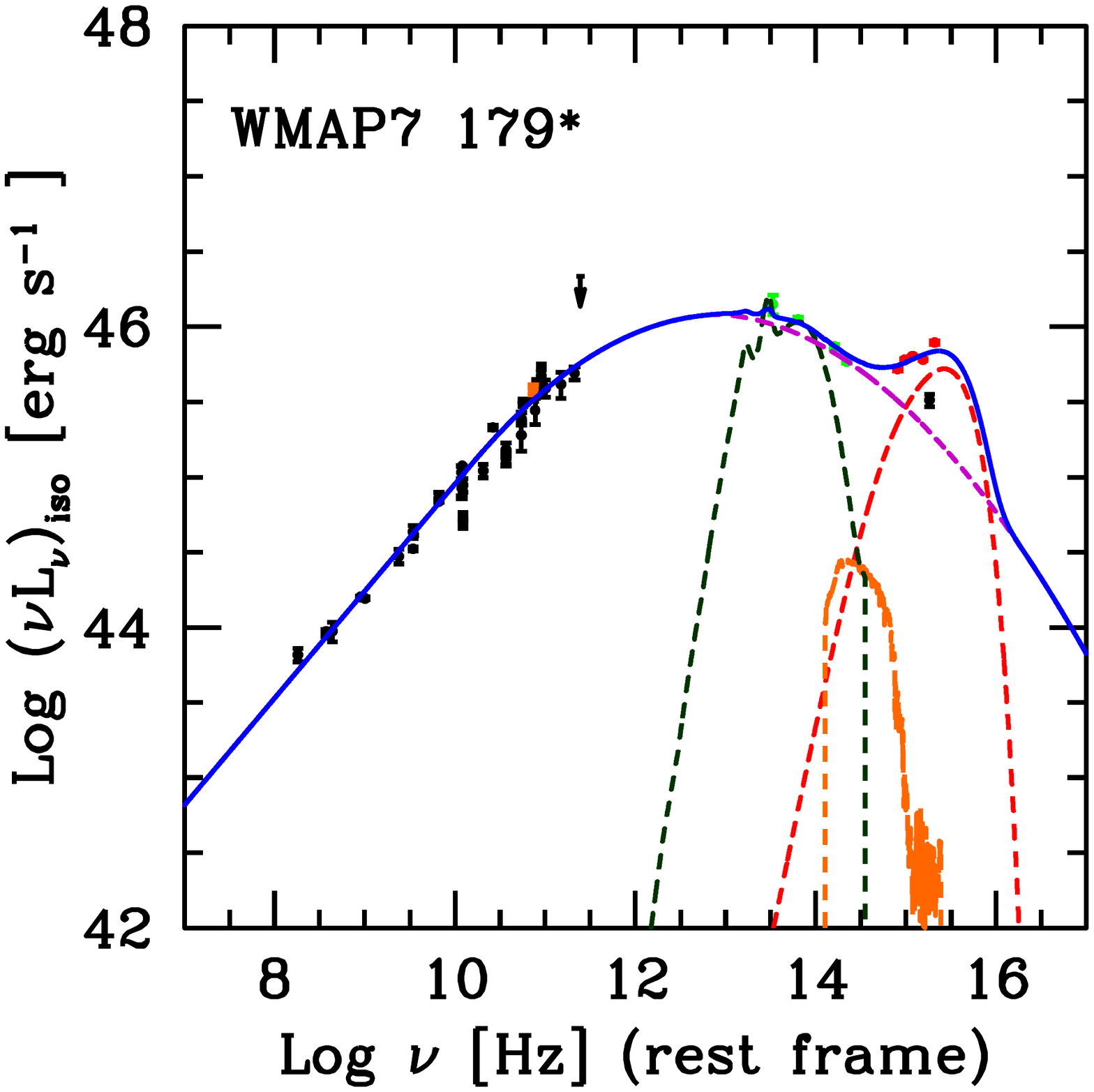}}
\subfloat{\includegraphics[width=0.32\textwidth,natwidth=610,natheight=642]{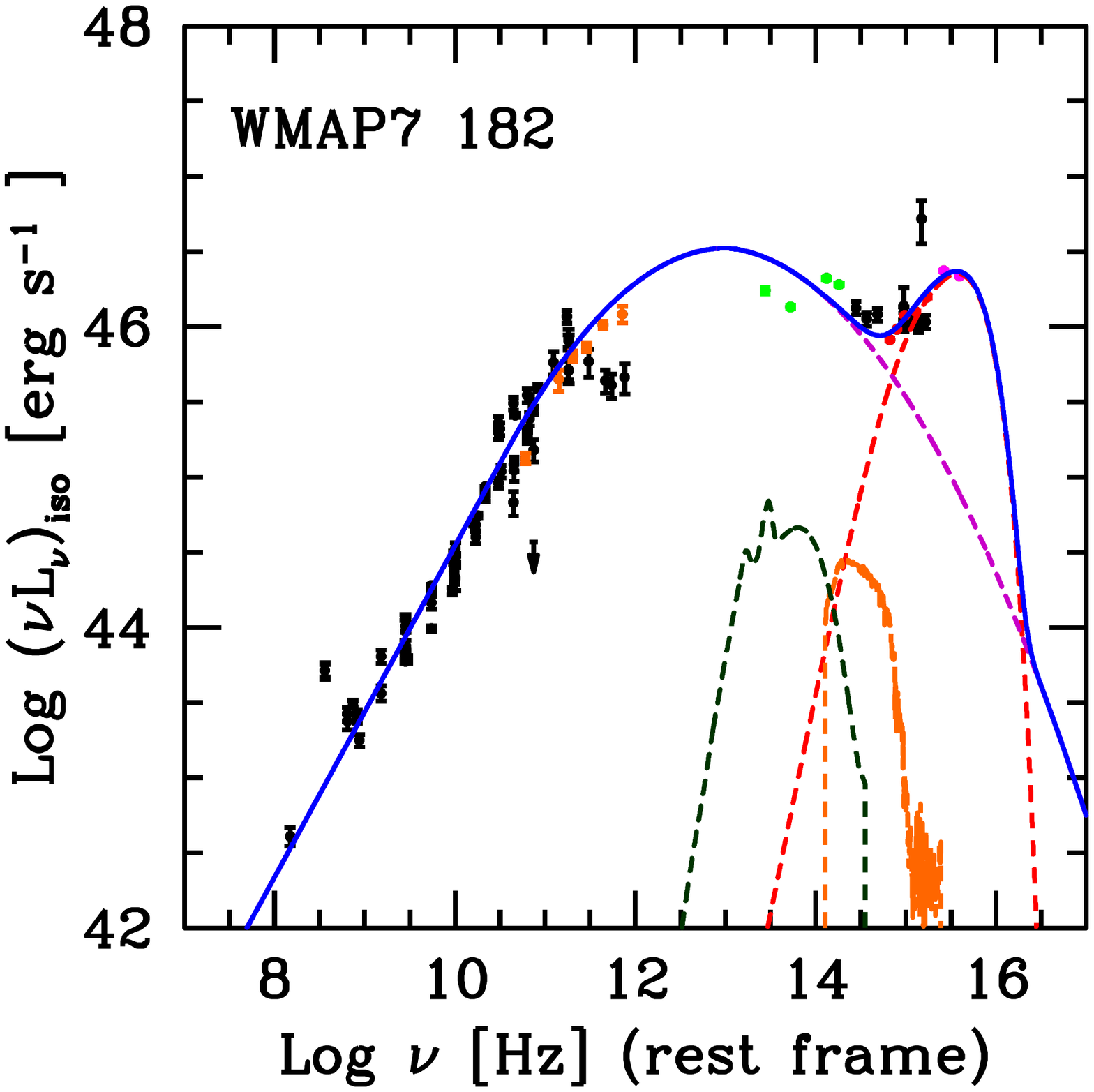}}
\subfloat{\includegraphics[width=0.32\textwidth,natwidth=610,natheight=642]{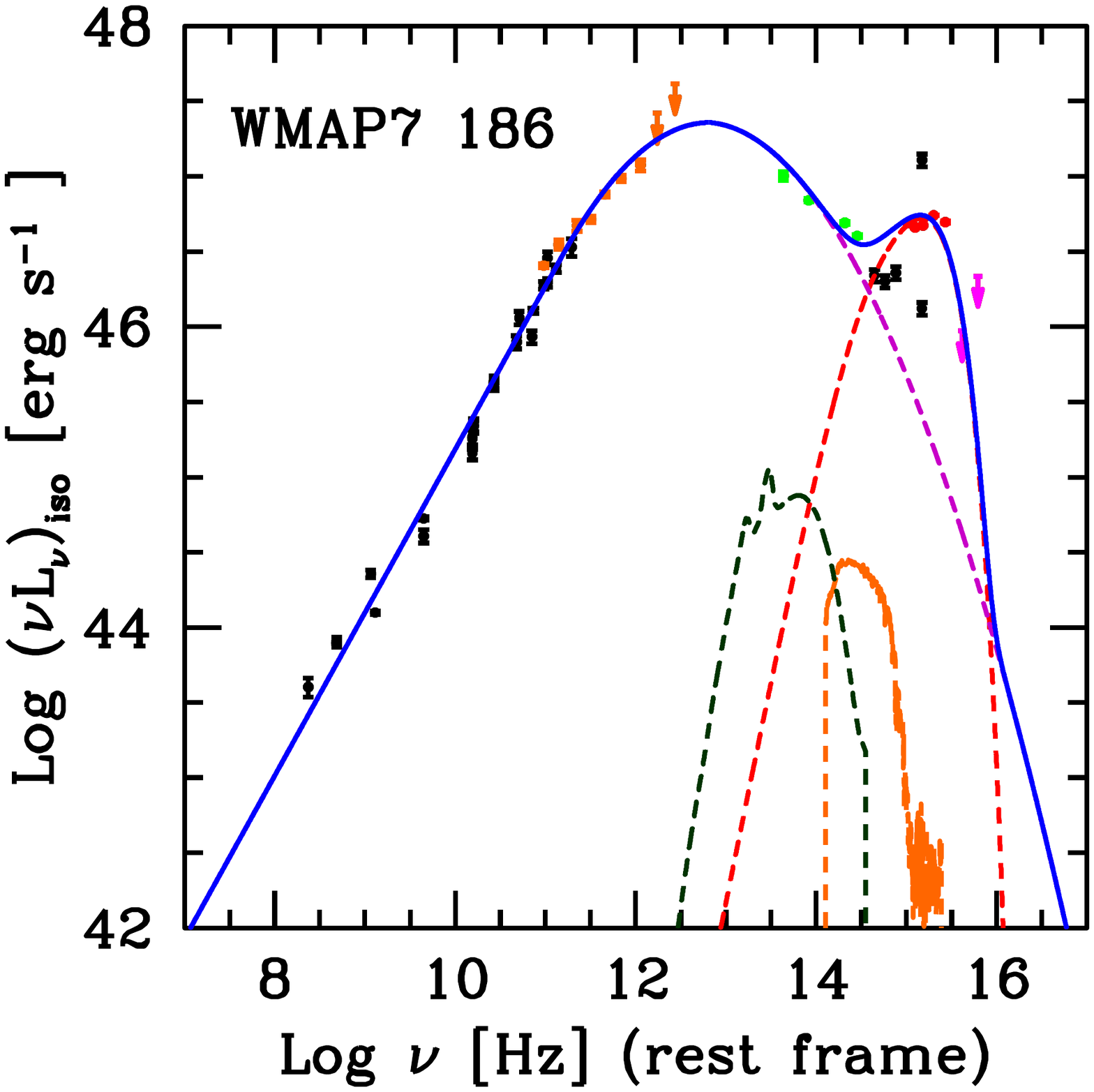}}\qquad
\subfloat{\includegraphics[width=0.32\textwidth,natwidth=610,natheight=642]{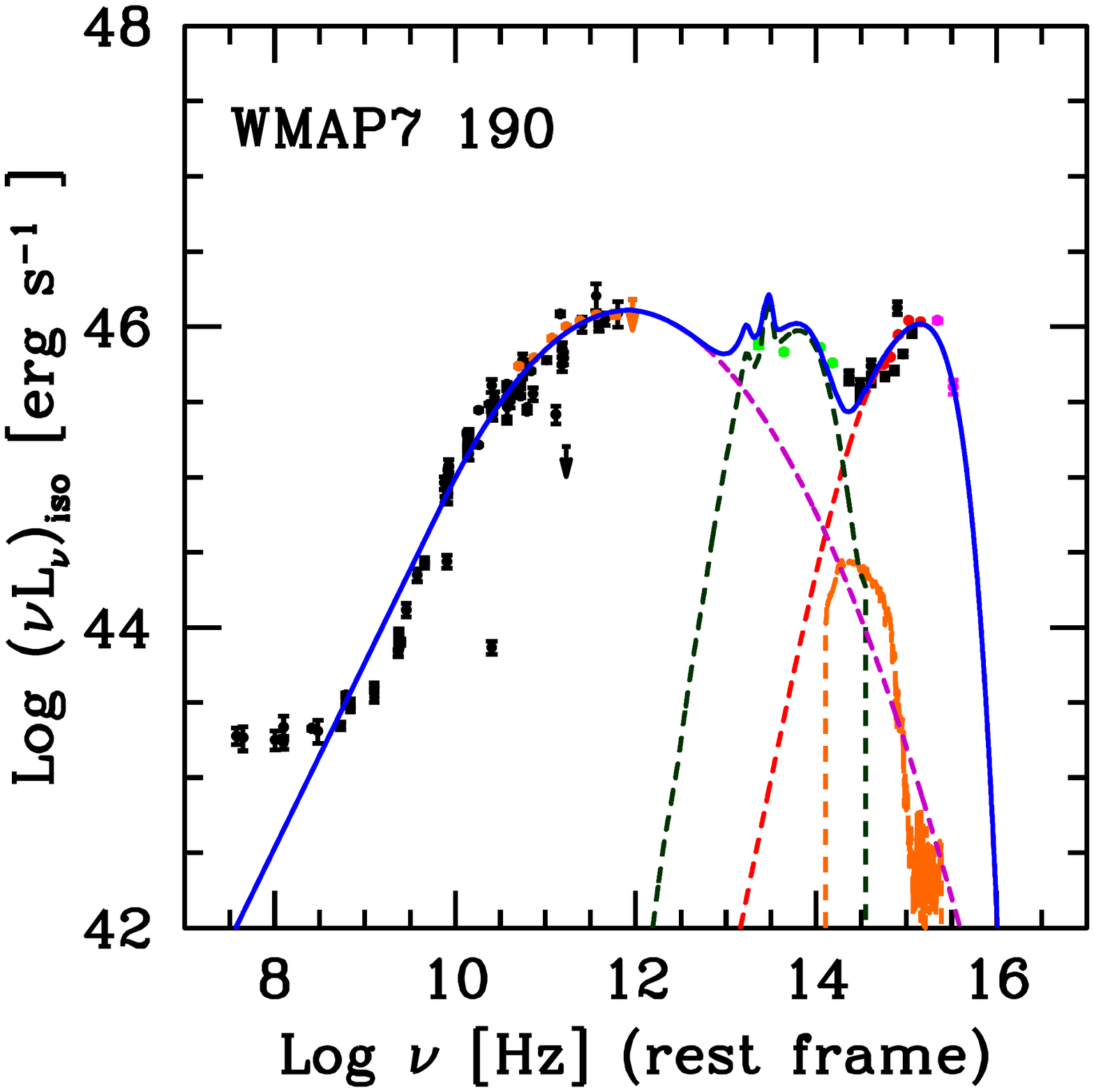}}
\subfloat{\includegraphics[width=0.32\textwidth,natwidth=610,natheight=642]{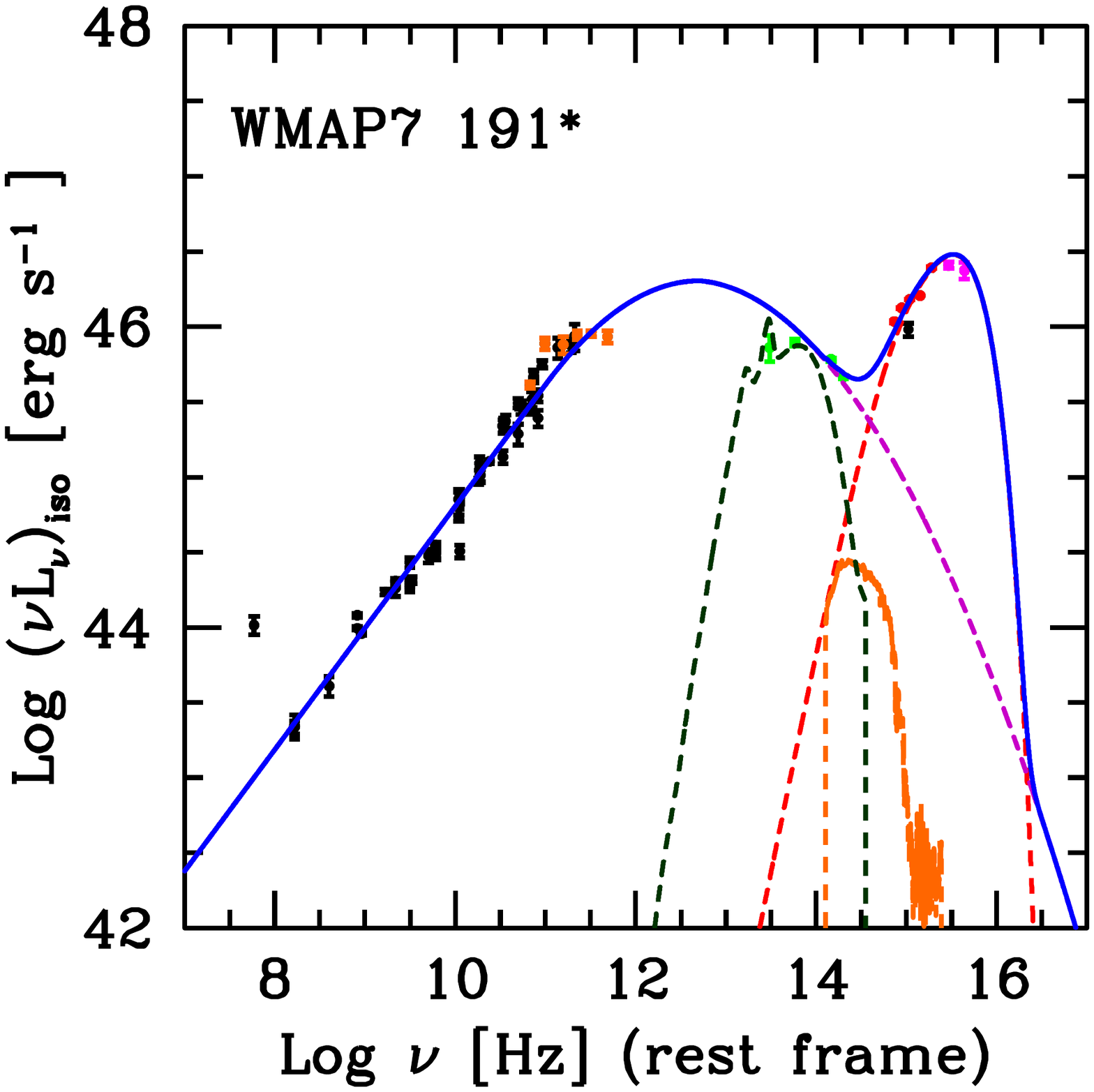}}
\subfloat{\includegraphics[width=0.32\textwidth,natwidth=610,natheight=642]{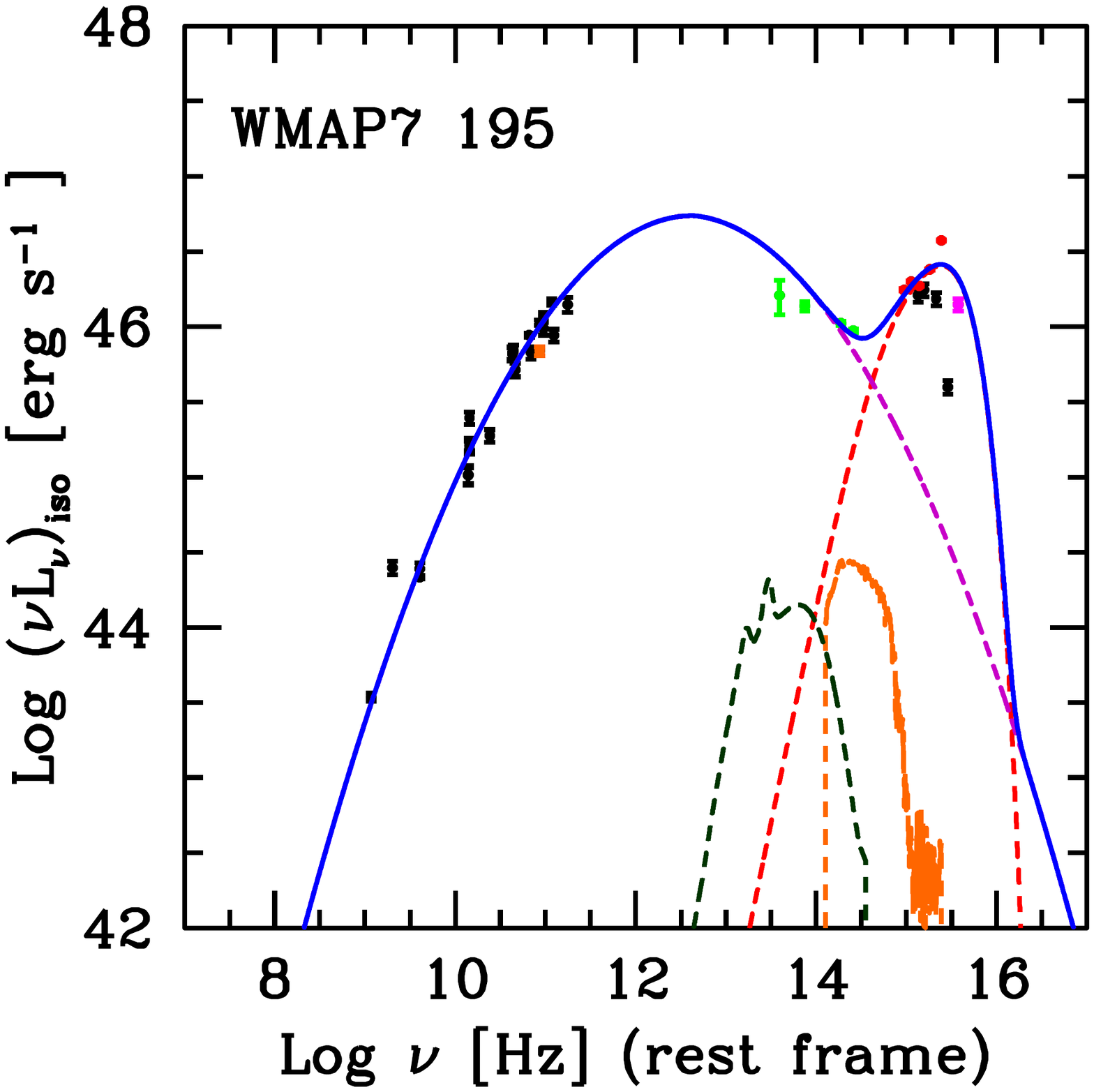}}\qquad
\subfloat{\includegraphics[width=0.32\textwidth,natwidth=610,natheight=642]{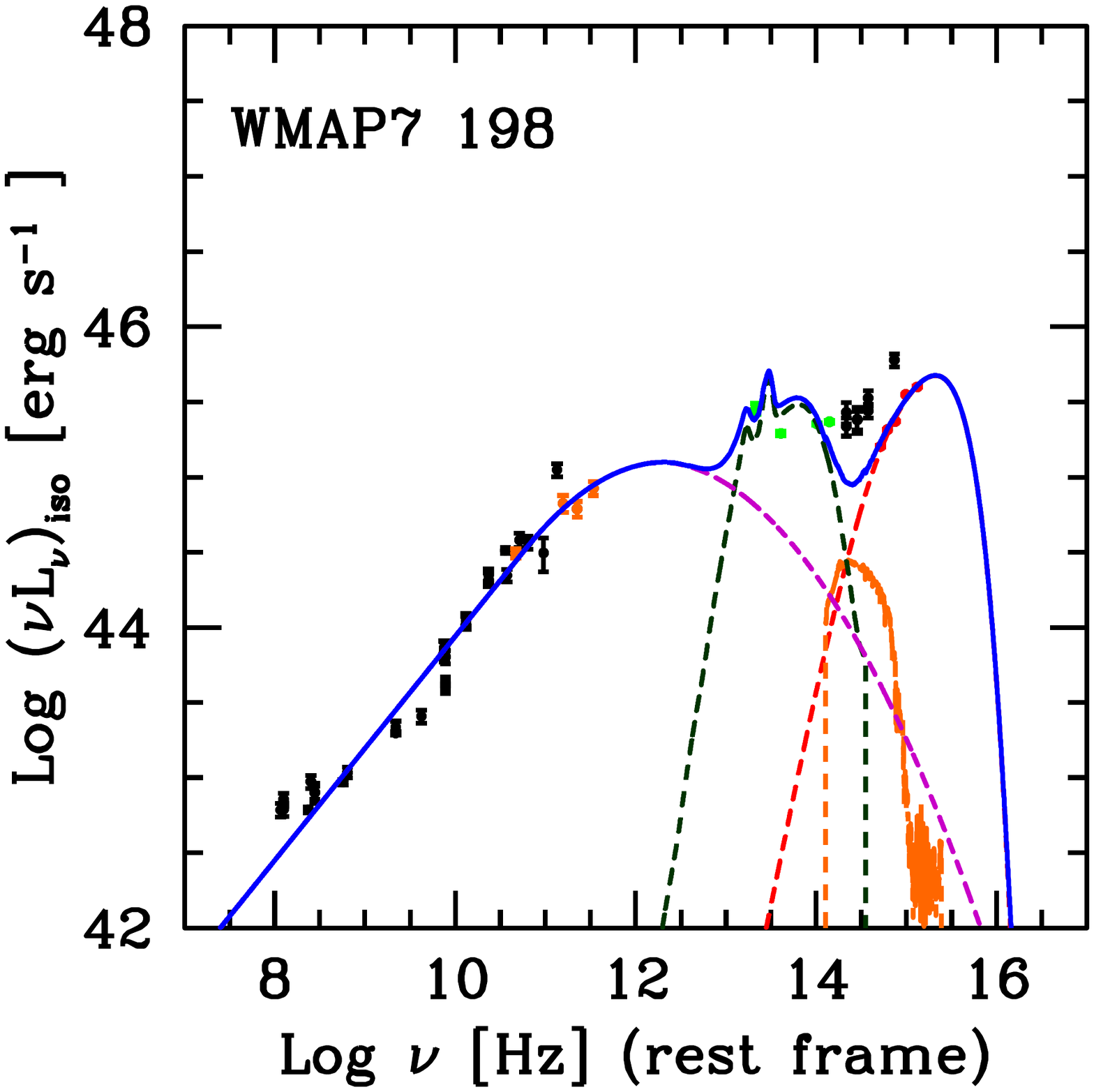}}
\subfloat{\includegraphics[width=0.32\textwidth,natwidth=610,natheight=642]{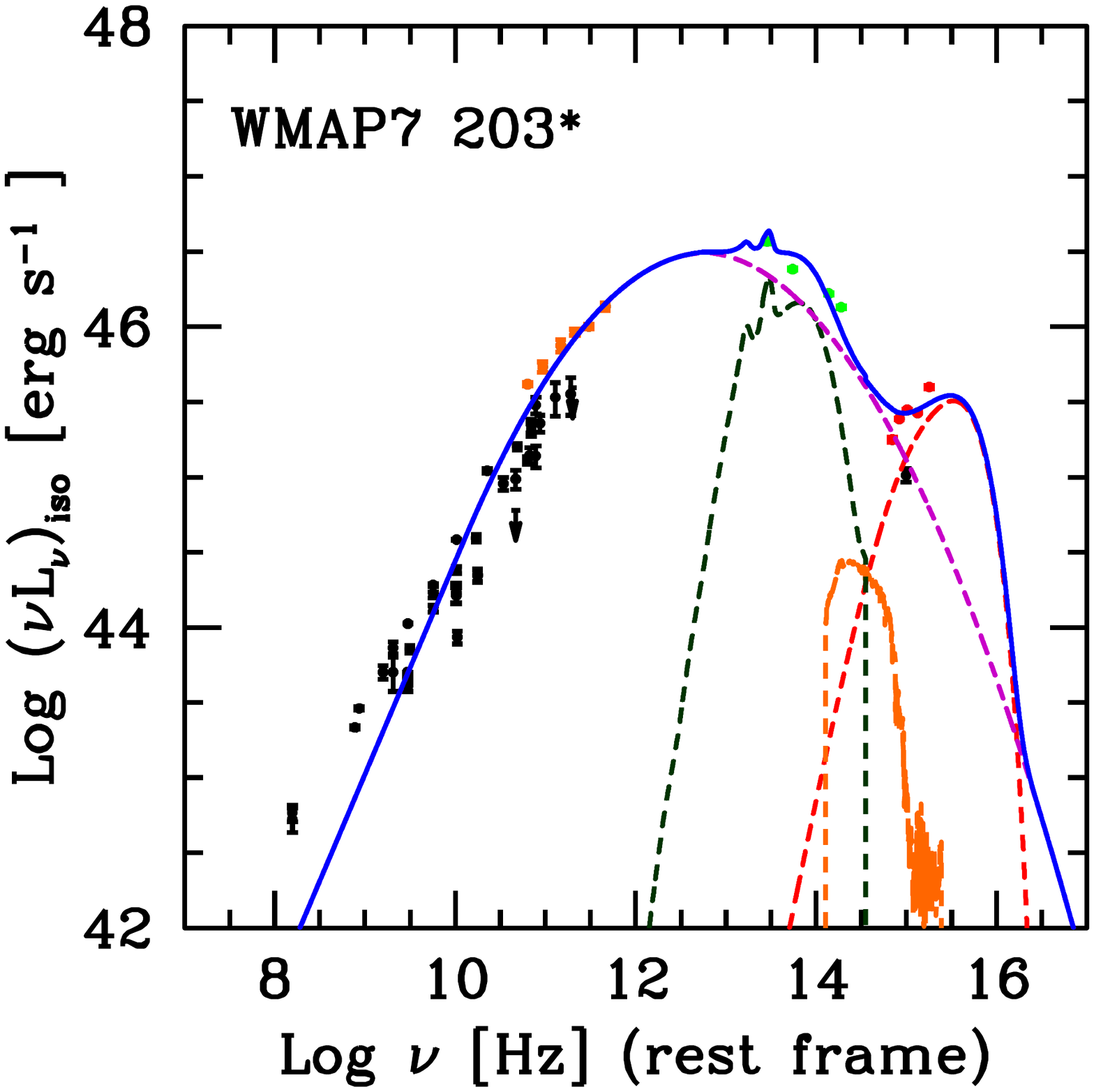}}
\subfloat{\includegraphics[width=0.32\textwidth,natwidth=610,natheight=642]{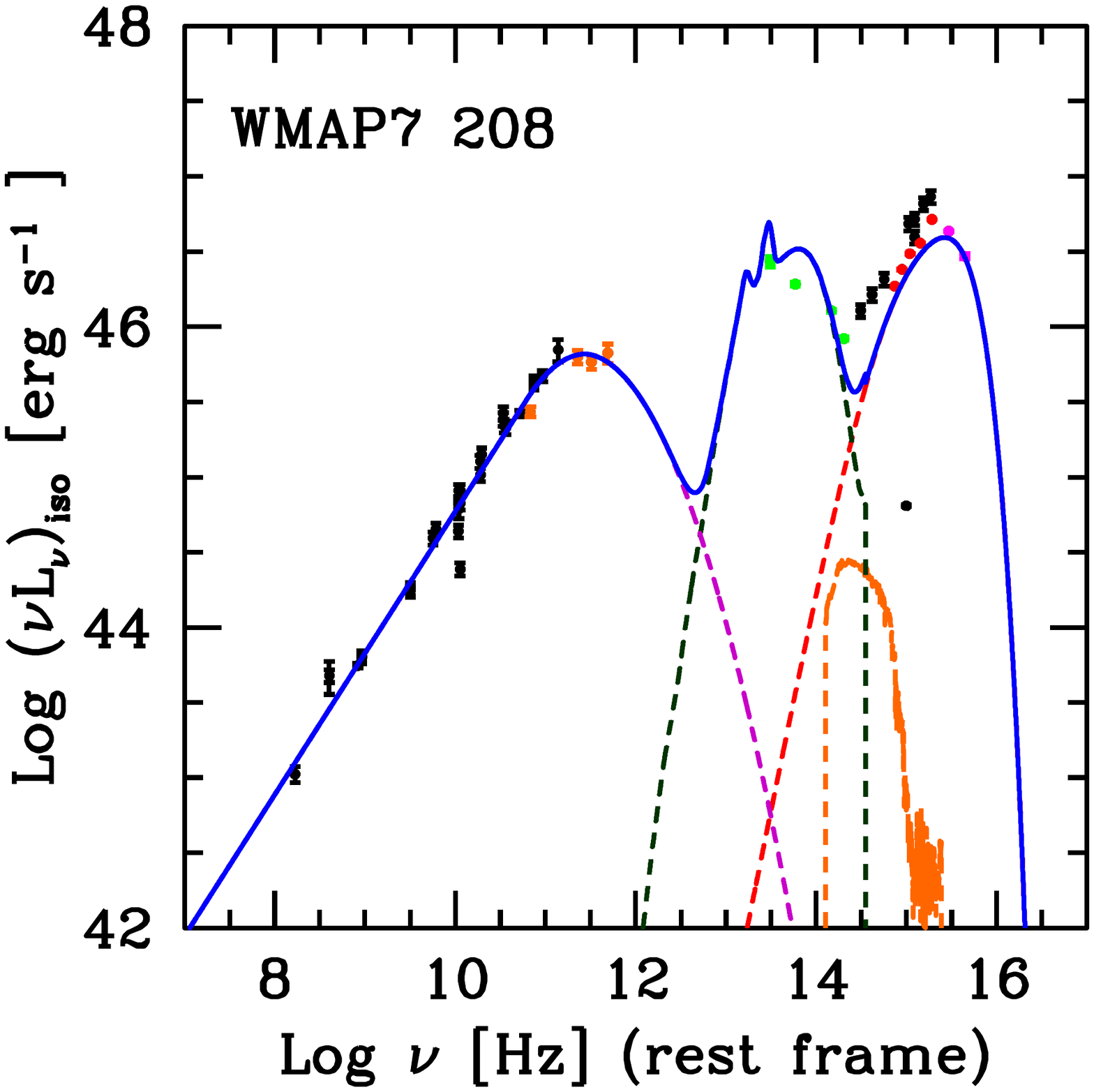}}\qquad
\caption{Continued.}
\end{figure*}

\begin{figure*} \centering
\ContinuedFloat
\subfloat{\includegraphics[width=0.32\textwidth,natwidth=610,natheight=642]{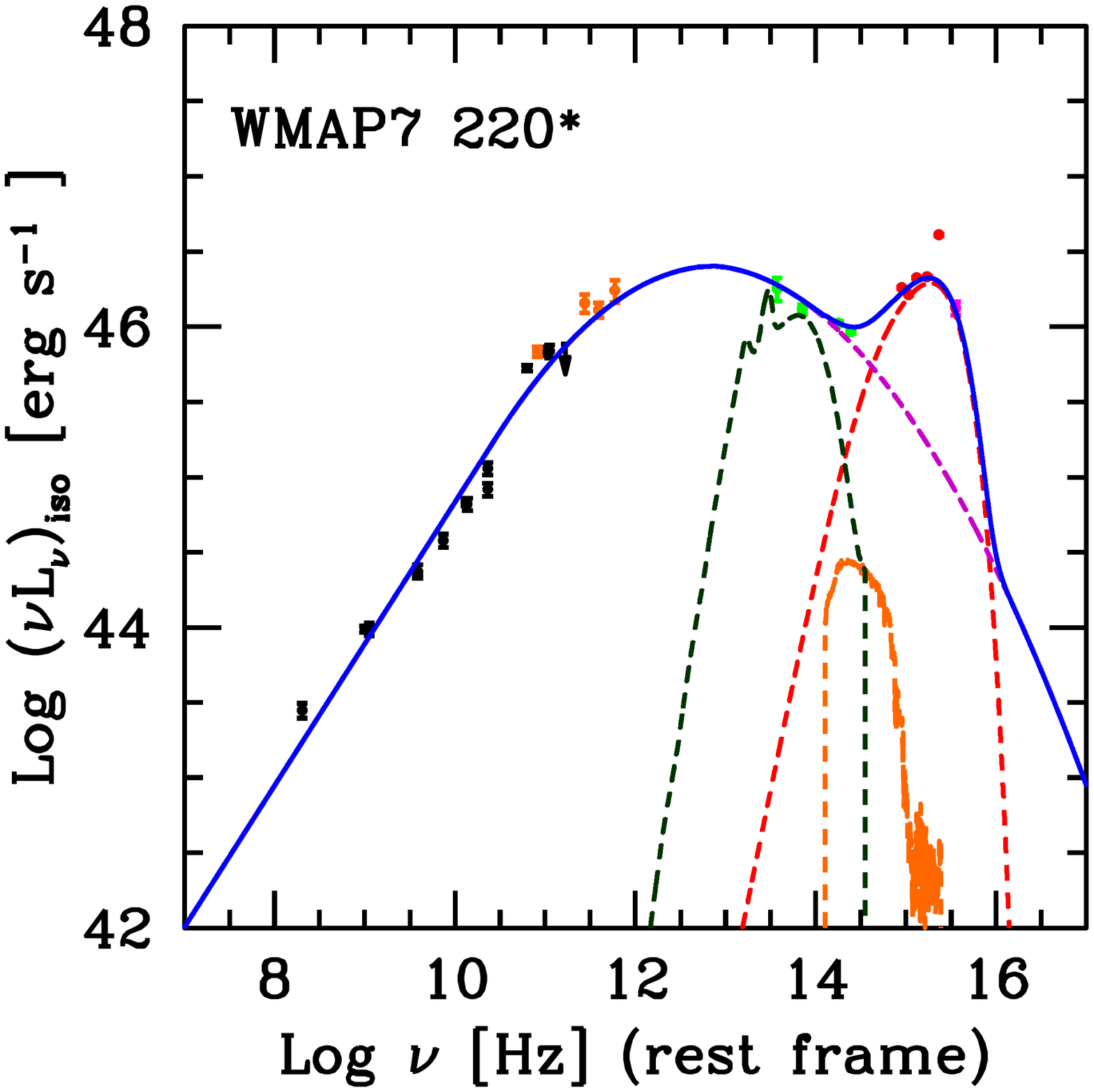}}
\subfloat{\includegraphics[width=0.32\textwidth,natwidth=610,natheight=642]{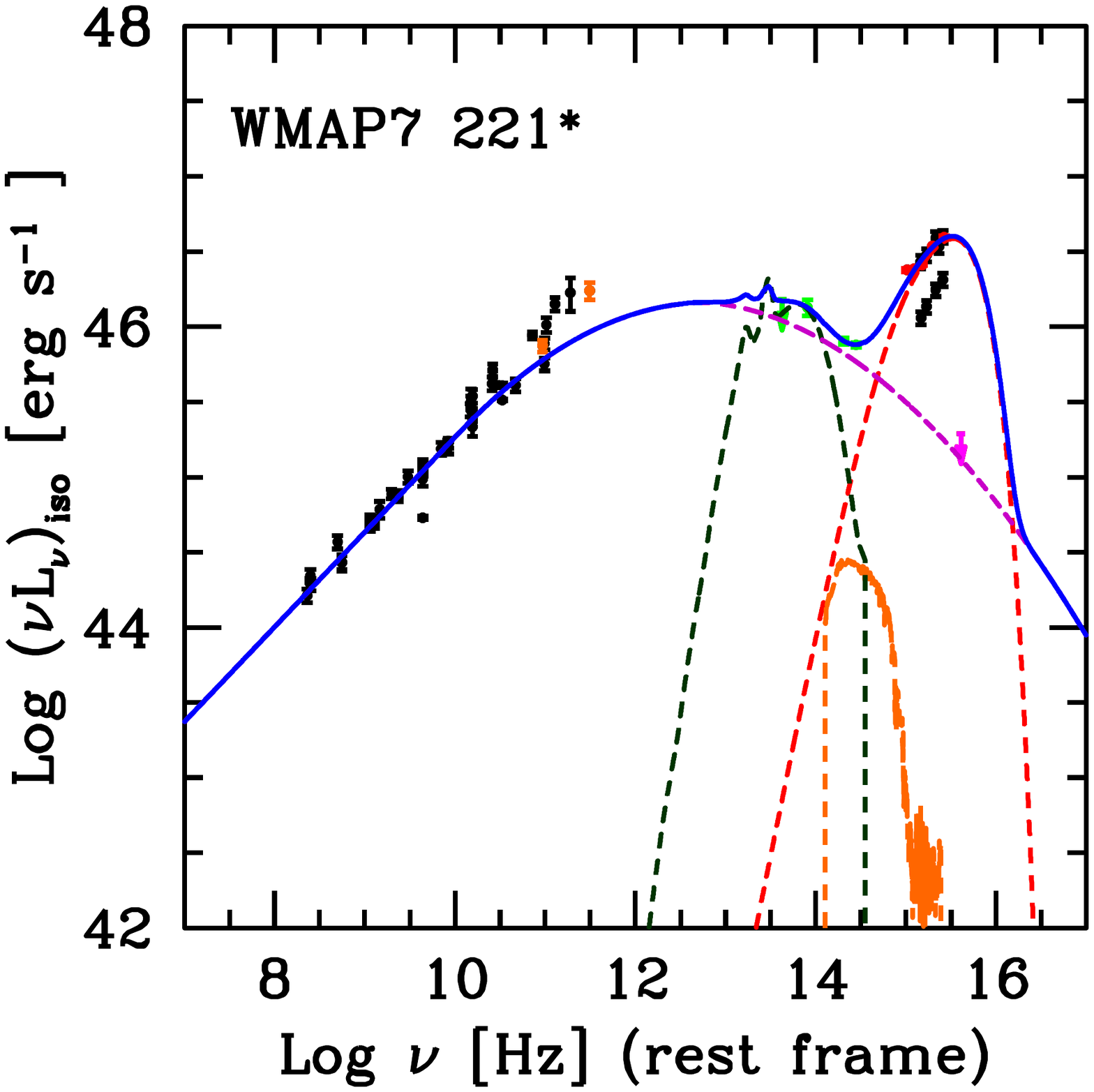}}
\subfloat{\includegraphics[width=0.32\textwidth,natwidth=610,natheight=642]{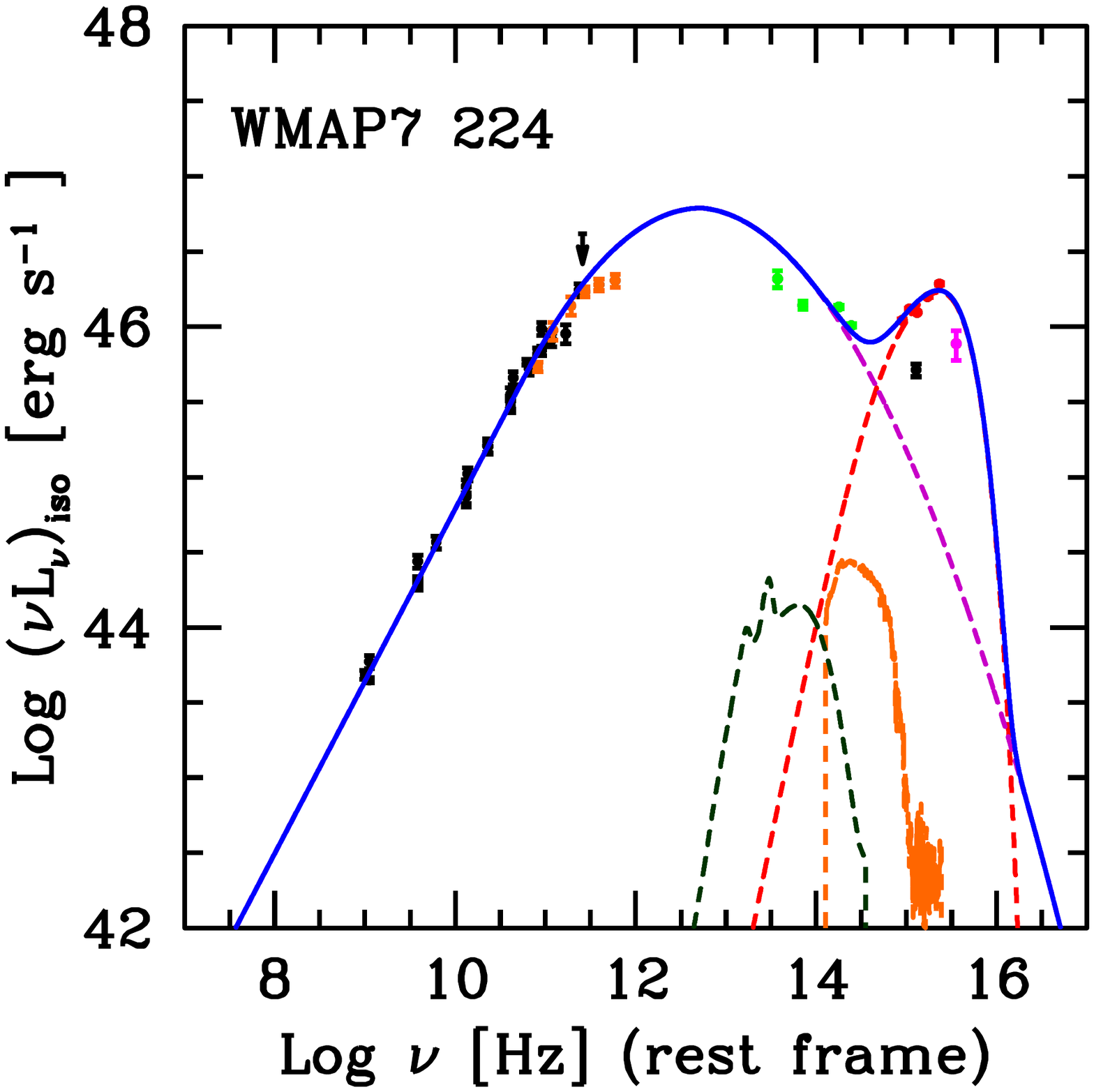}}\qquad
\subfloat{\includegraphics[width=0.32\textwidth,natwidth=610,natheight=642]{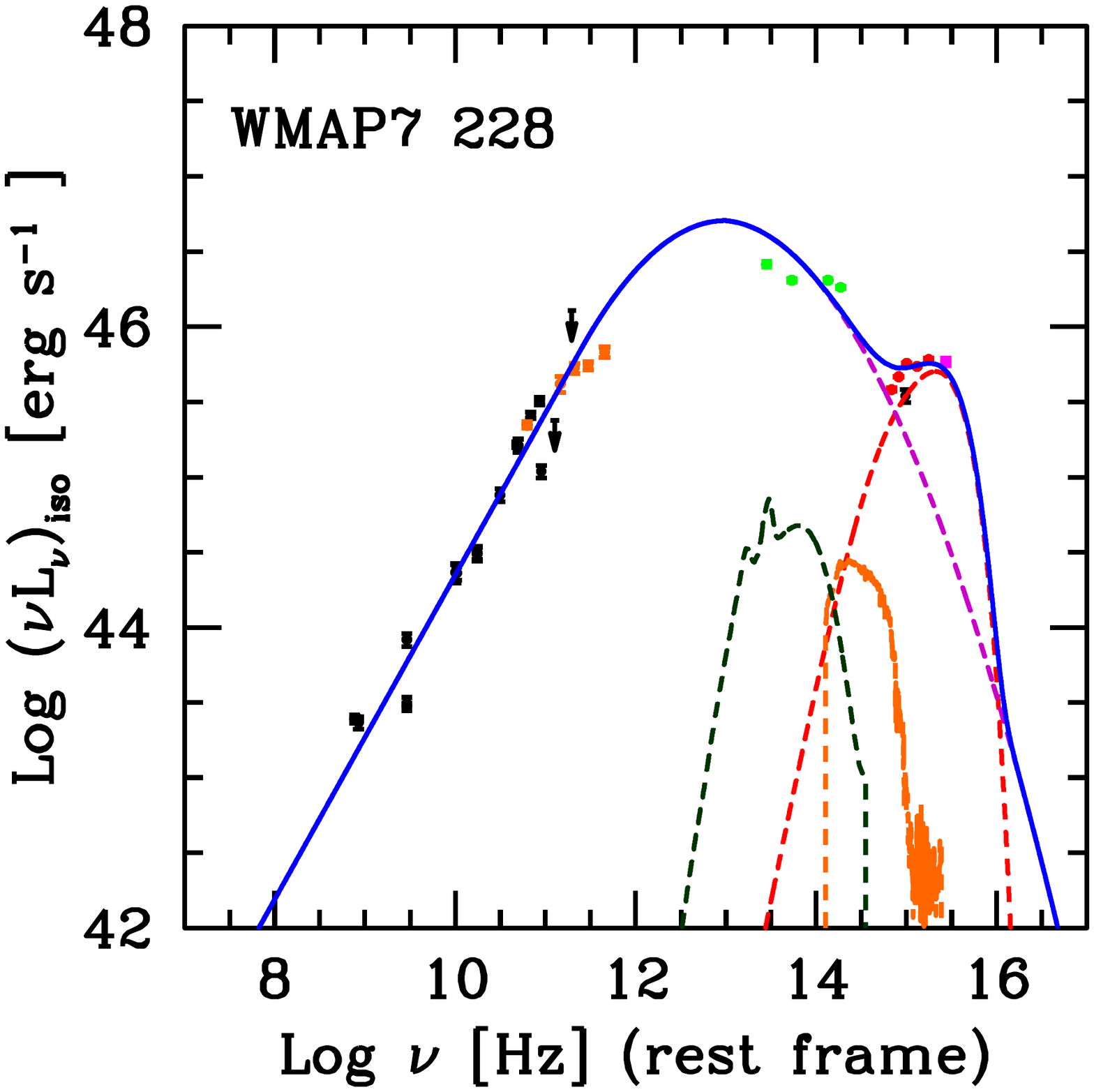}}
\subfloat{\includegraphics[width=0.32\textwidth,natwidth=610,natheight=642]{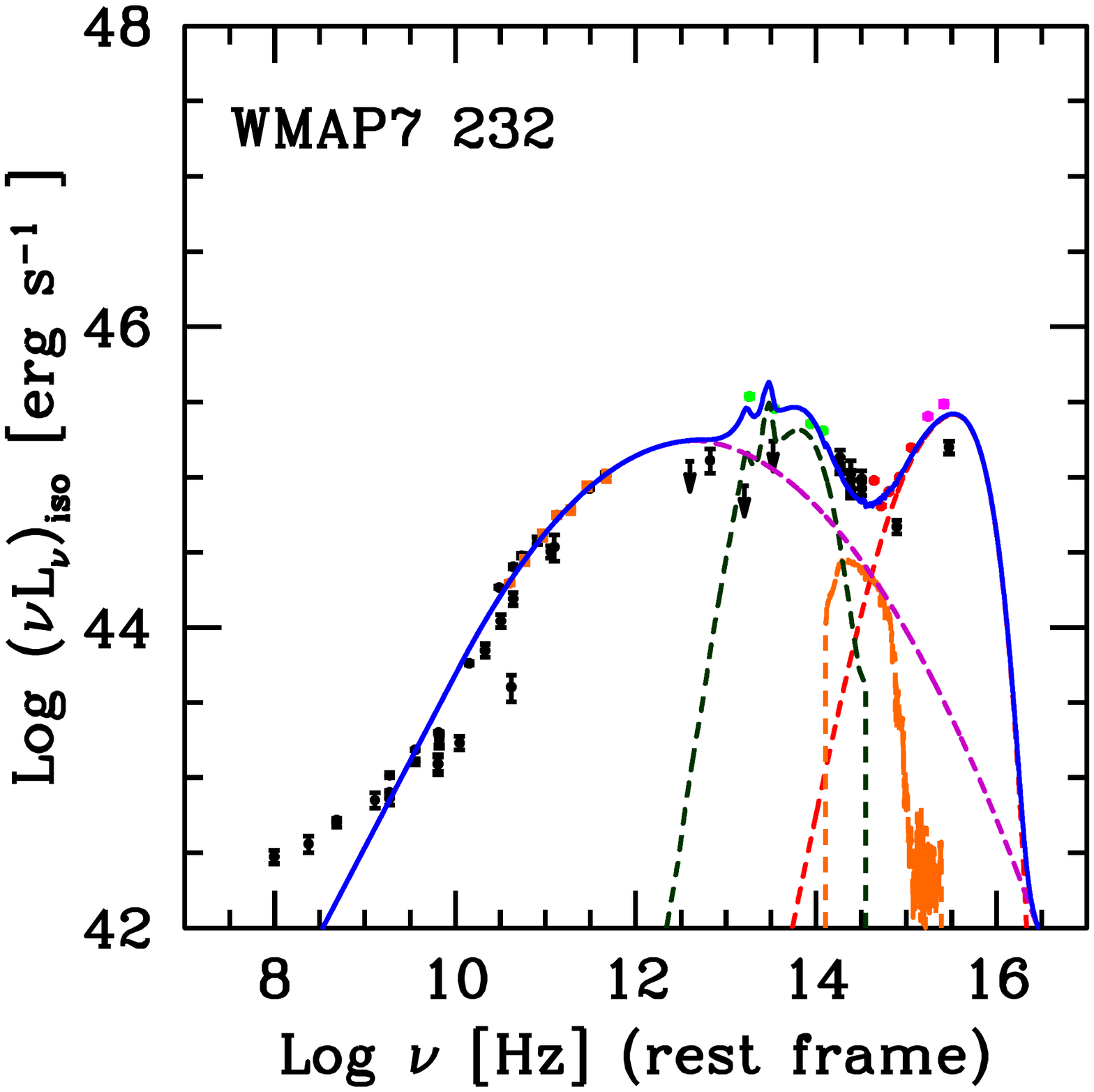}}
\subfloat{\includegraphics[width=0.32\textwidth,natwidth=610,natheight=642]{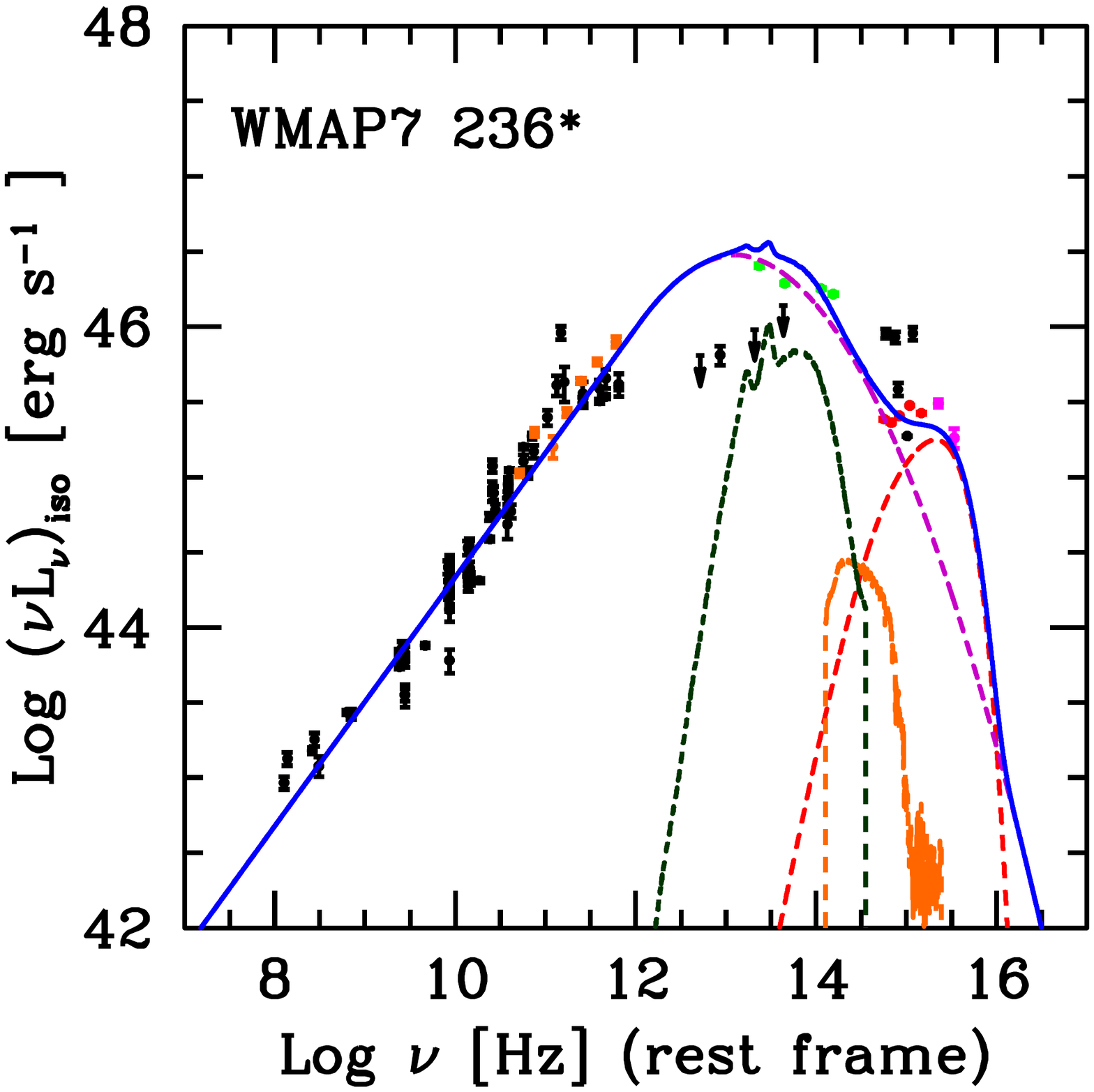}}\qquad
\subfloat{\includegraphics[width=0.32\textwidth,natwidth=610,natheight=642]{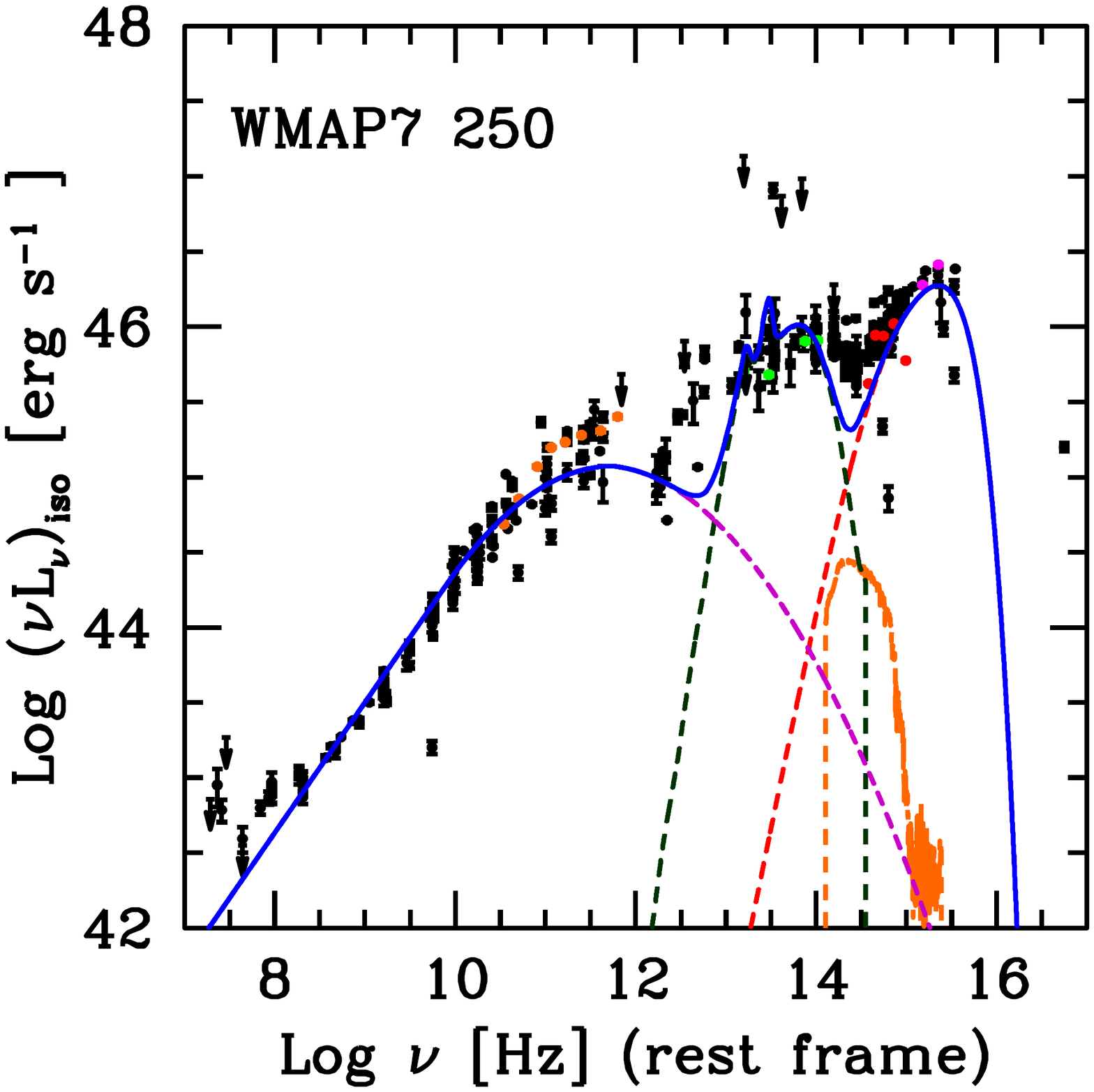}}
\subfloat{\includegraphics[width=0.32\textwidth,natwidth=610,natheight=642]{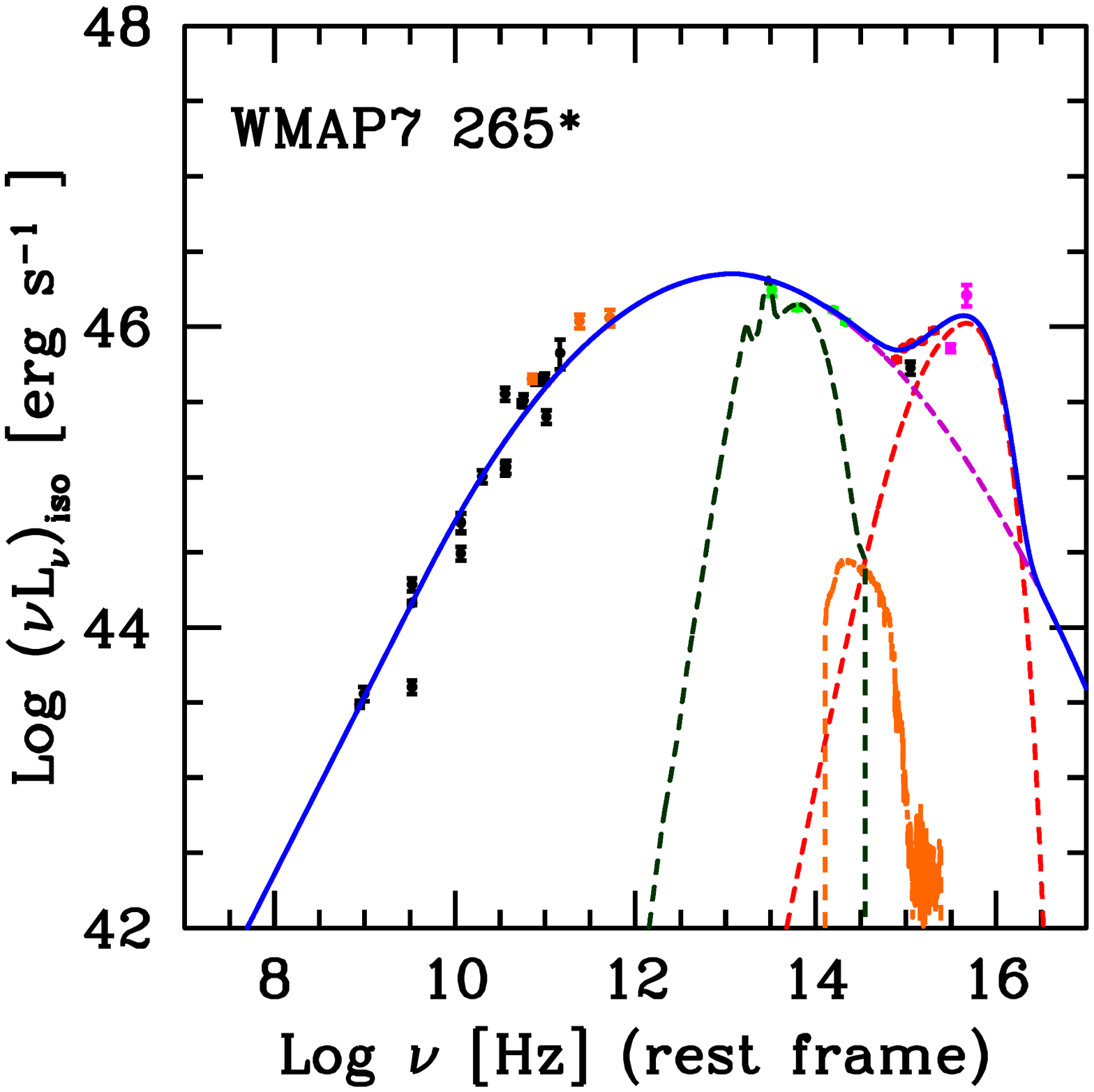}}
\subfloat{\includegraphics[width=0.32\textwidth,natwidth=610,natheight=642]{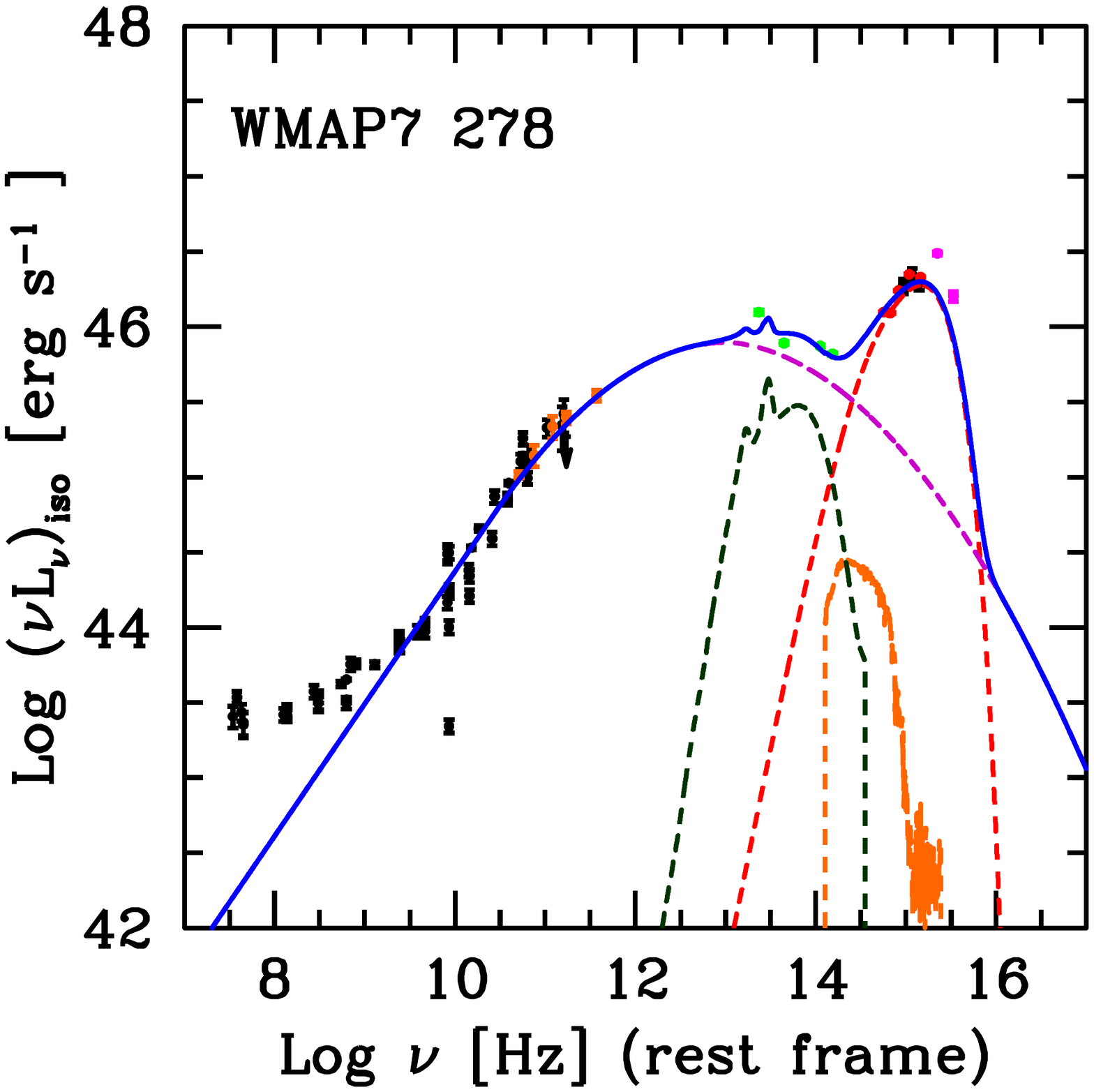}}\qquad
\caption{Continued.}
\end{figure*}

\begin{figure*} \centering
\ContinuedFloat
\subfloat{\includegraphics[width=0.32\textwidth,natwidth=610,natheight=642]{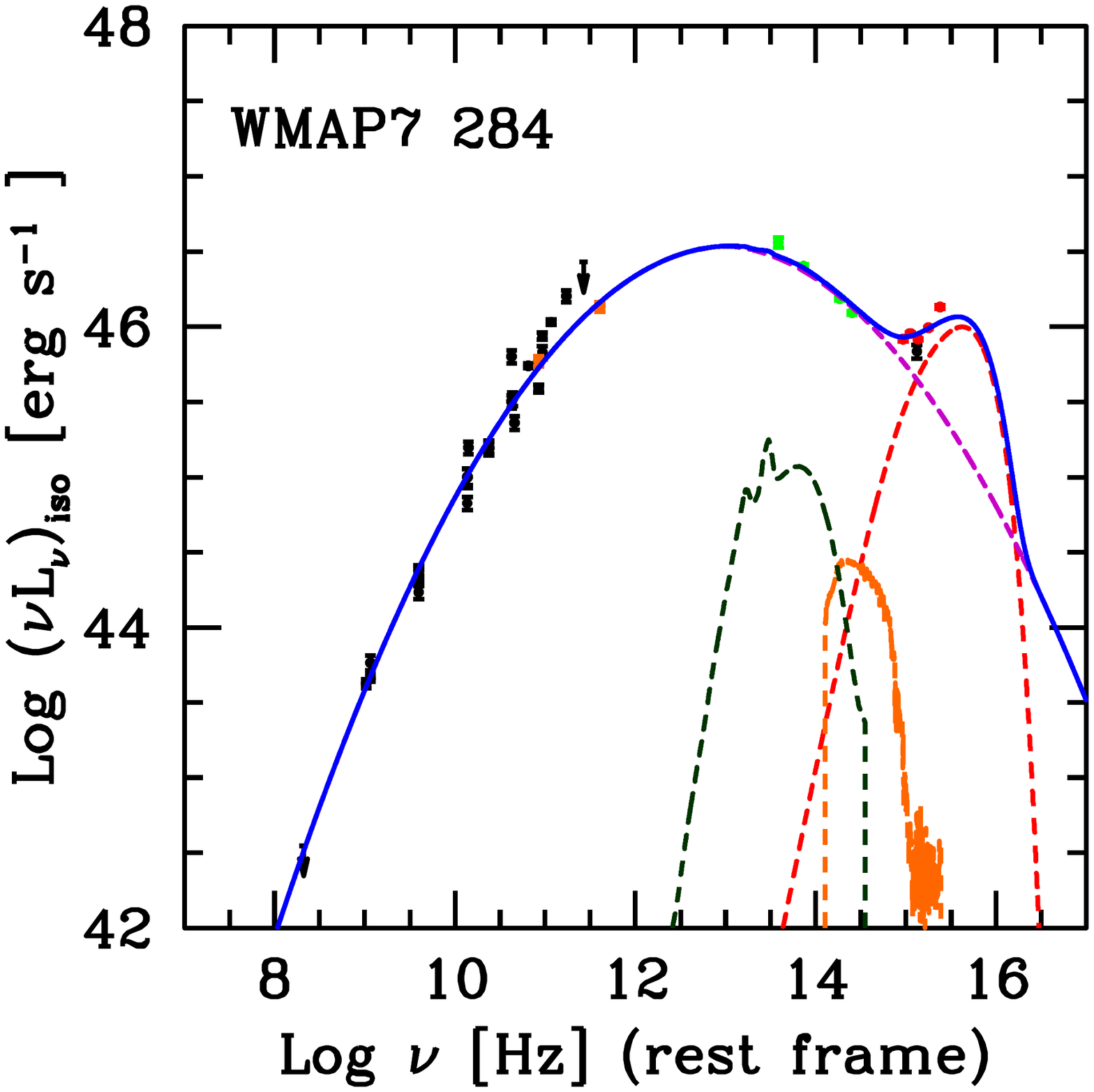}}
\subfloat{\includegraphics[width=0.32\textwidth,natwidth=610,natheight=642]{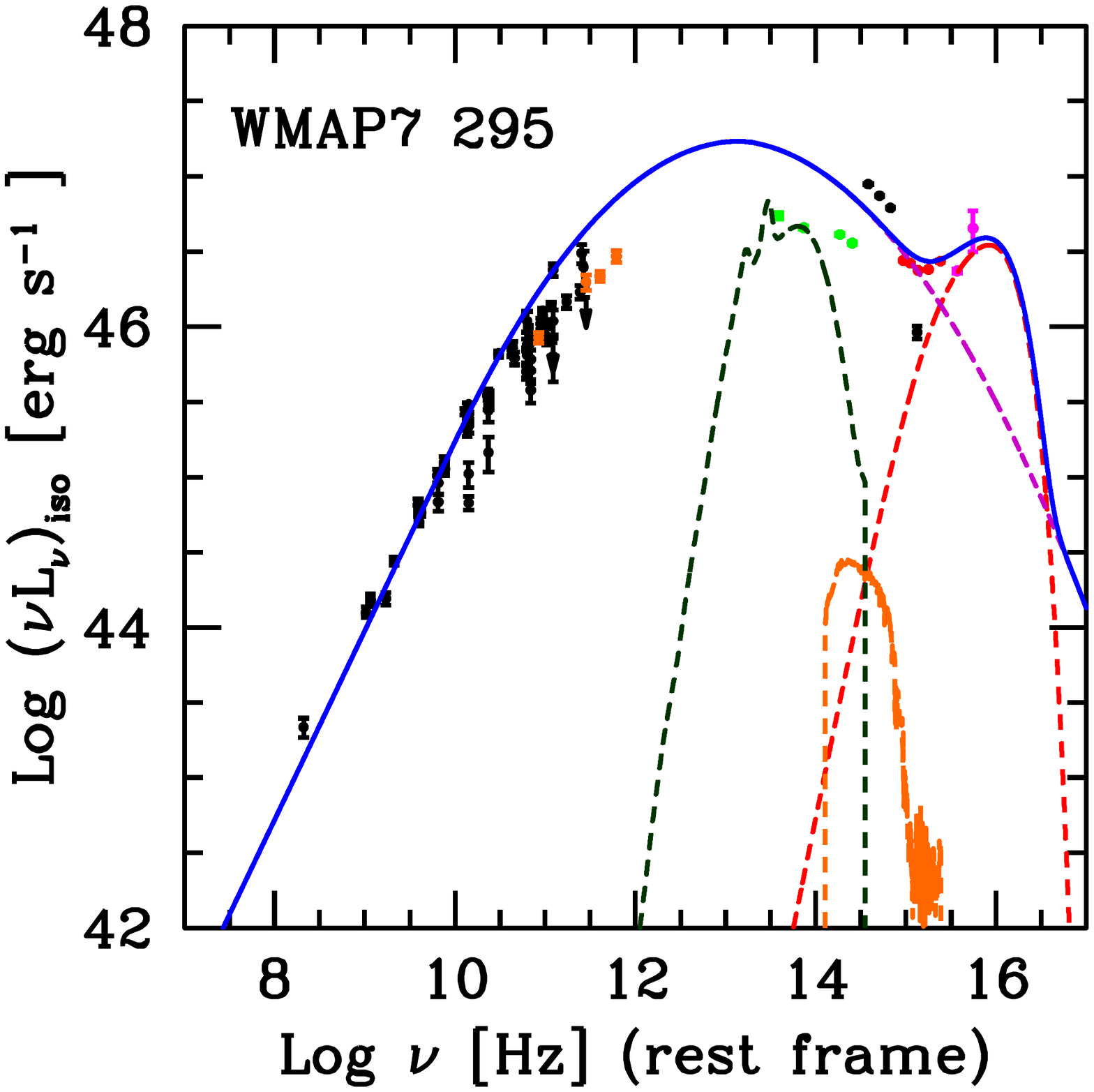}}
\subfloat{\includegraphics[width=0.32\textwidth,natwidth=610,natheight=642]{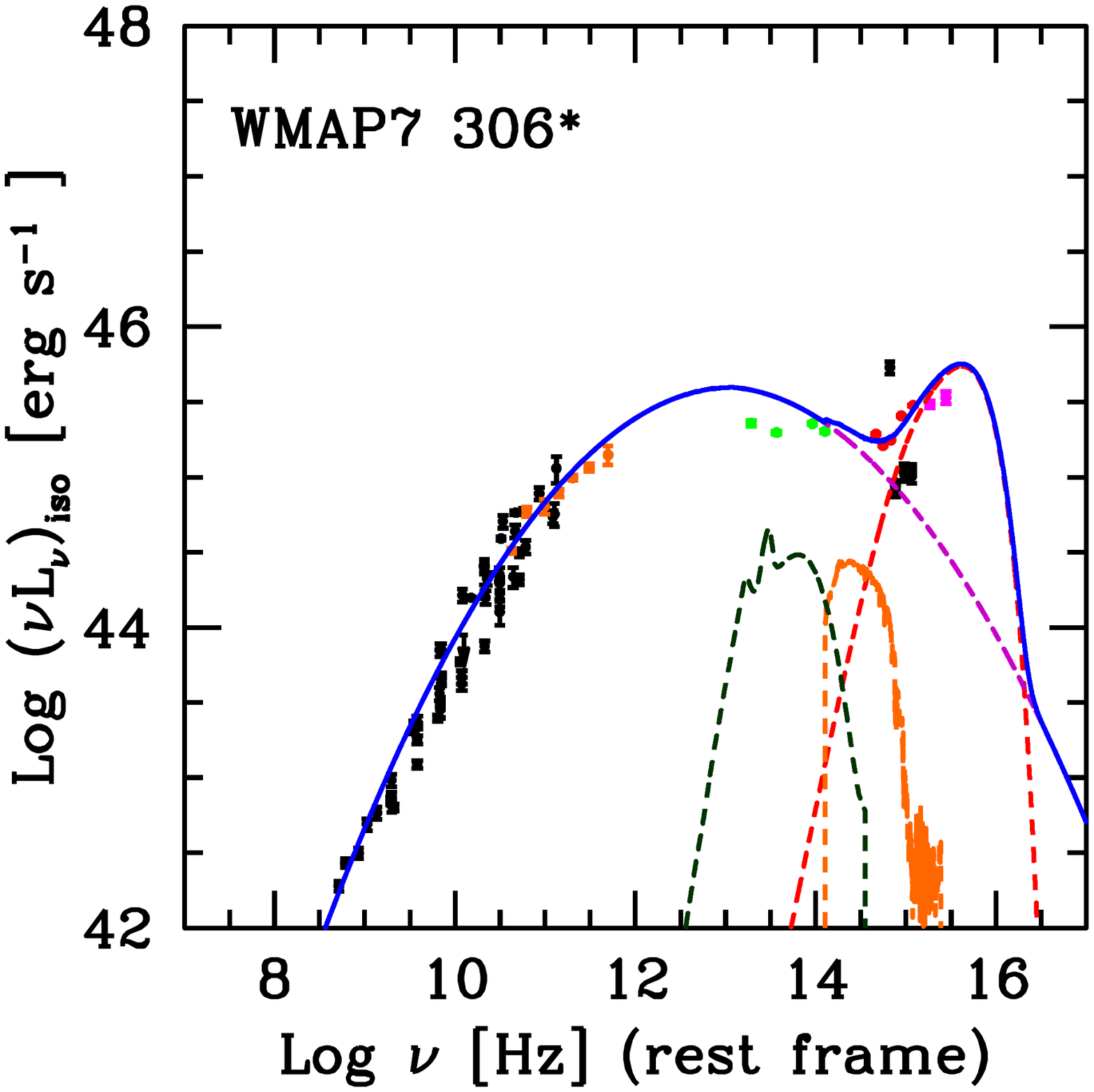}}\qquad
\subfloat{\includegraphics[width=0.32\textwidth,natwidth=610,natheight=642]{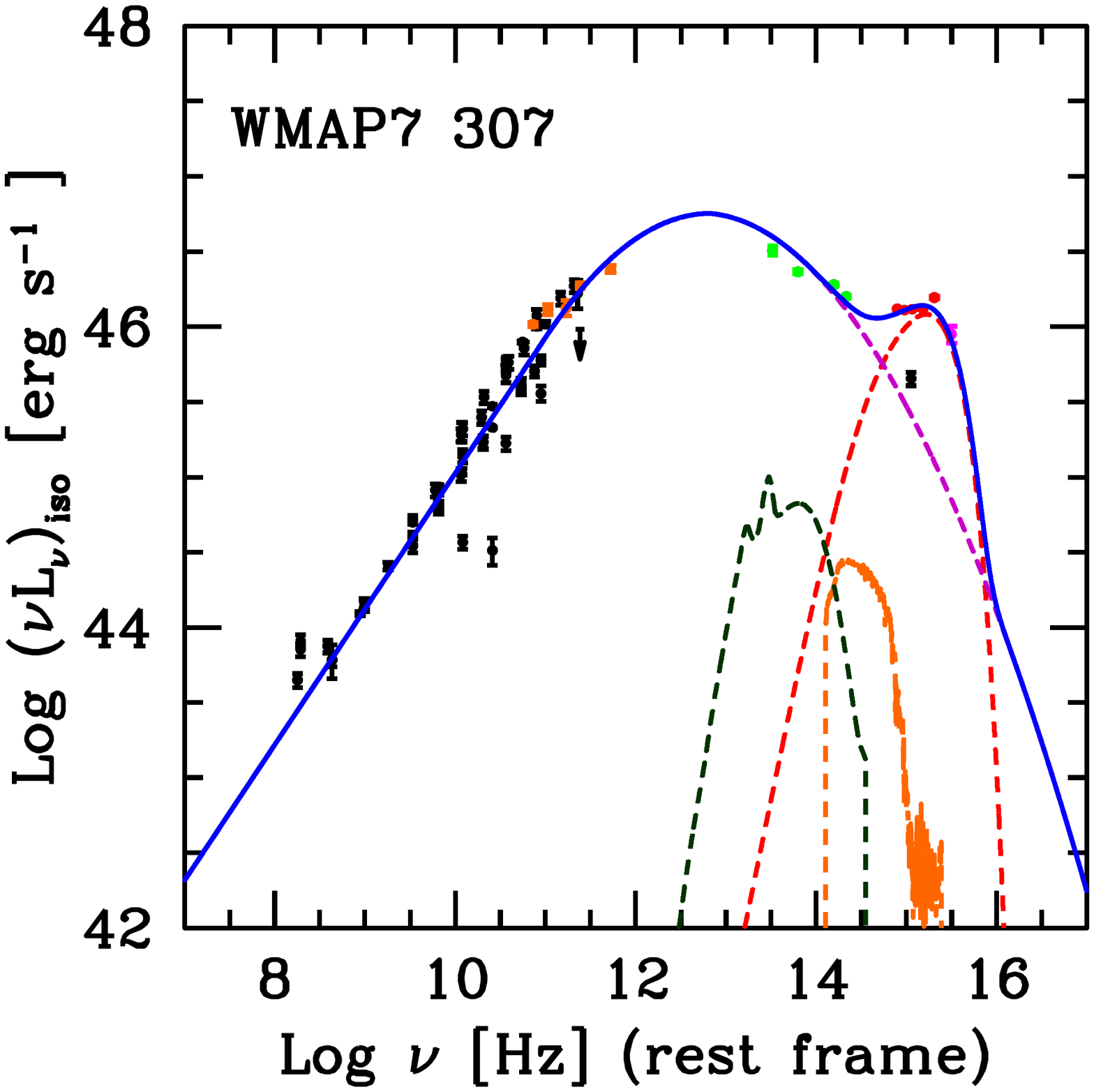}}
\subfloat{\includegraphics[width=0.32\textwidth,natwidth=610,natheight=642]{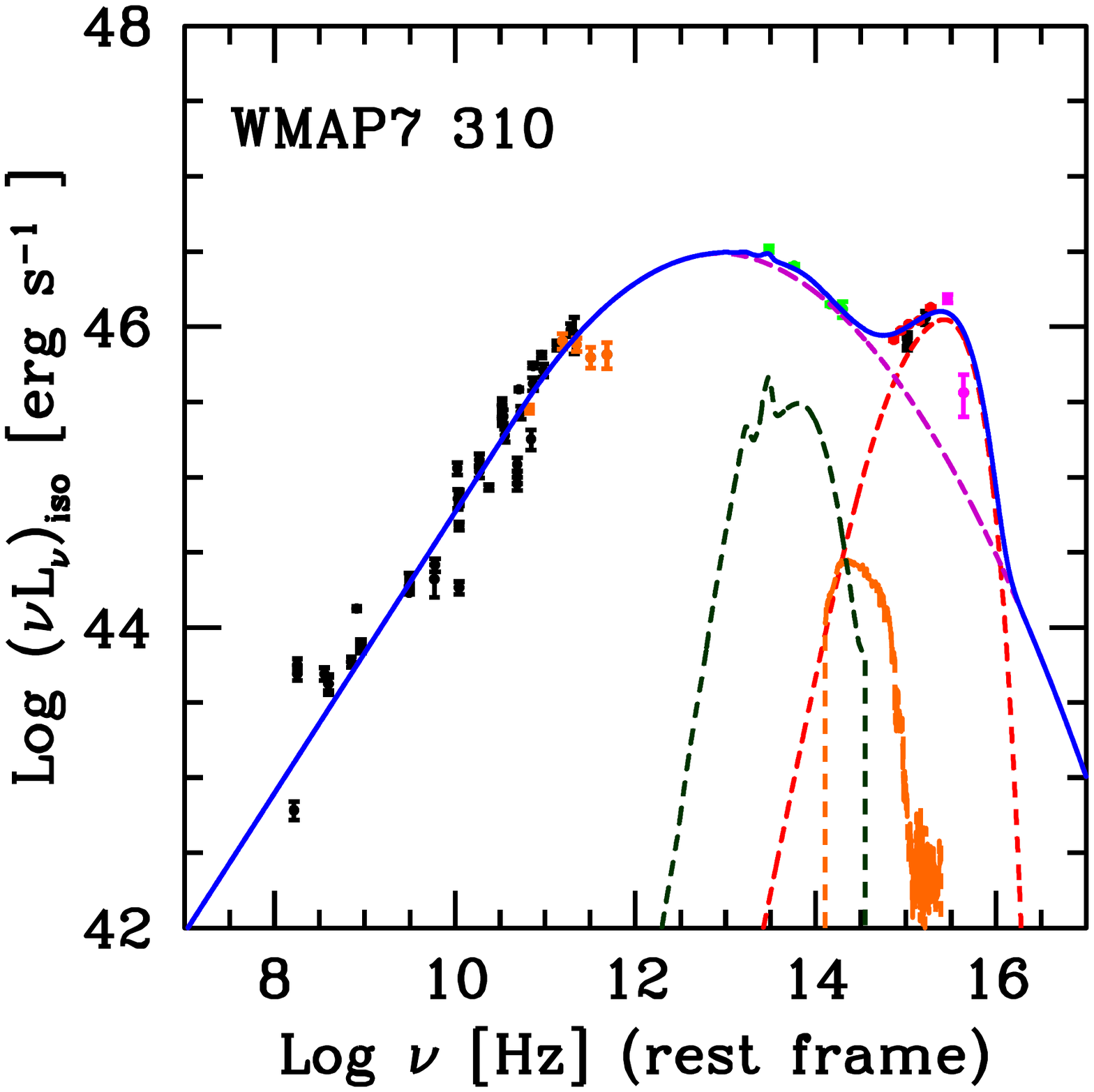}}
\subfloat{\includegraphics[width=0.32\textwidth,natwidth=610,natheight=642]{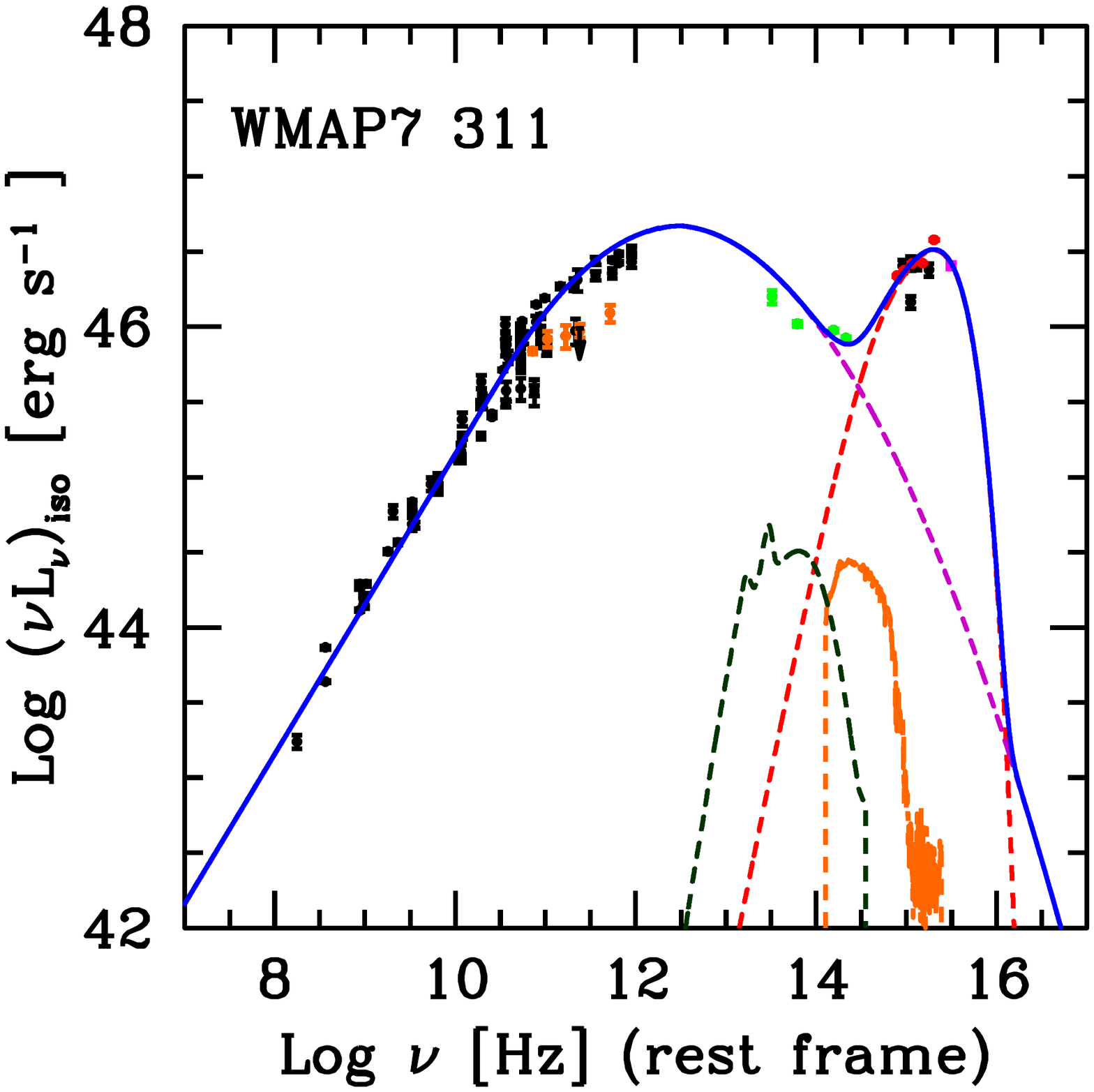}}\qquad
\subfloat{\includegraphics[width=0.32\textwidth,natwidth=610,natheight=642]{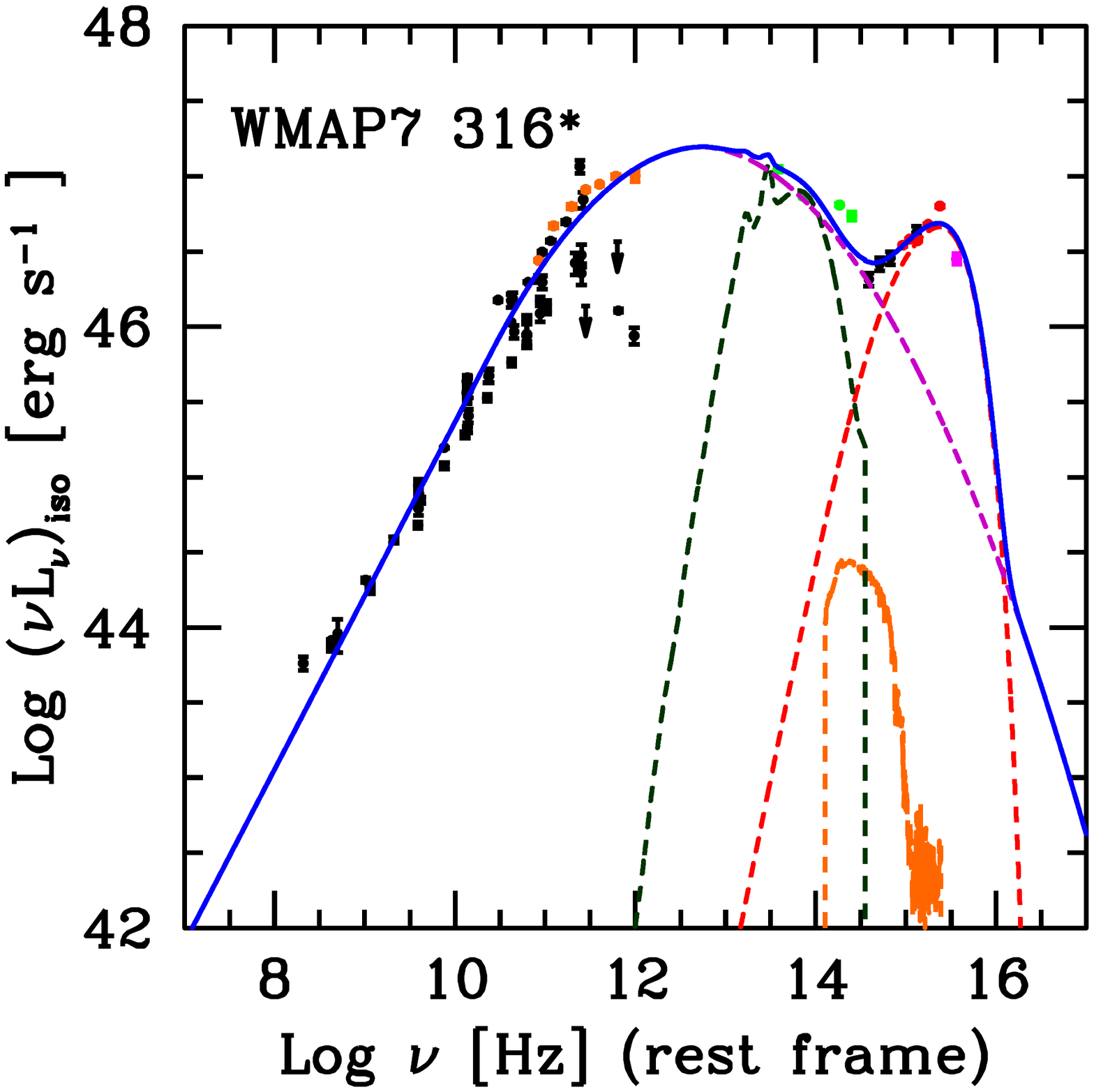}}
\subfloat{\includegraphics[width=0.32\textwidth,natwidth=610,natheight=642]{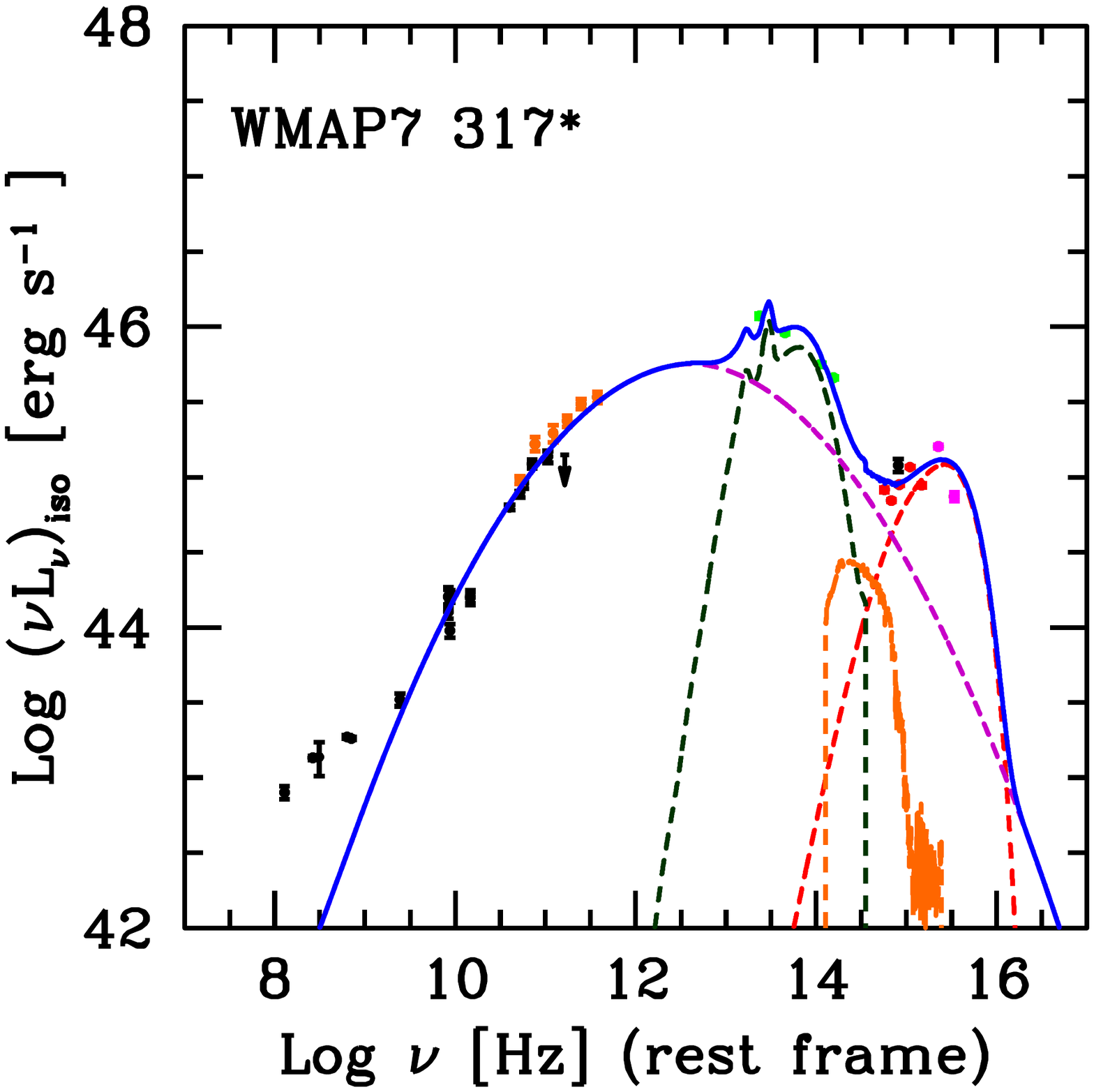}}
\subfloat{\includegraphics[width=0.32\textwidth,natwidth=610,natheight=642]{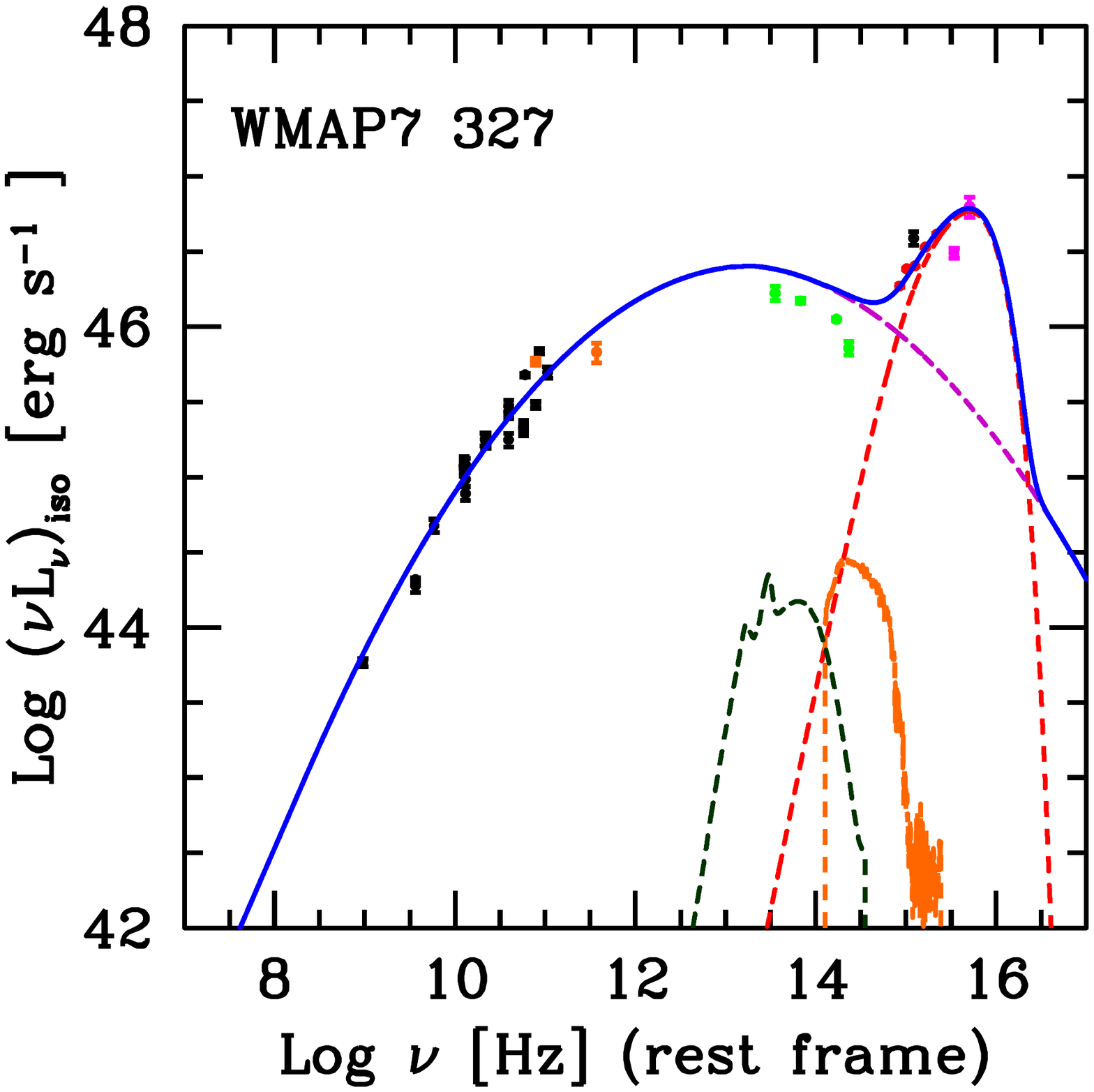}}\qquad
\caption{Continued.}
\end{figure*}

\begin{figure*} \centering
\ContinuedFloat
\subfloat{\includegraphics[width=0.32\textwidth,natwidth=610,natheight=642]{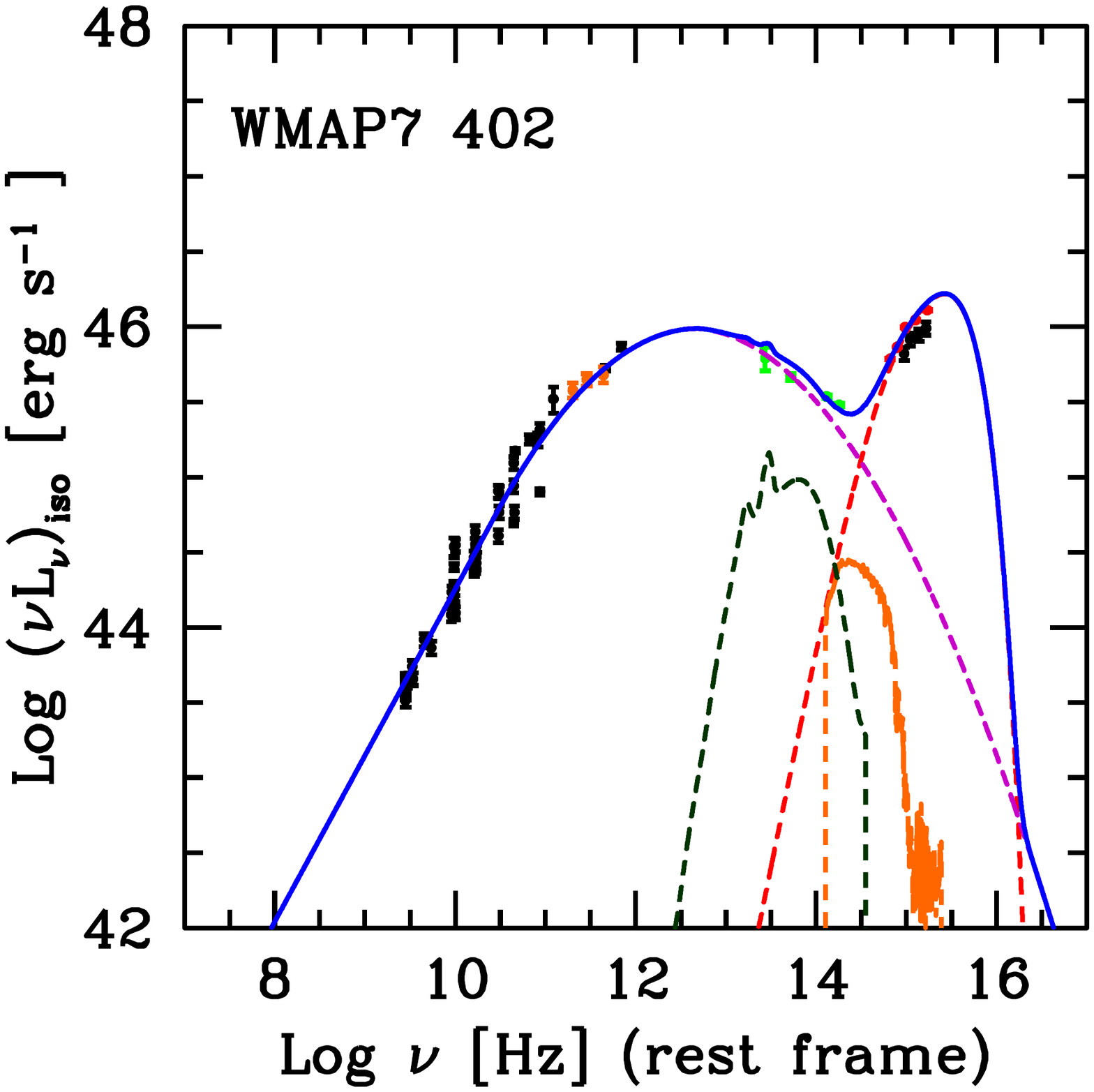}}
\subfloat{\includegraphics[width=0.32\textwidth,natwidth=610,natheight=642]{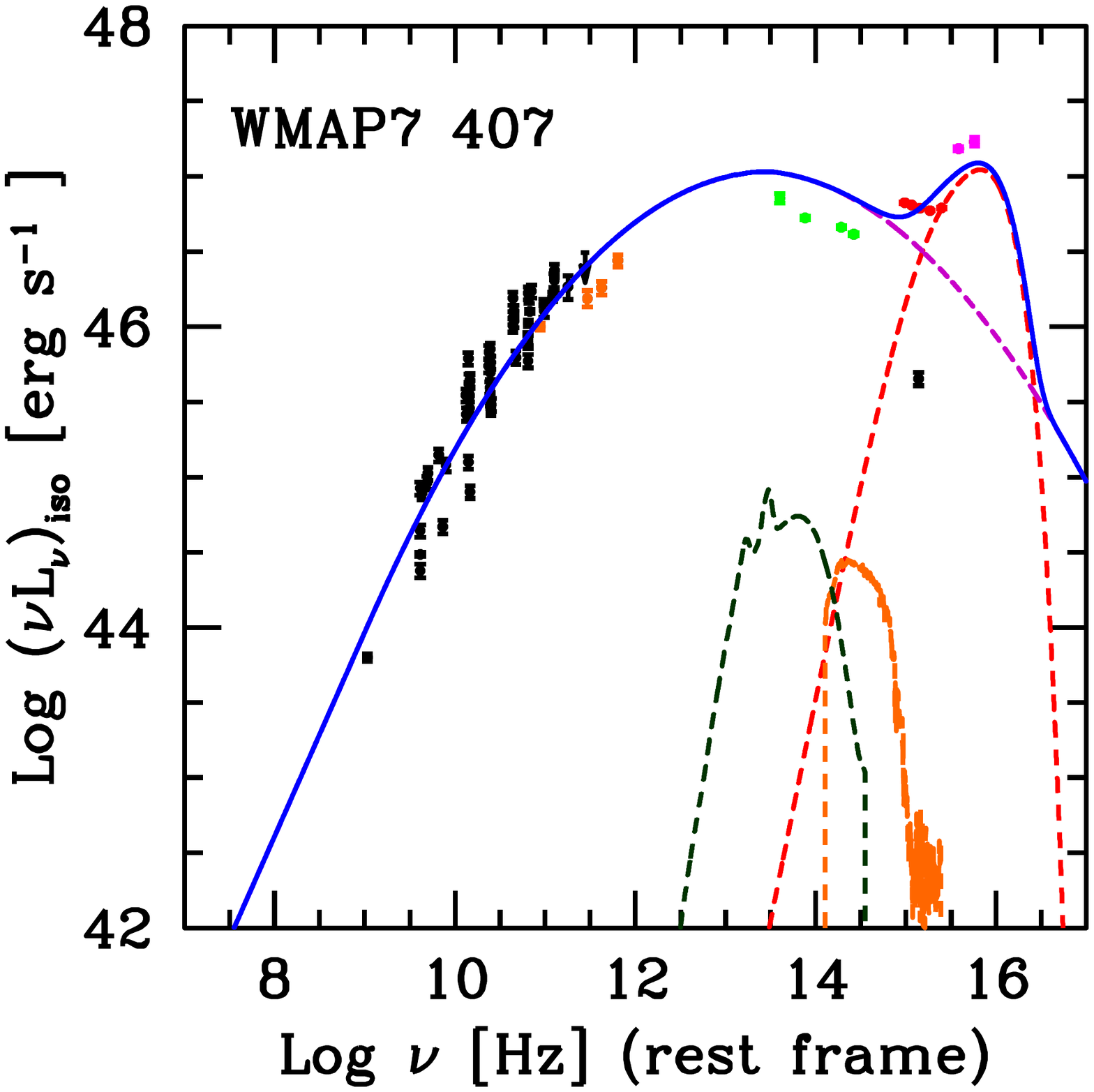}}
\subfloat{\includegraphics[width=0.32\textwidth,natwidth=610,natheight=642]{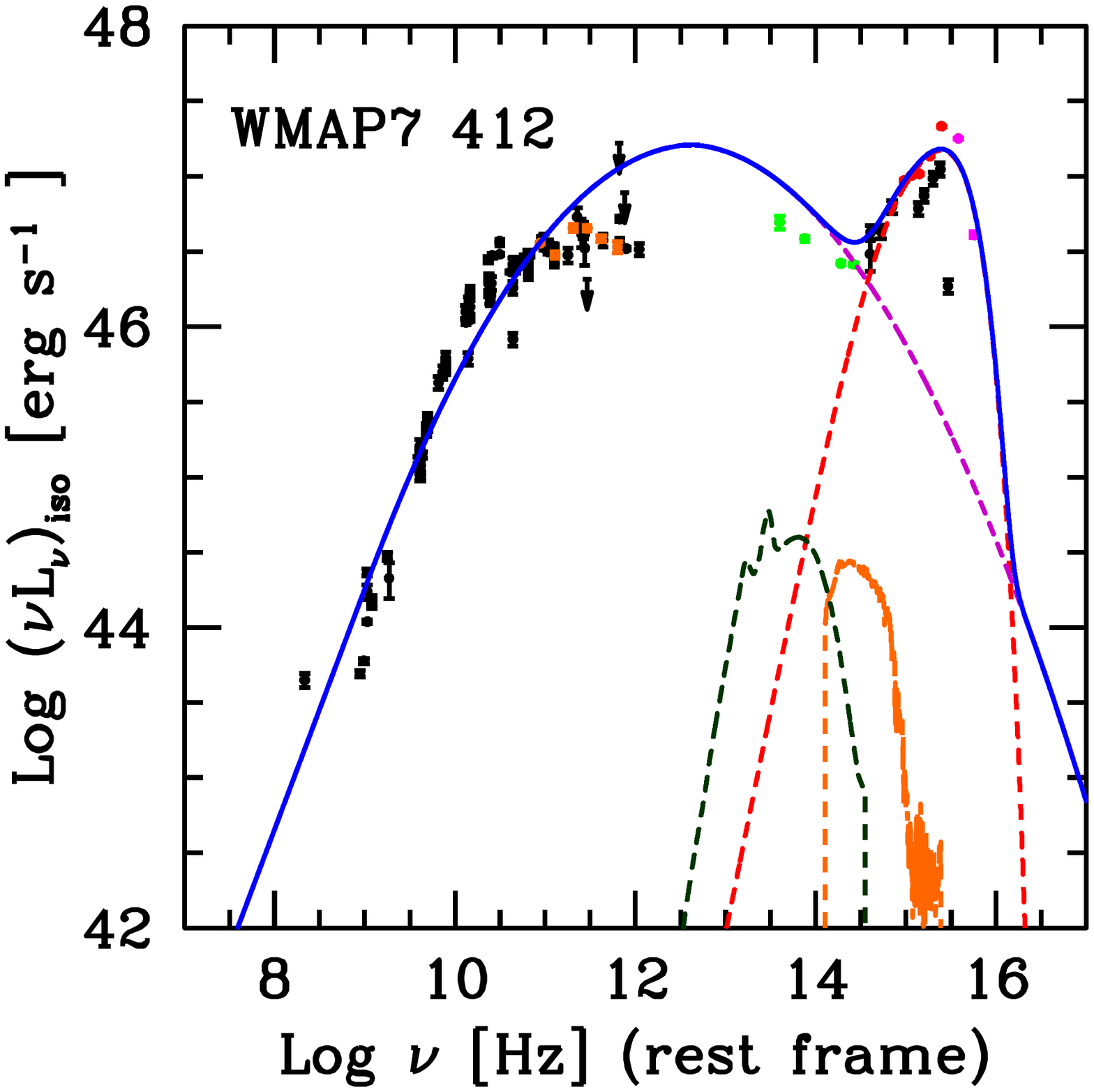}}\qquad
\subfloat{\includegraphics[width=0.32\textwidth,natwidth=610,natheight=642]{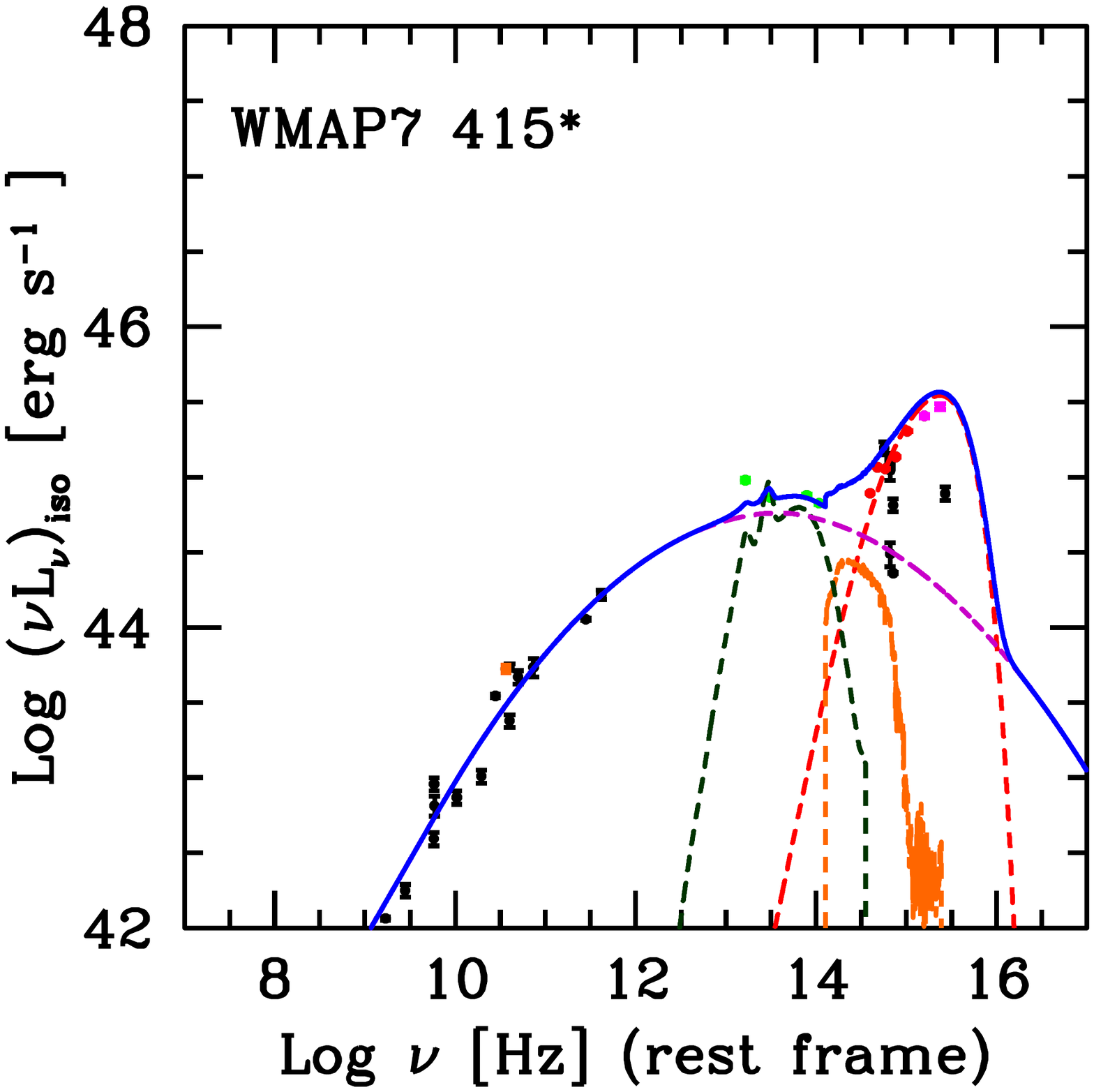}}
\subfloat{\includegraphics[width=0.32\textwidth,natwidth=610,natheight=642]{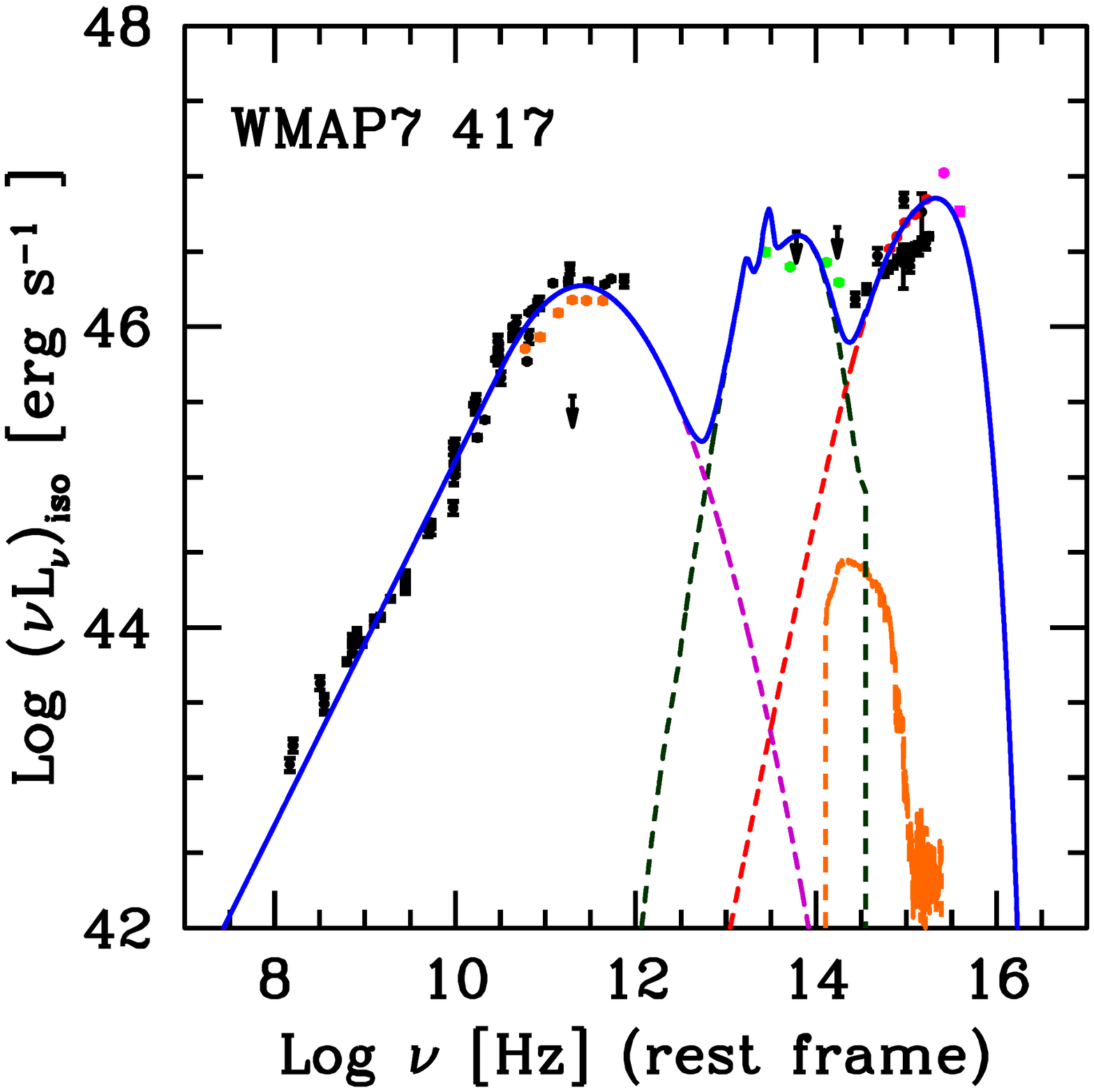}}
\subfloat{\includegraphics[width=0.32\textwidth,natwidth=610,natheight=642]{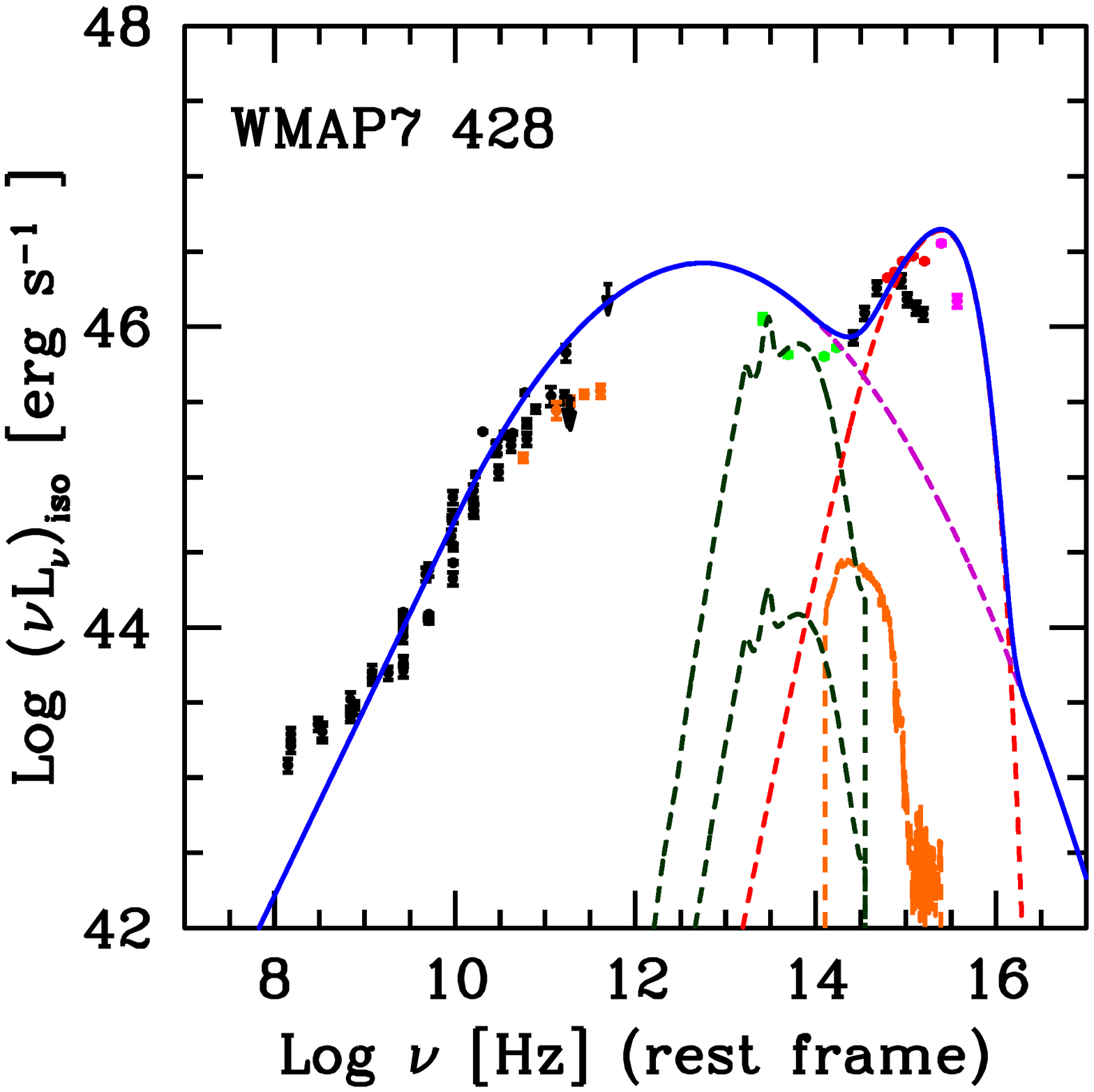}}\qquad
\subfloat{\includegraphics[width=0.32\textwidth,natwidth=610,natheight=642]{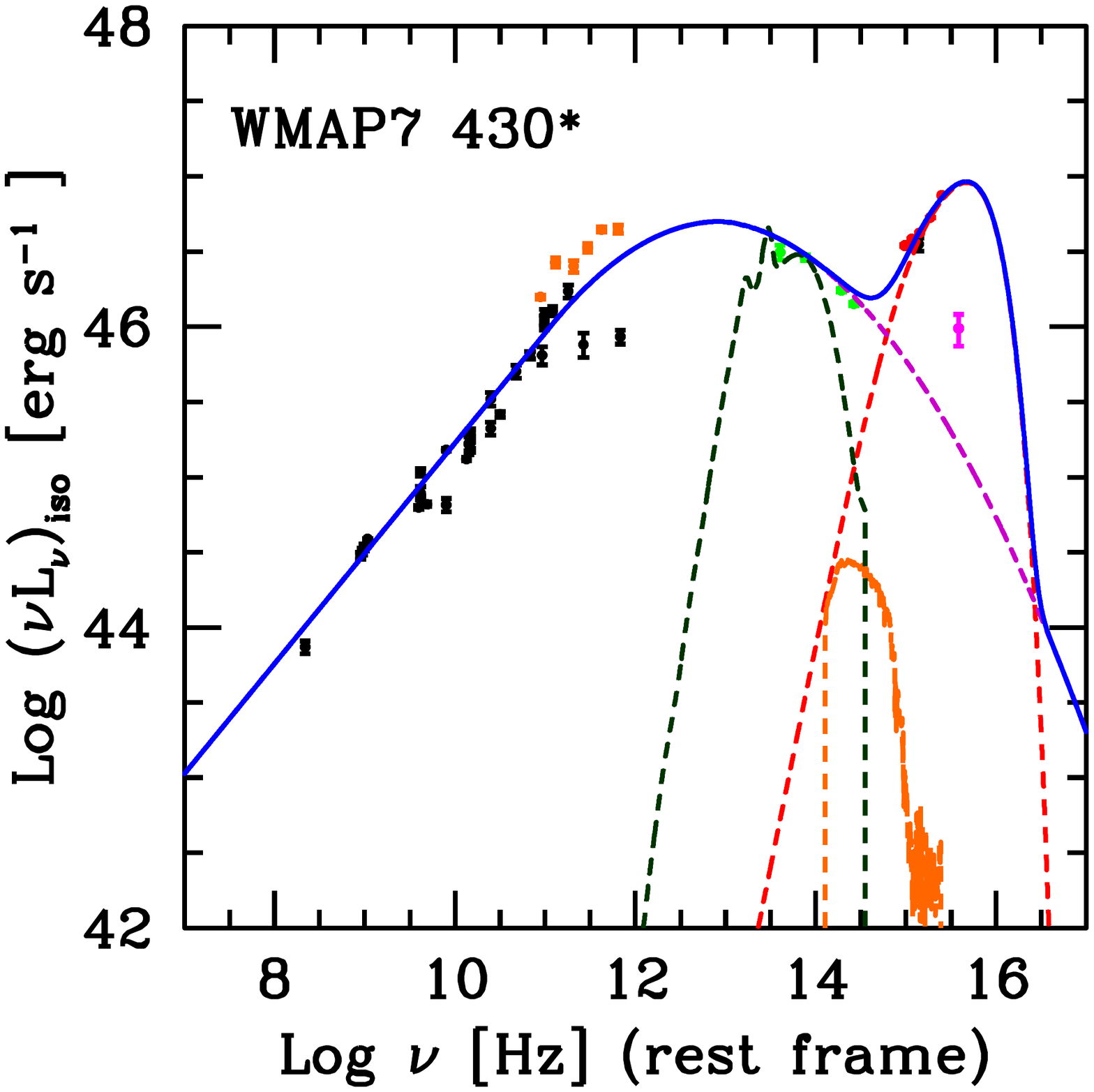}}
\subfloat{\includegraphics[width=0.32\textwidth,natwidth=610,natheight=642]{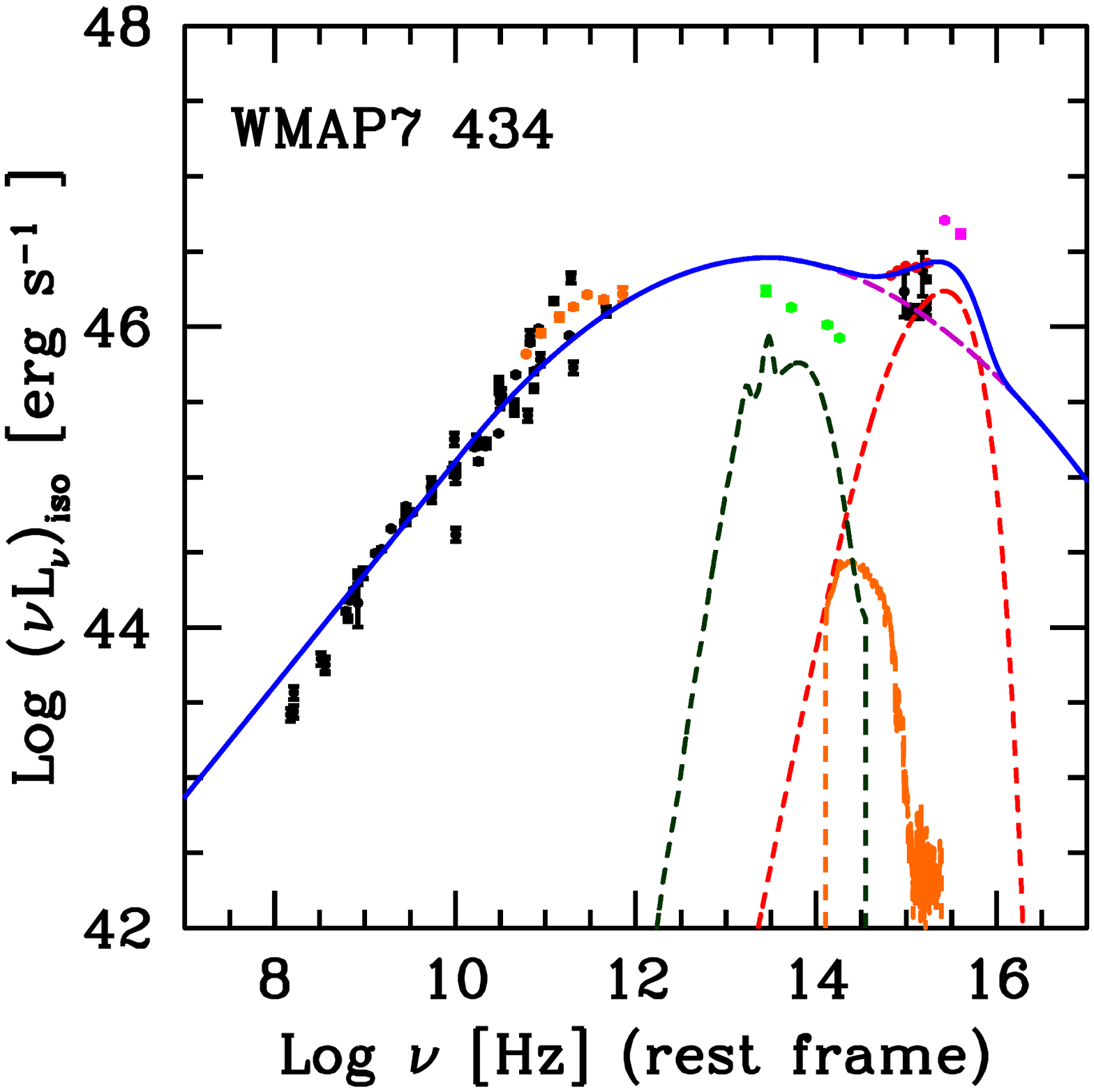}}
\subfloat{\includegraphics[width=0.32\textwidth,natwidth=610,natheight=642]{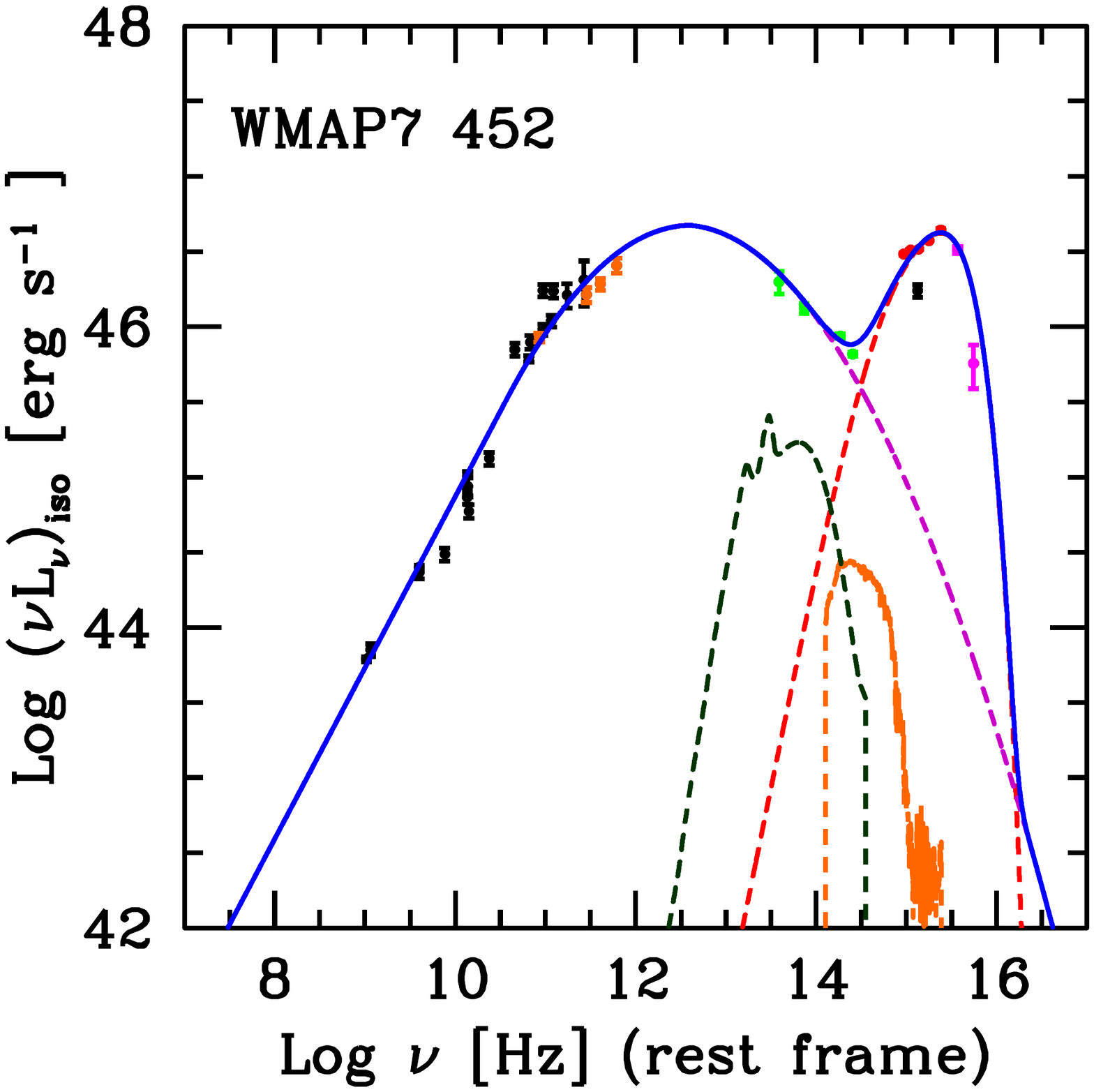}}\qquad
\caption{Continued.}
\end{figure*}

\begin{figure*} \centering
\ContinuedFloat
\subfloat{\includegraphics[width=0.32\textwidth,natwidth=610,natheight=642]{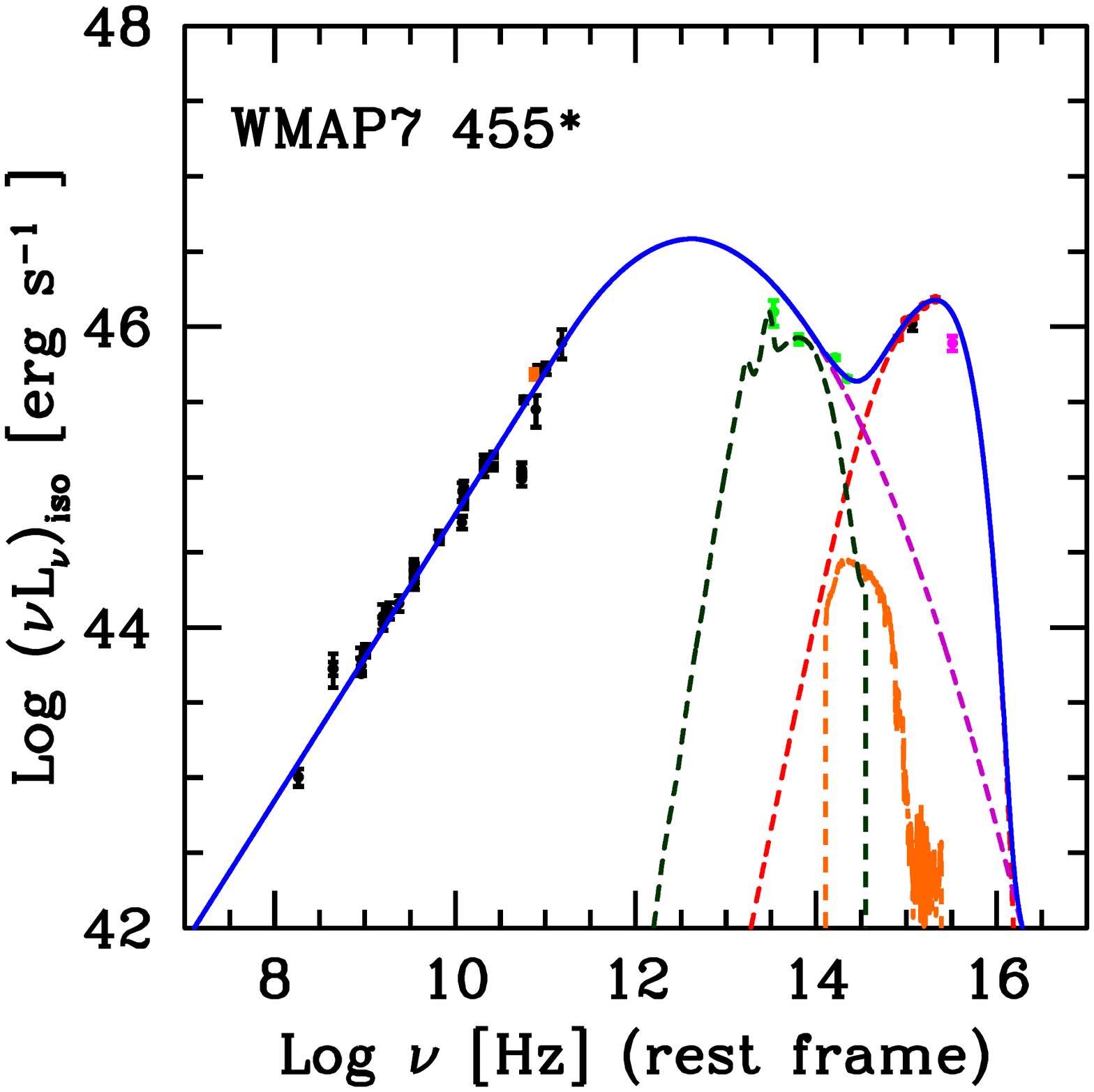}}
\subfloat{\includegraphics[width=0.32\textwidth,natwidth=610,natheight=642]{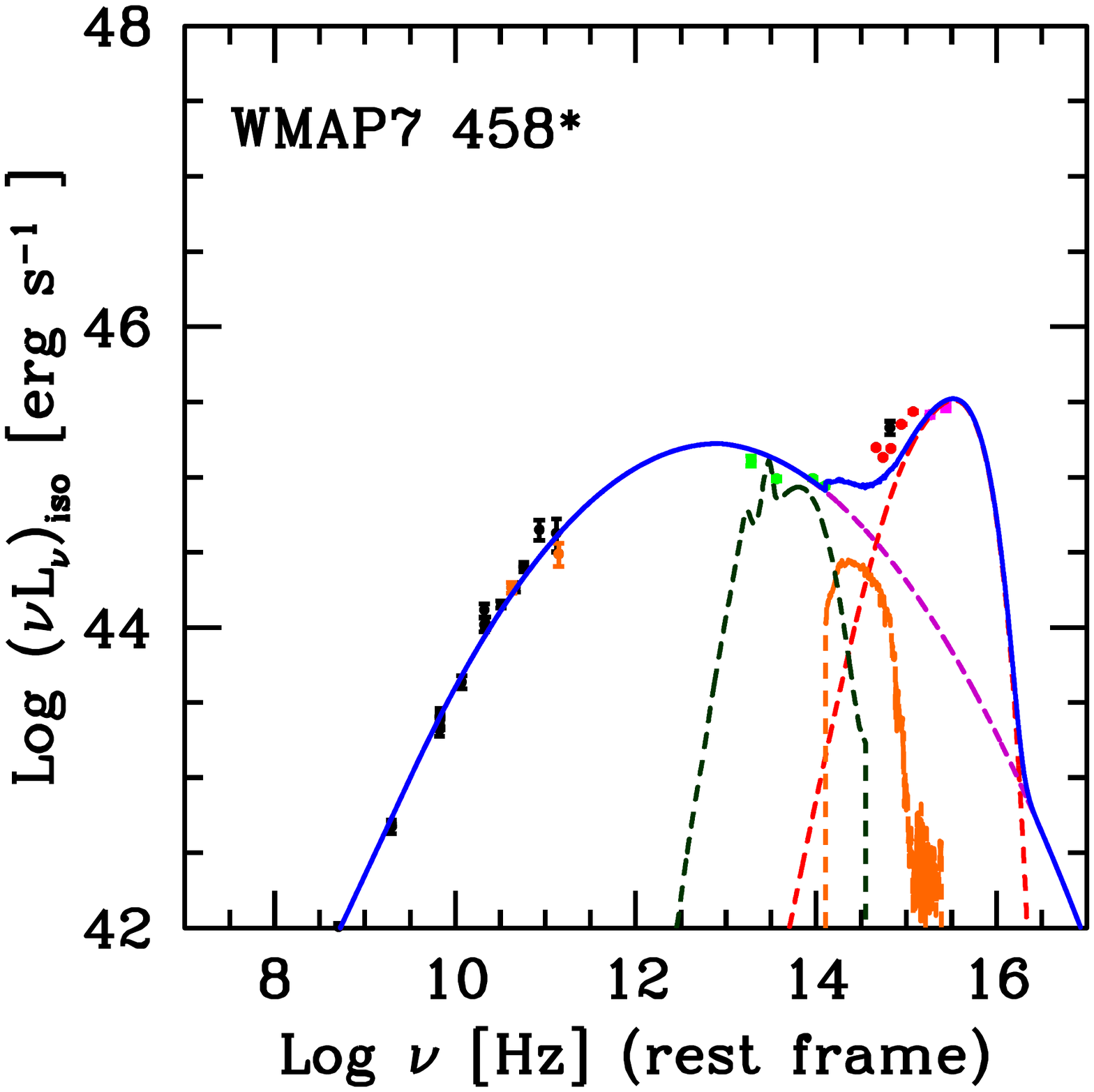}}\qquad
\subfloat{\includegraphics[width=0.32\textwidth,natwidth=610,natheight=642]{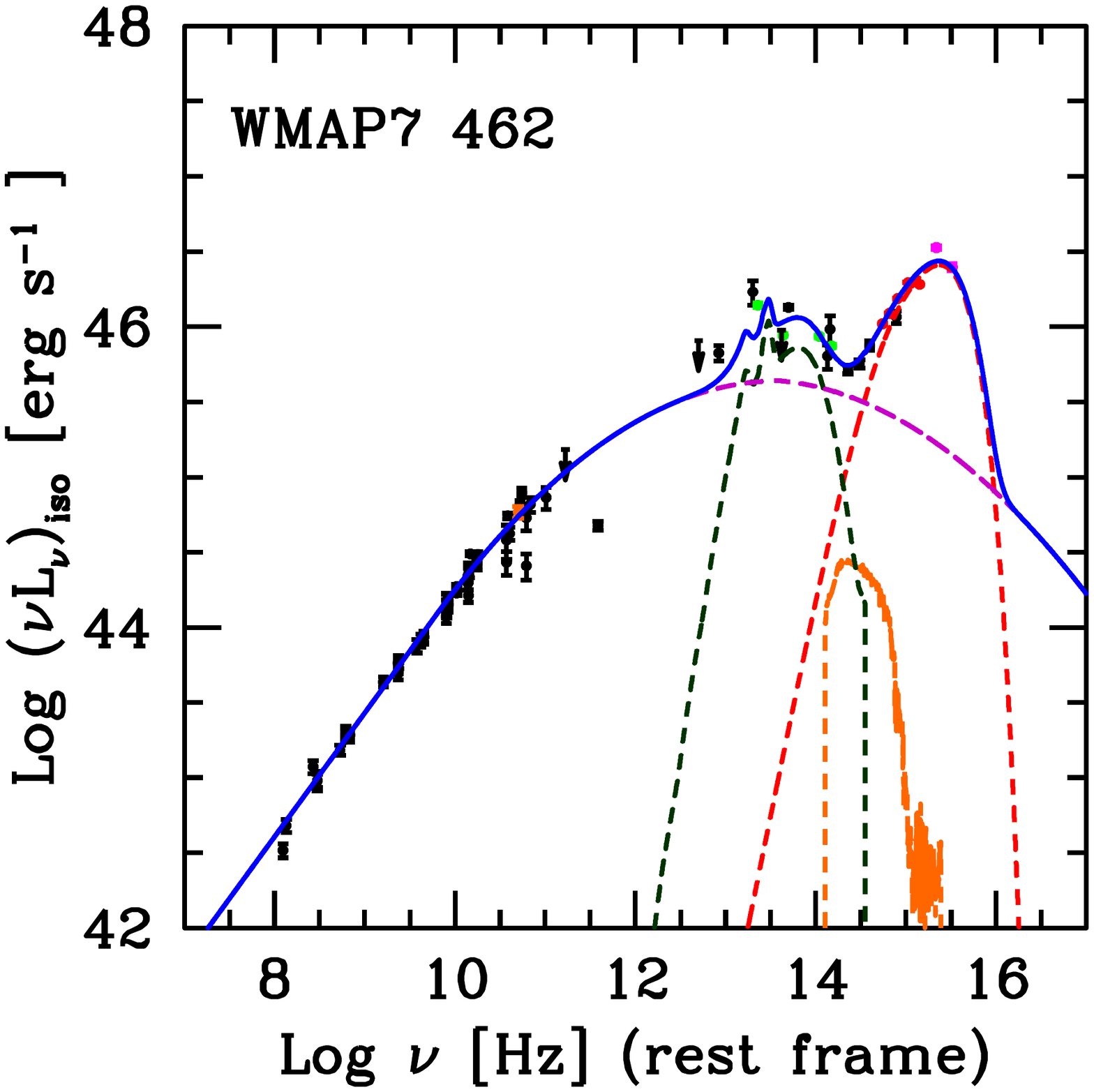}}
\subfloat{\includegraphics[width=0.32\textwidth,natwidth=610,natheight=642]{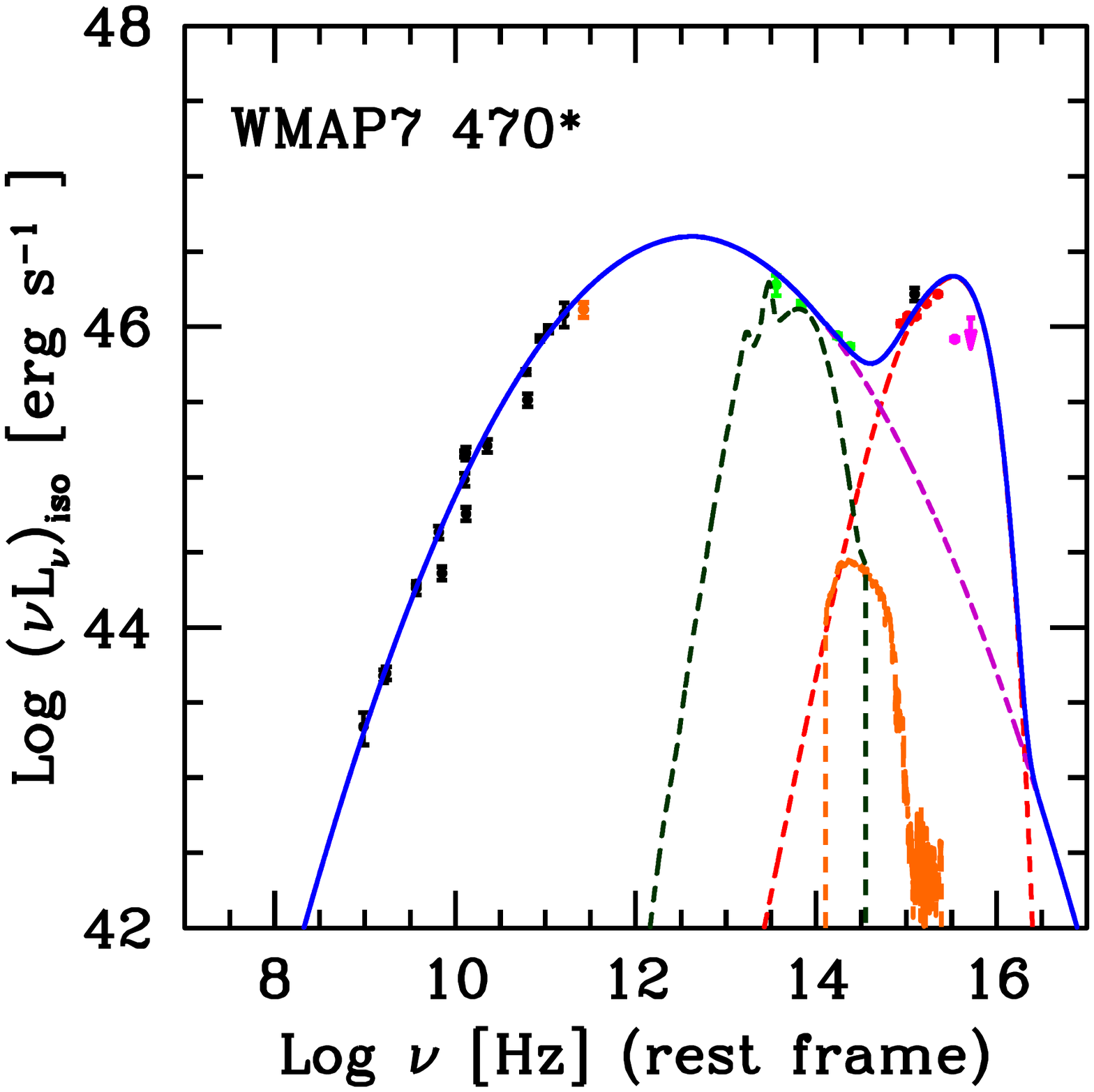}}
\caption{Continued.}
\end{figure*}

\begin{figure*} \centering
\subfloat{\includegraphics[width=0.32\textwidth,natwidth=610,natheight=642]{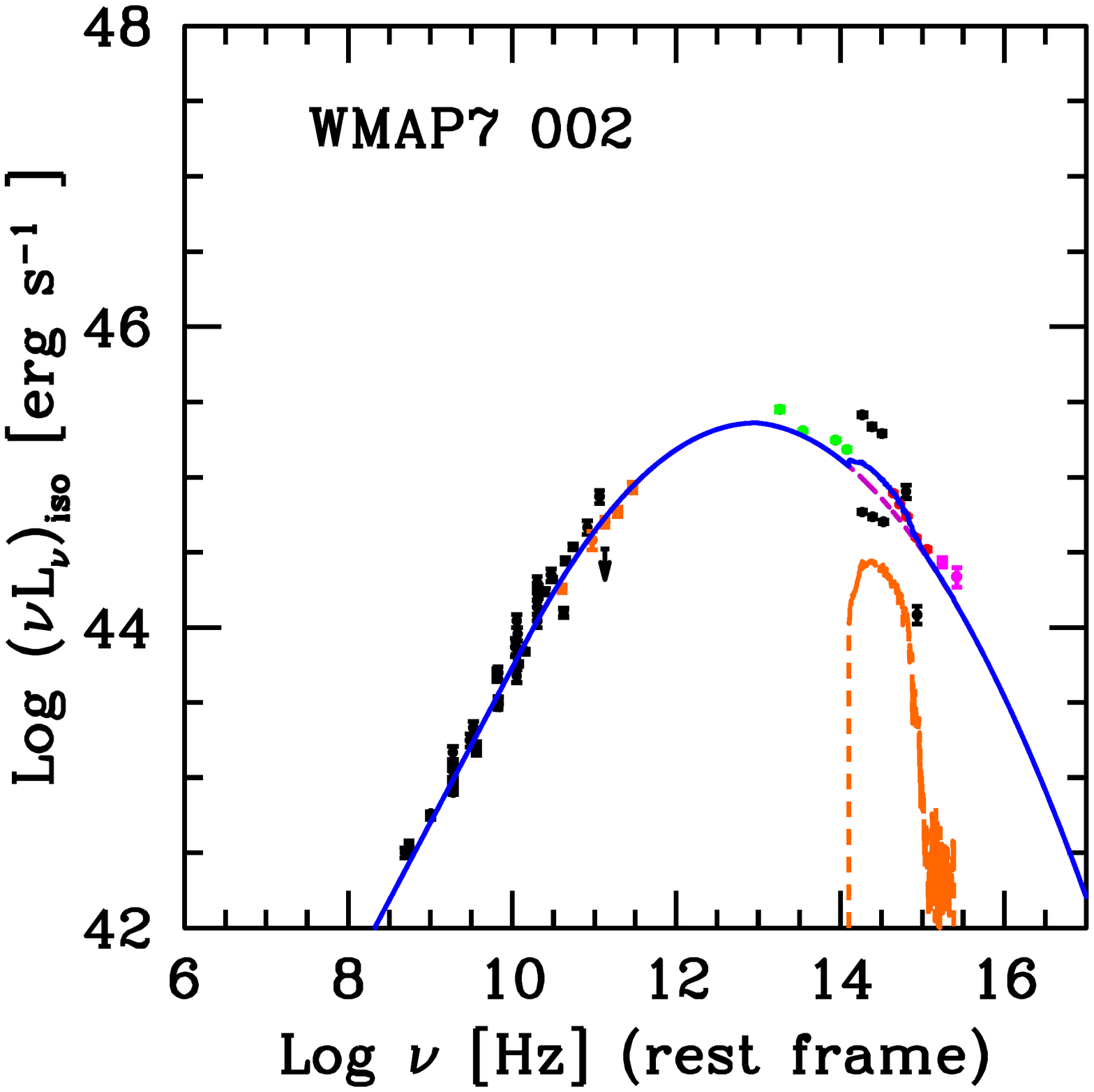}}
\subfloat{\includegraphics[width=0.32\textwidth,natwidth=610,natheight=642]{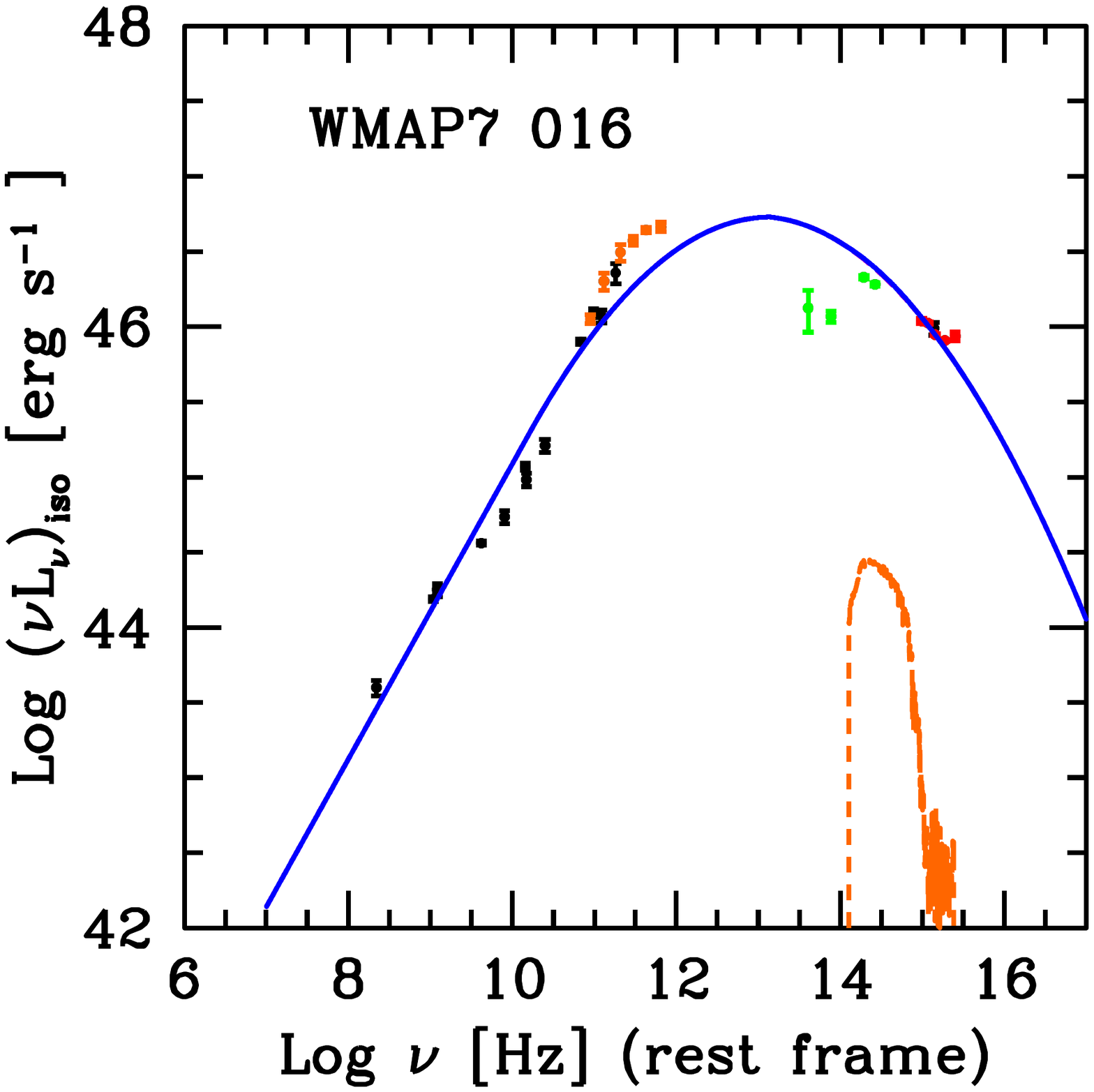}}
\subfloat{\includegraphics[width=0.32\textwidth,natwidth=610,natheight=642]{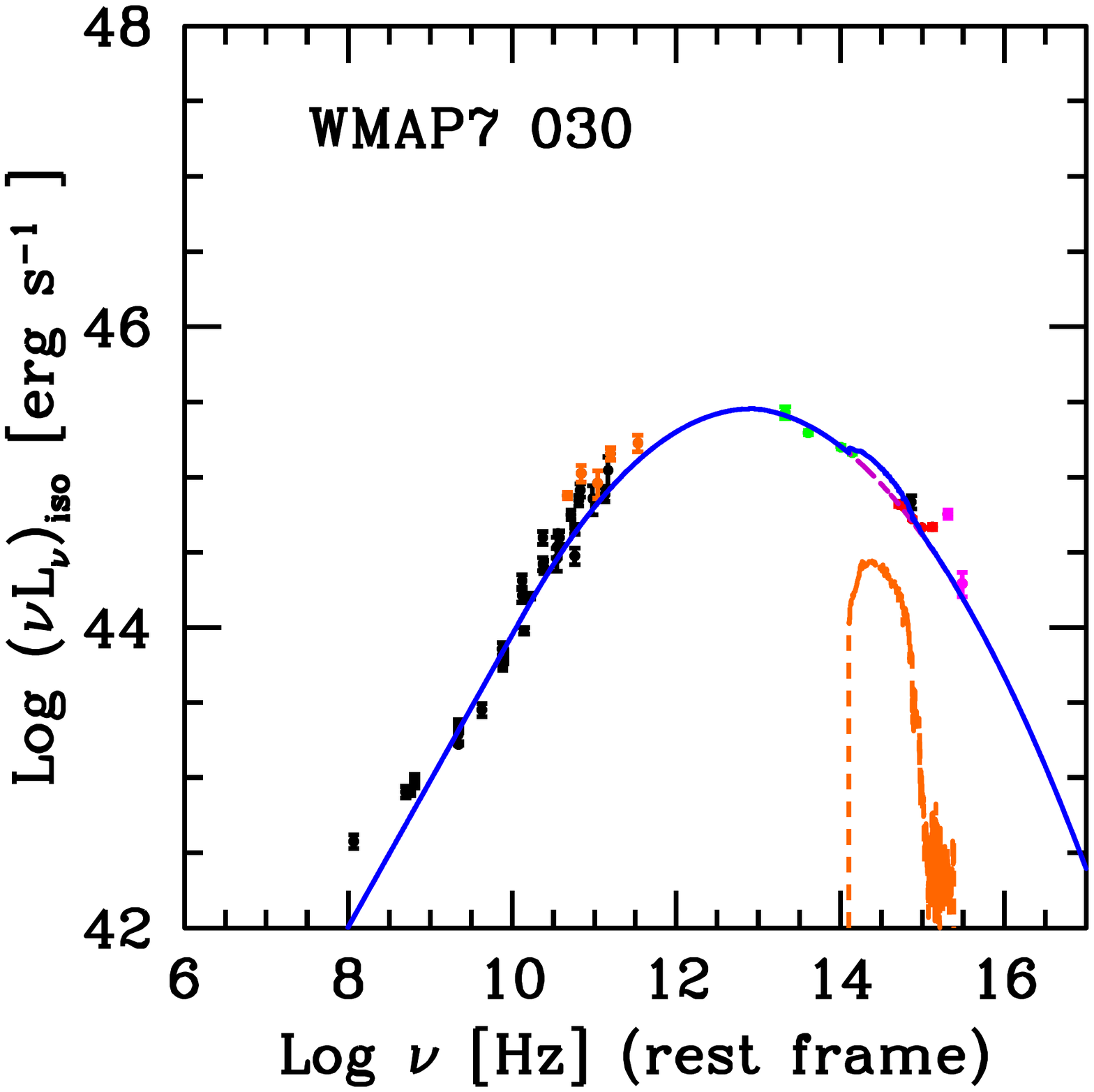}}\qquad
\subfloat{\includegraphics[width=0.32\textwidth,natwidth=610,natheight=642]{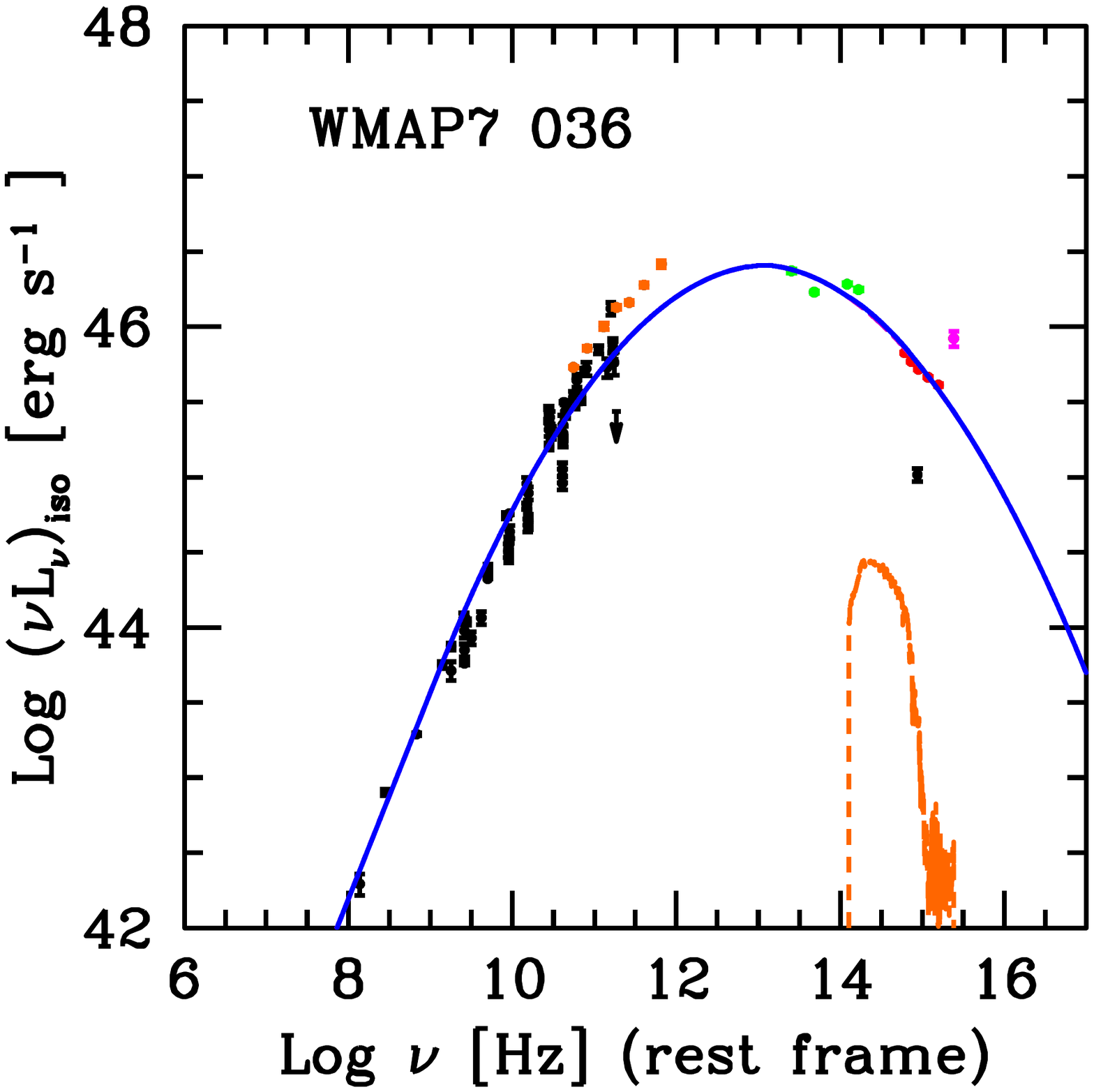}}
\subfloat{\includegraphics[width=0.32\textwidth,natwidth=610,natheight=642]{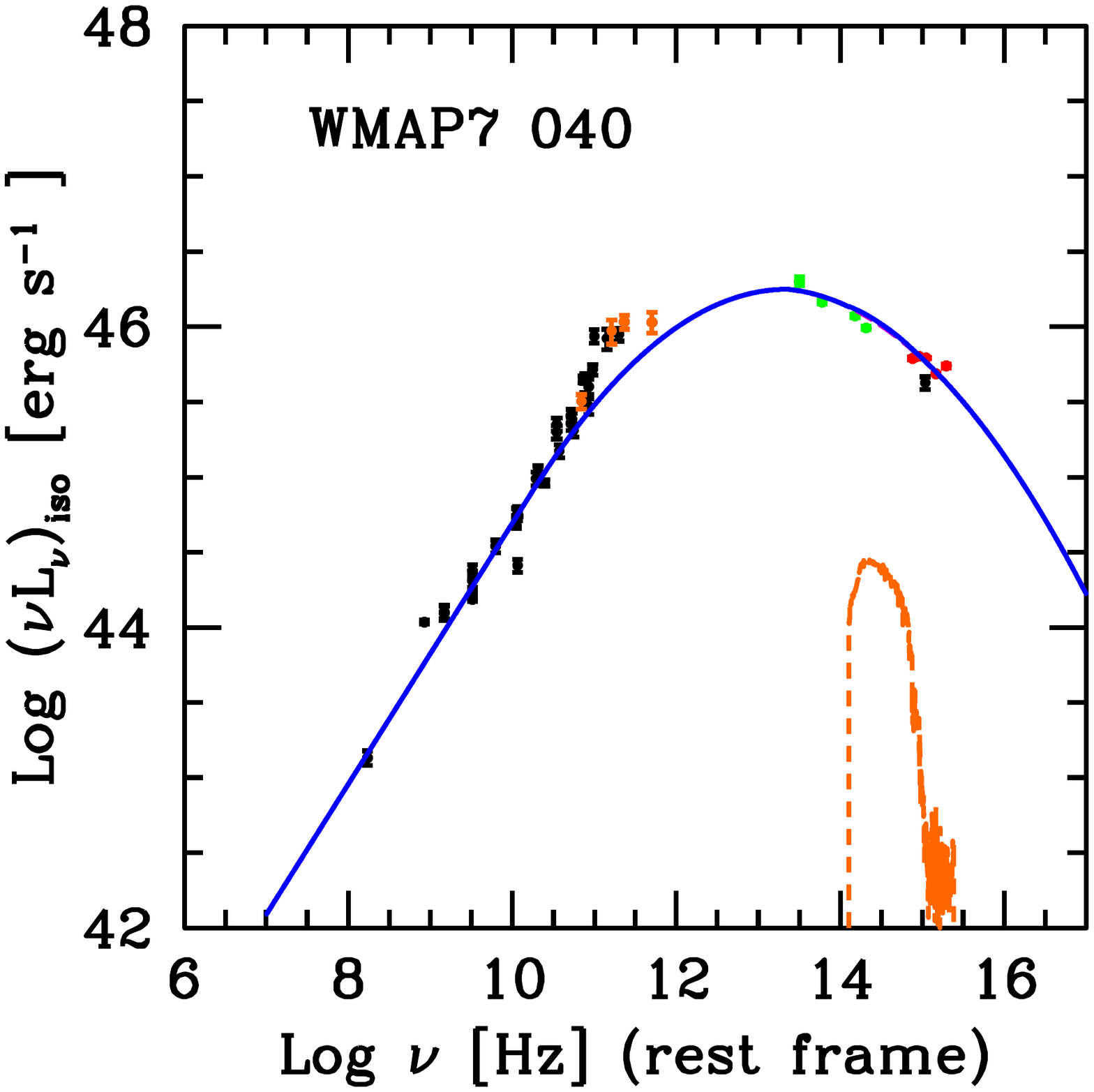}}
\subfloat{\includegraphics[width=0.32\textwidth,natwidth=610,natheight=642]{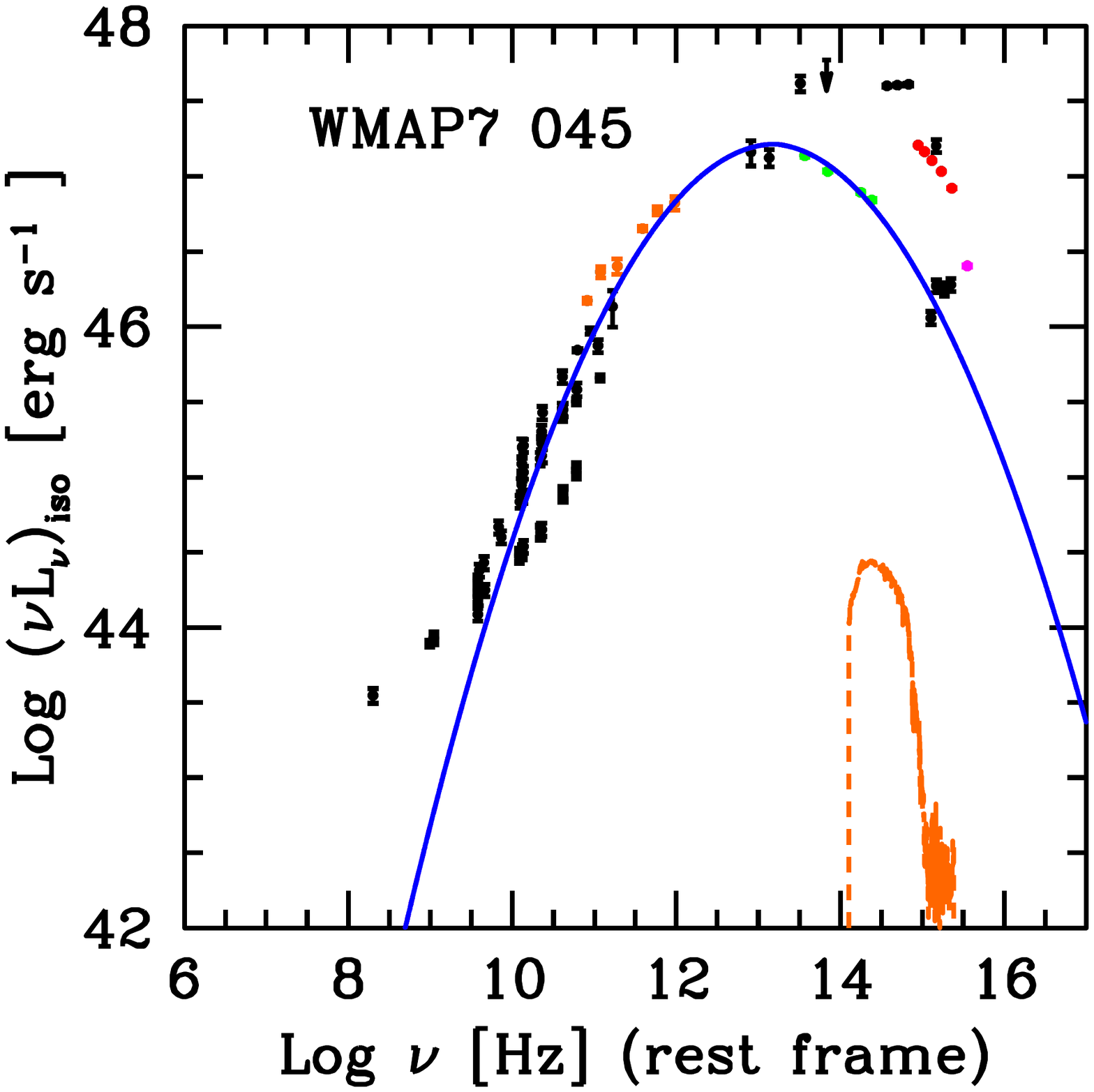}}\qquad
\subfloat{\includegraphics[width=0.32\textwidth,natwidth=610,natheight=642]{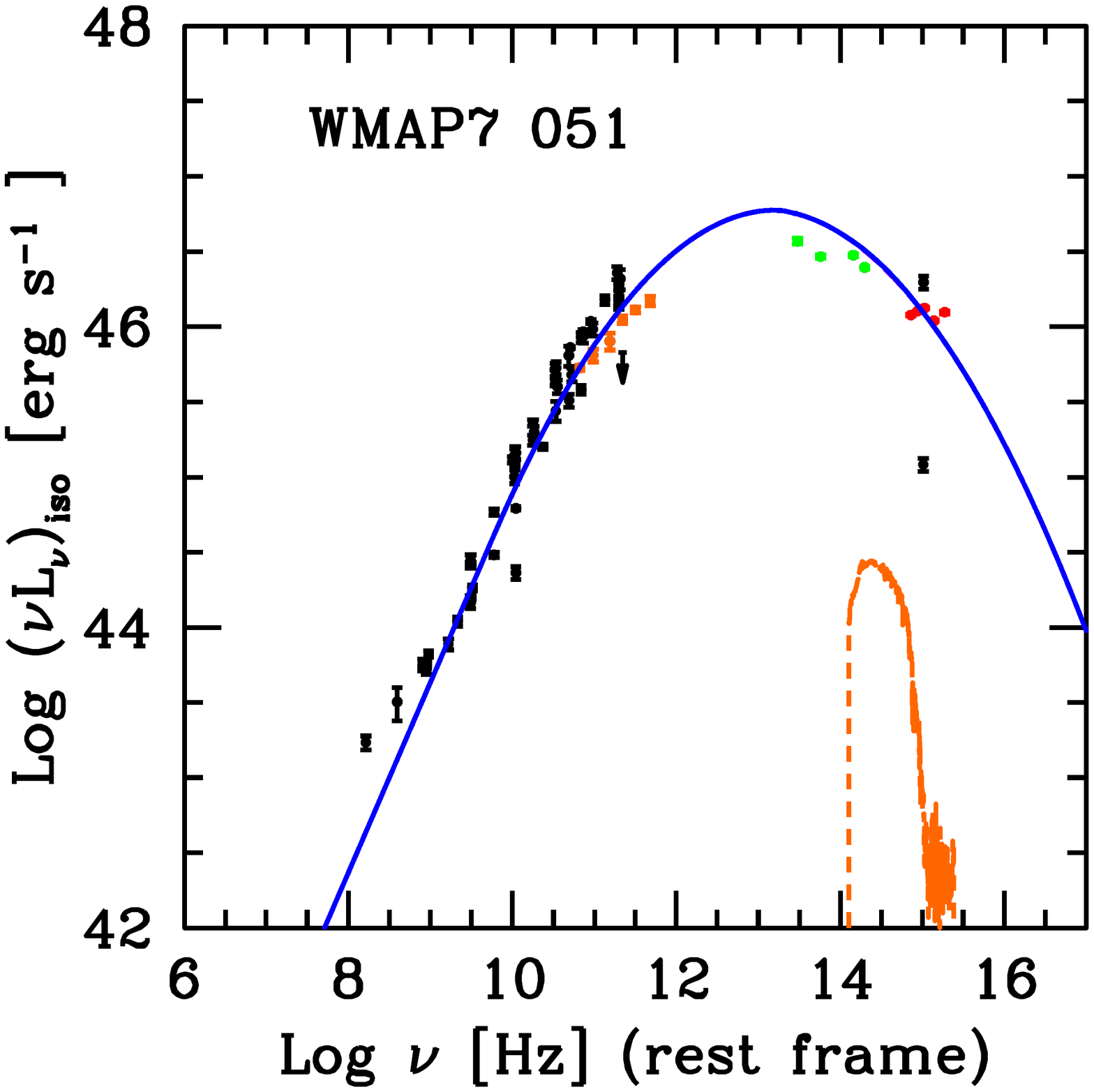}}
\subfloat{\includegraphics[width=0.32\textwidth,natwidth=610,natheight=642]{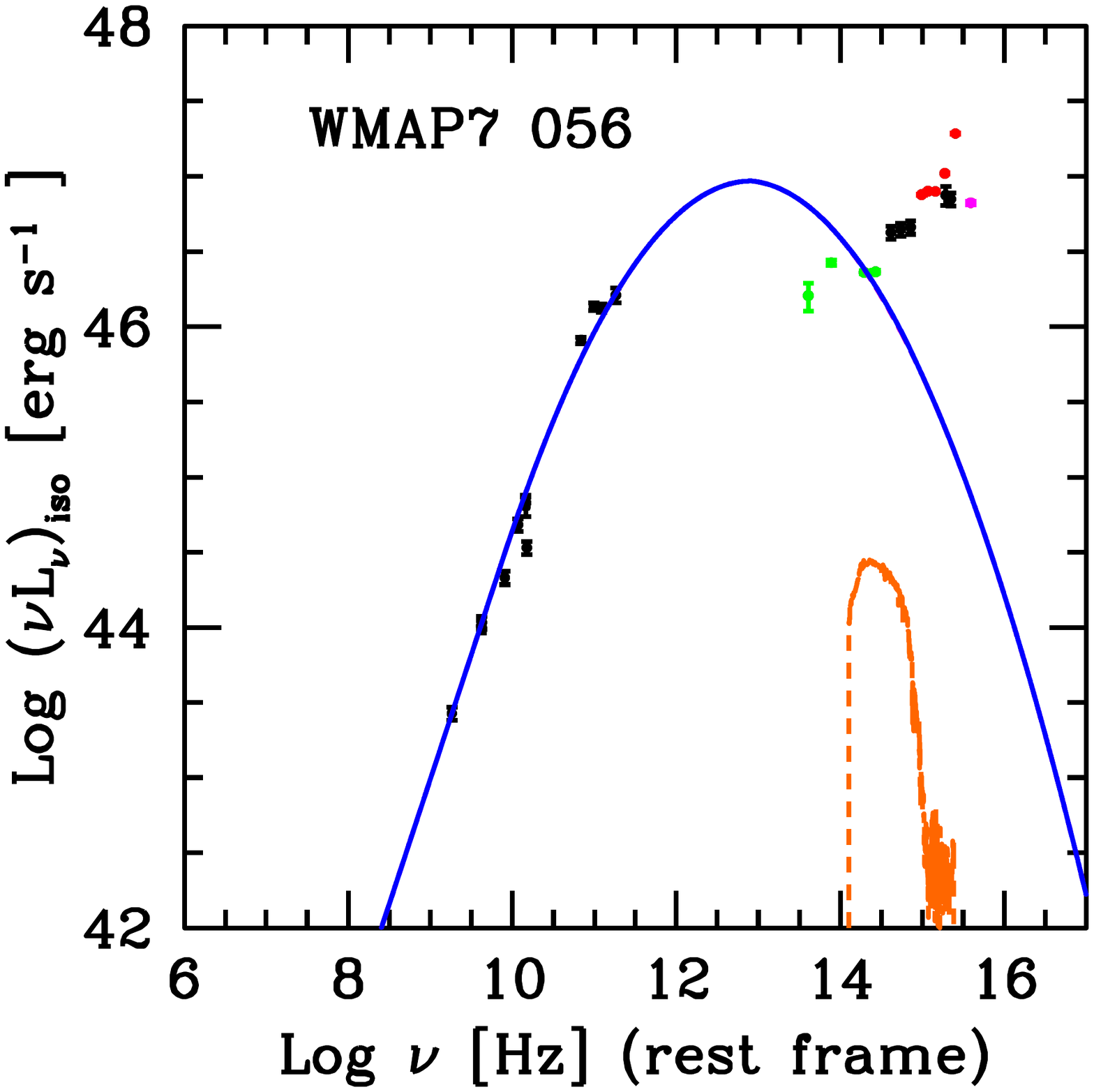}}
\subfloat{\includegraphics[width=0.32\textwidth,natwidth=610,natheight=642]{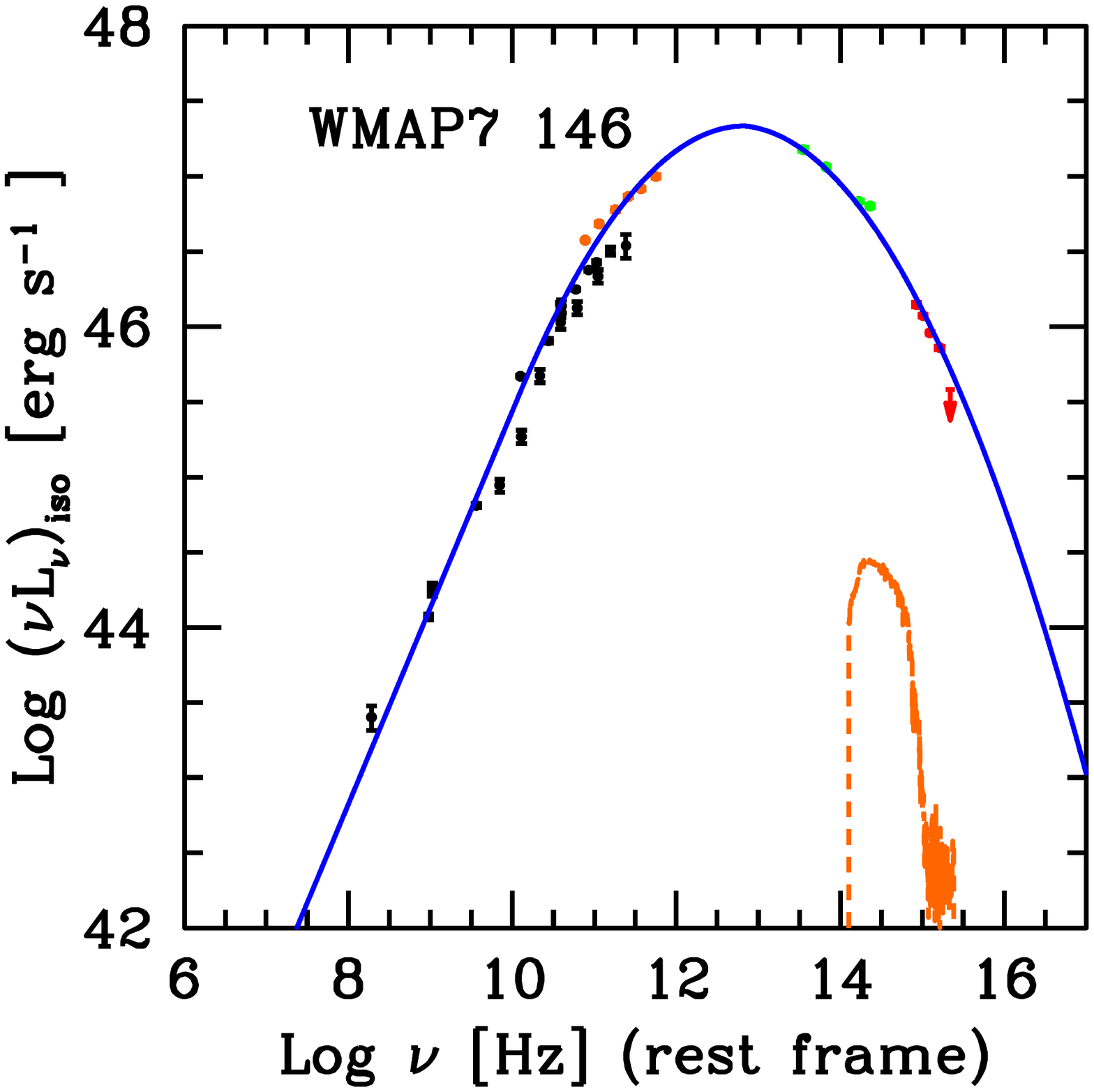}}\qquad
\caption{ SEDs of the 25 FSRQs in the sample with no clear evidence of optical/UV bump. The meaning of lines and data points is the same as in Fig.~\ref{fig:SEDs_FSRQs_with_torus}. Torus and disc templates are not included in the modeling.}
\label{fig:SEDs_FSRQs_SDSS_NOdisc}
\end{figure*}

\begin{figure*} \centering
\ContinuedFloat
\subfloat{\includegraphics[width=0.32\textwidth,natwidth=610,natheight=642]{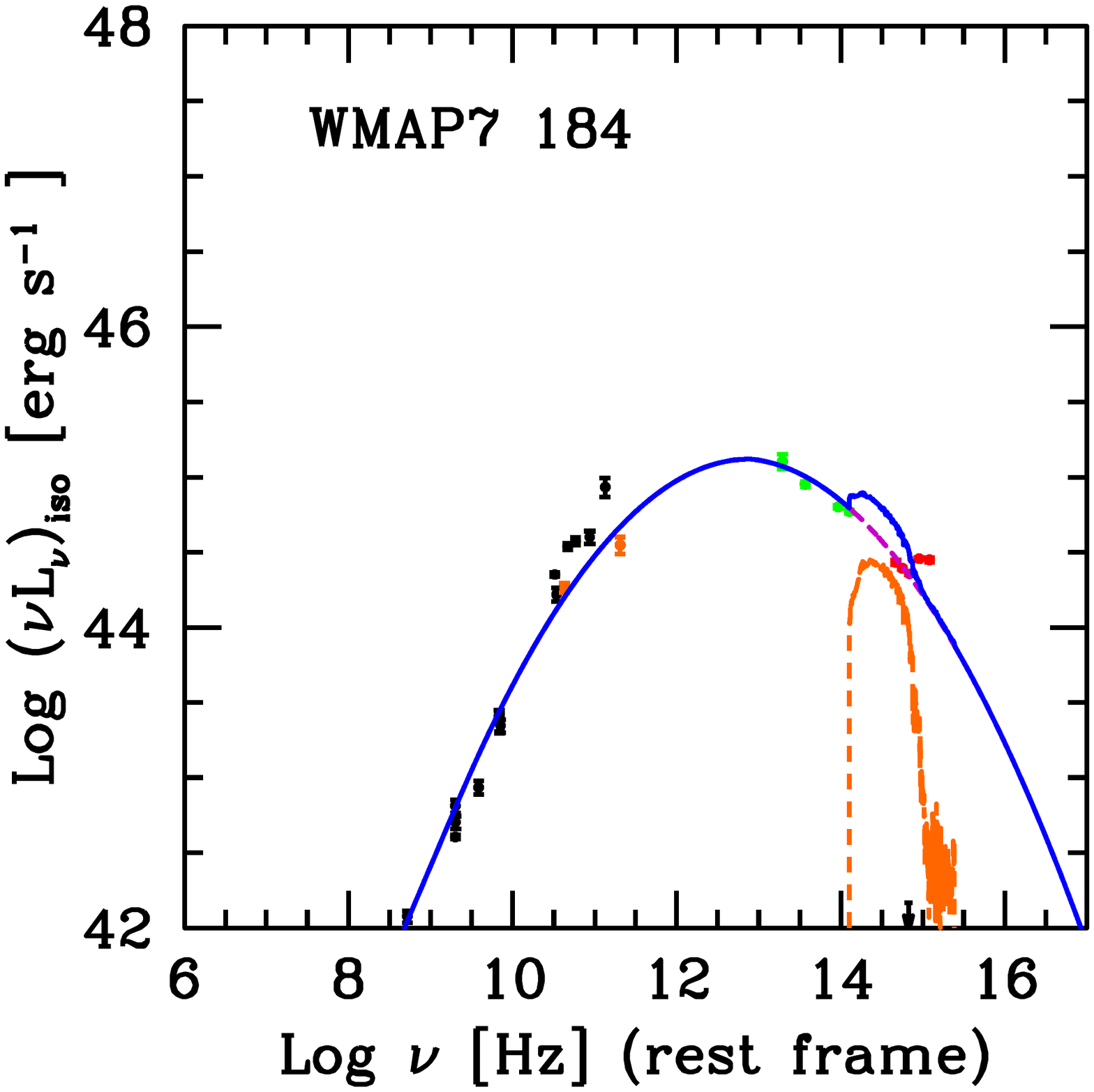}}
\subfloat{\includegraphics[width=0.32\textwidth,natwidth=610,natheight=642]{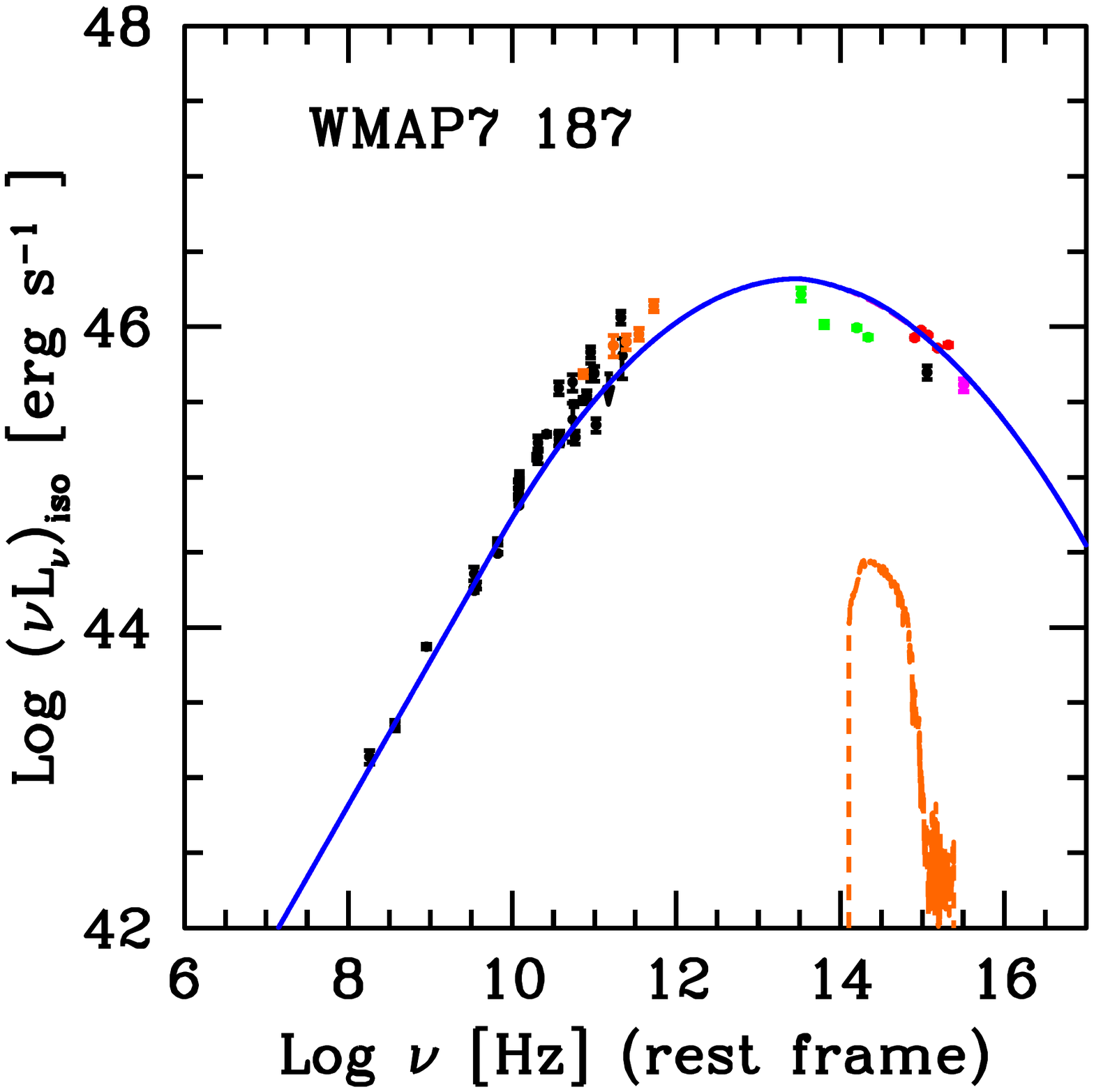}}
\subfloat{\includegraphics[width=0.32\textwidth,natwidth=610,natheight=642]{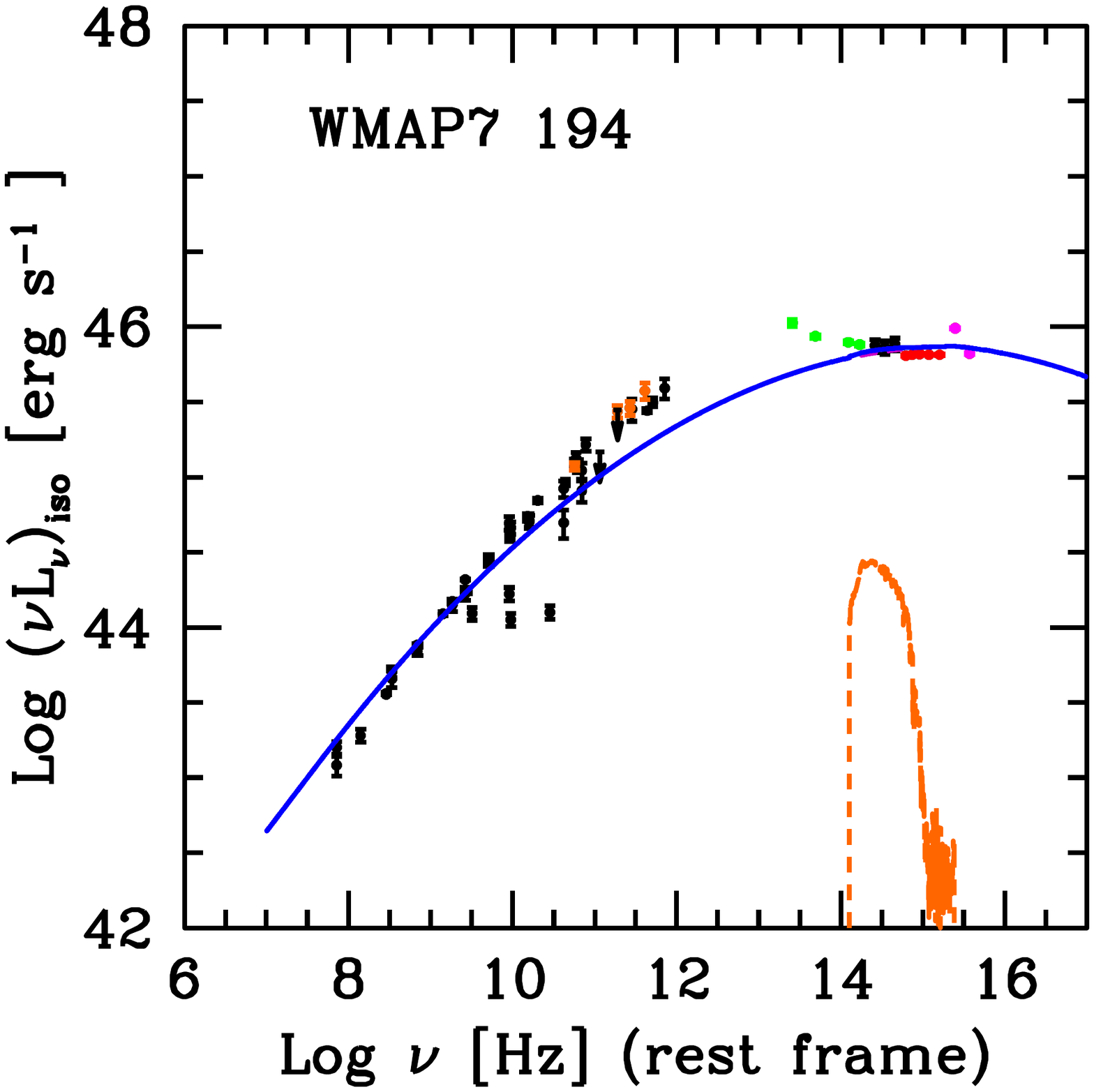}}\qquad
\subfloat{\includegraphics[width=0.32\textwidth,natwidth=610,natheight=642]{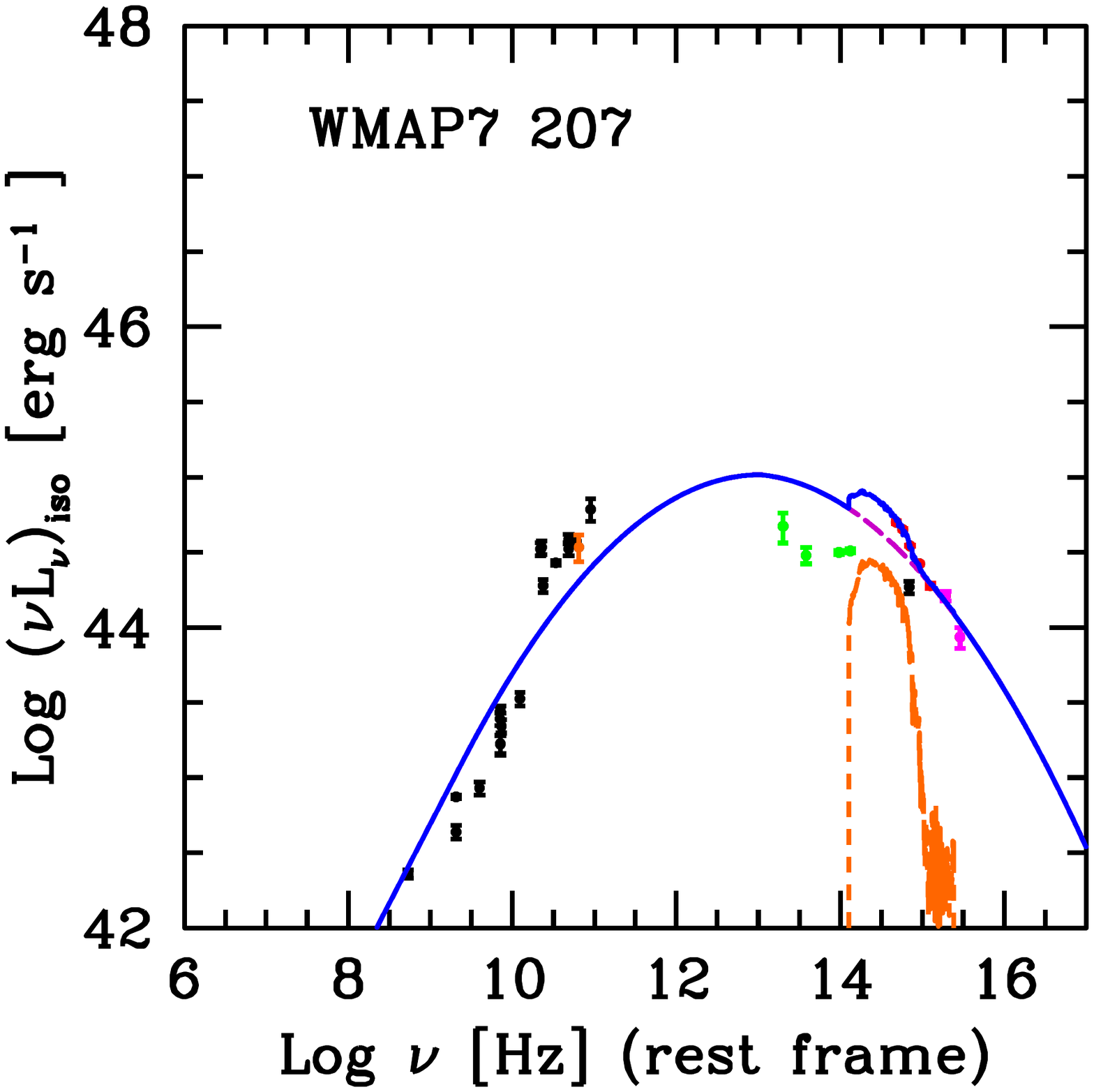}}
\subfloat{\includegraphics[width=0.32\textwidth,natwidth=610,natheight=642]{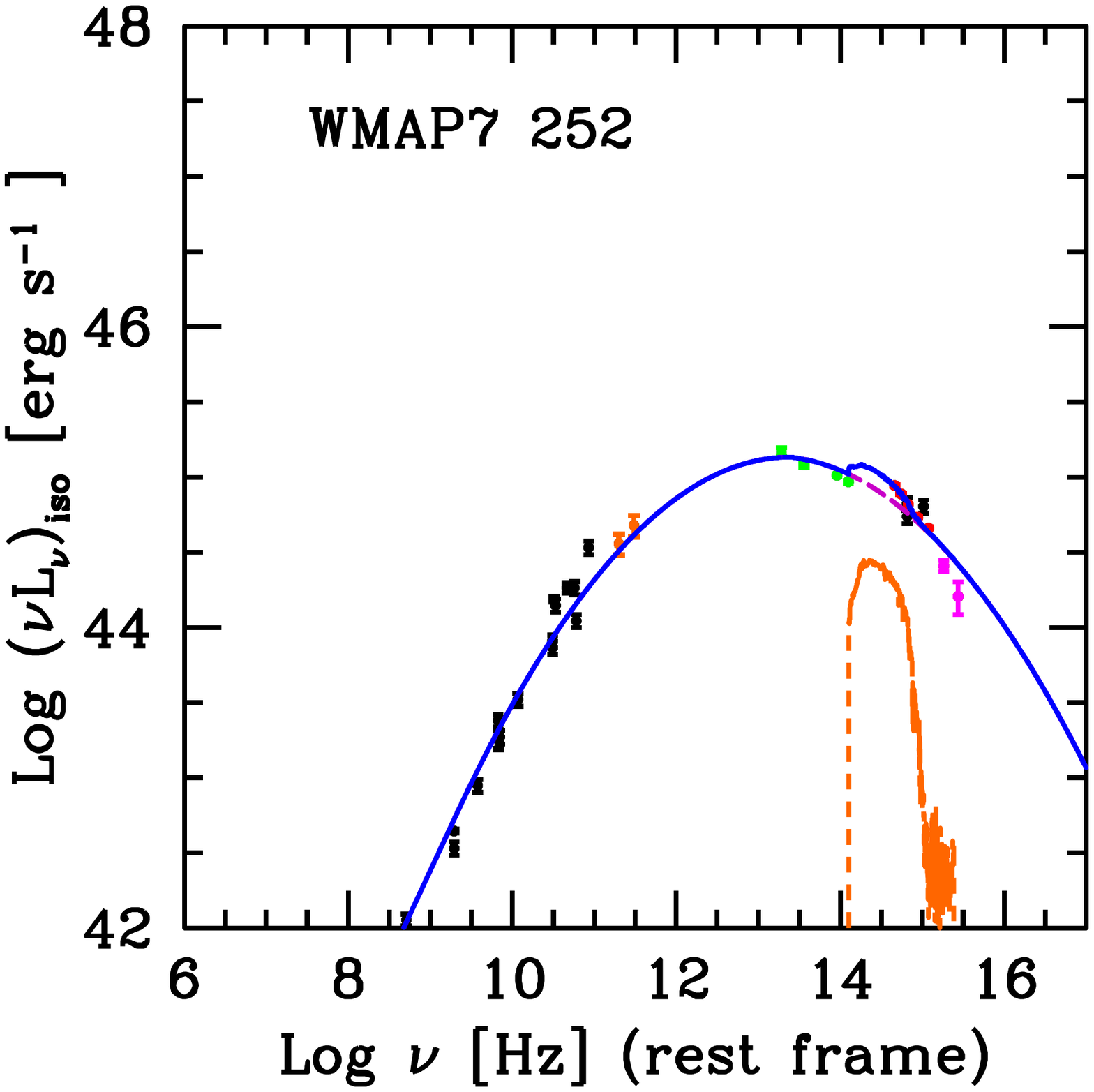}}
\subfloat{\includegraphics[width=0.32\textwidth,natwidth=610,natheight=642]{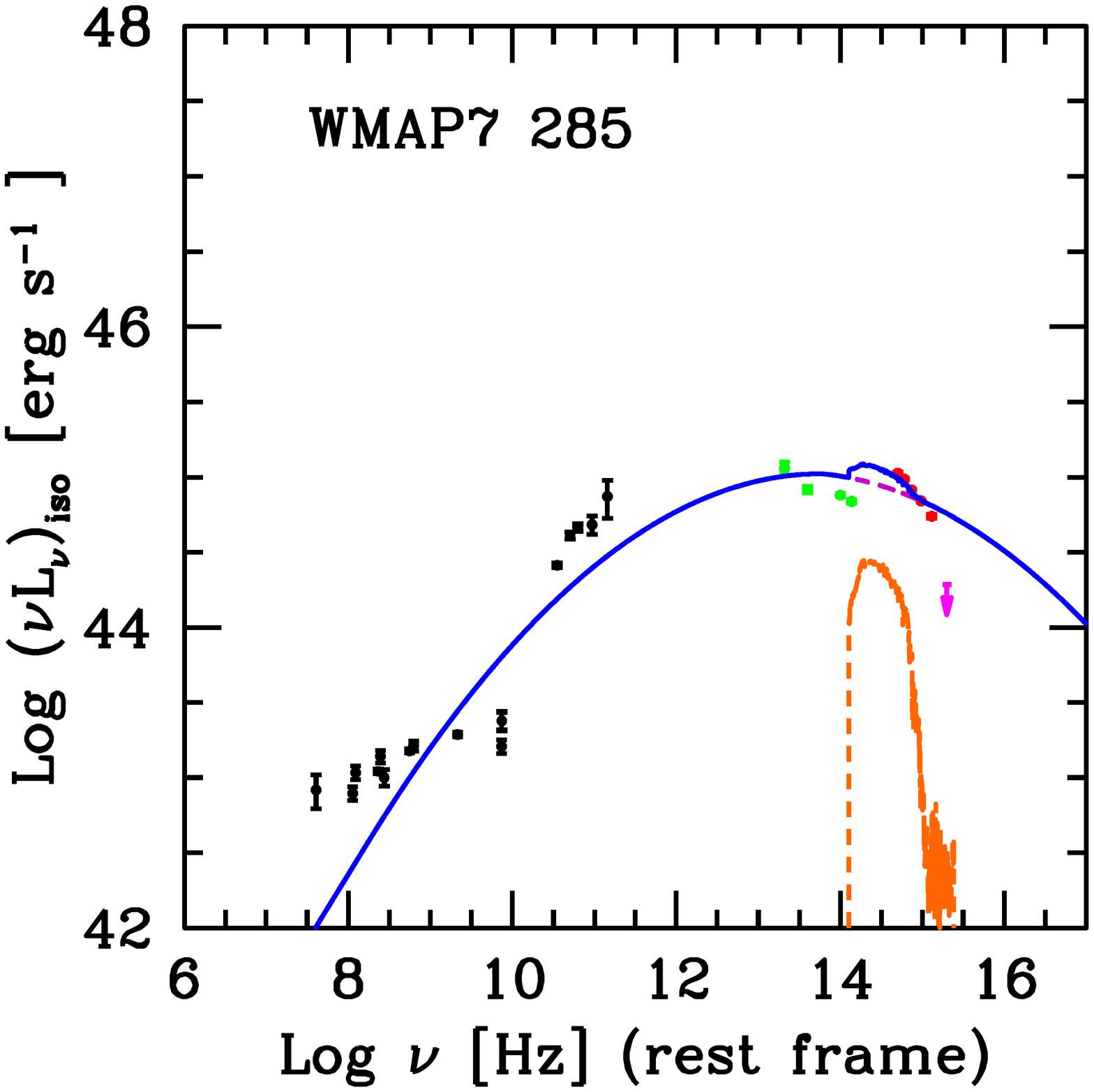}}\qquad
\subfloat{\includegraphics[width=0.32\textwidth,natwidth=610,natheight=642]{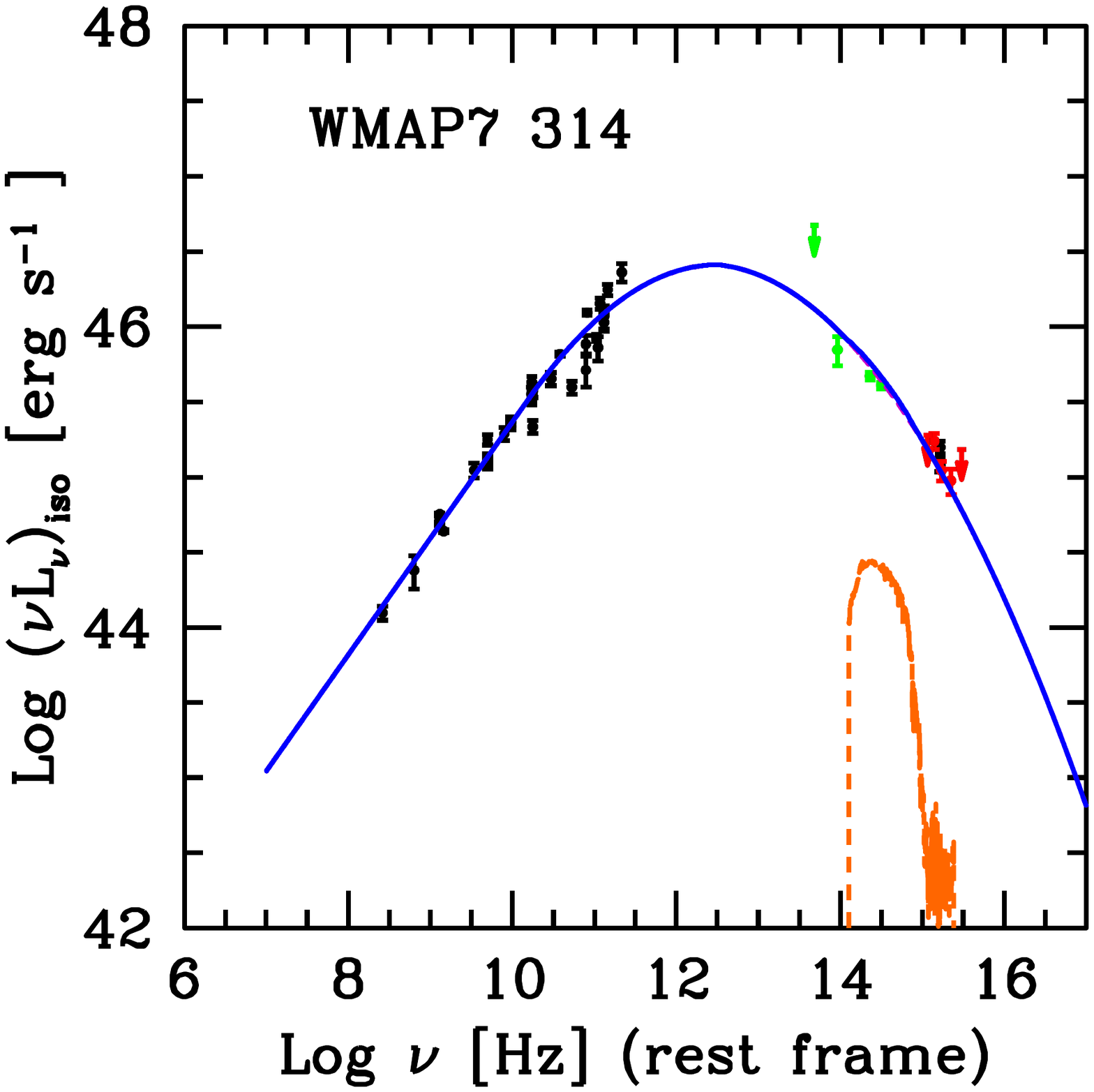}}
\subfloat{\includegraphics[width=0.32\textwidth,natwidth=610,natheight=642]{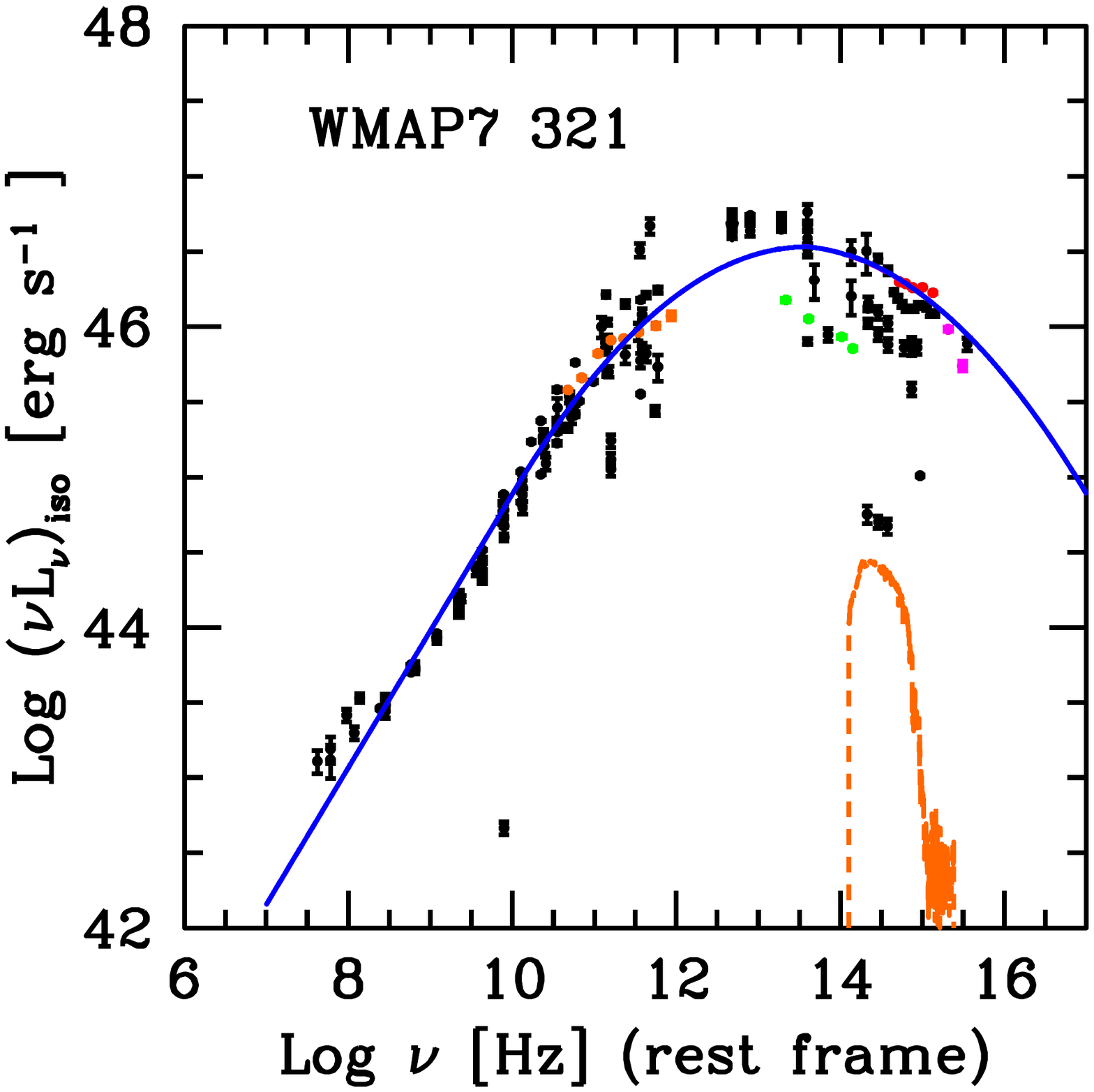}}
\subfloat{\includegraphics[width=0.32\textwidth,natwidth=610,natheight=642]{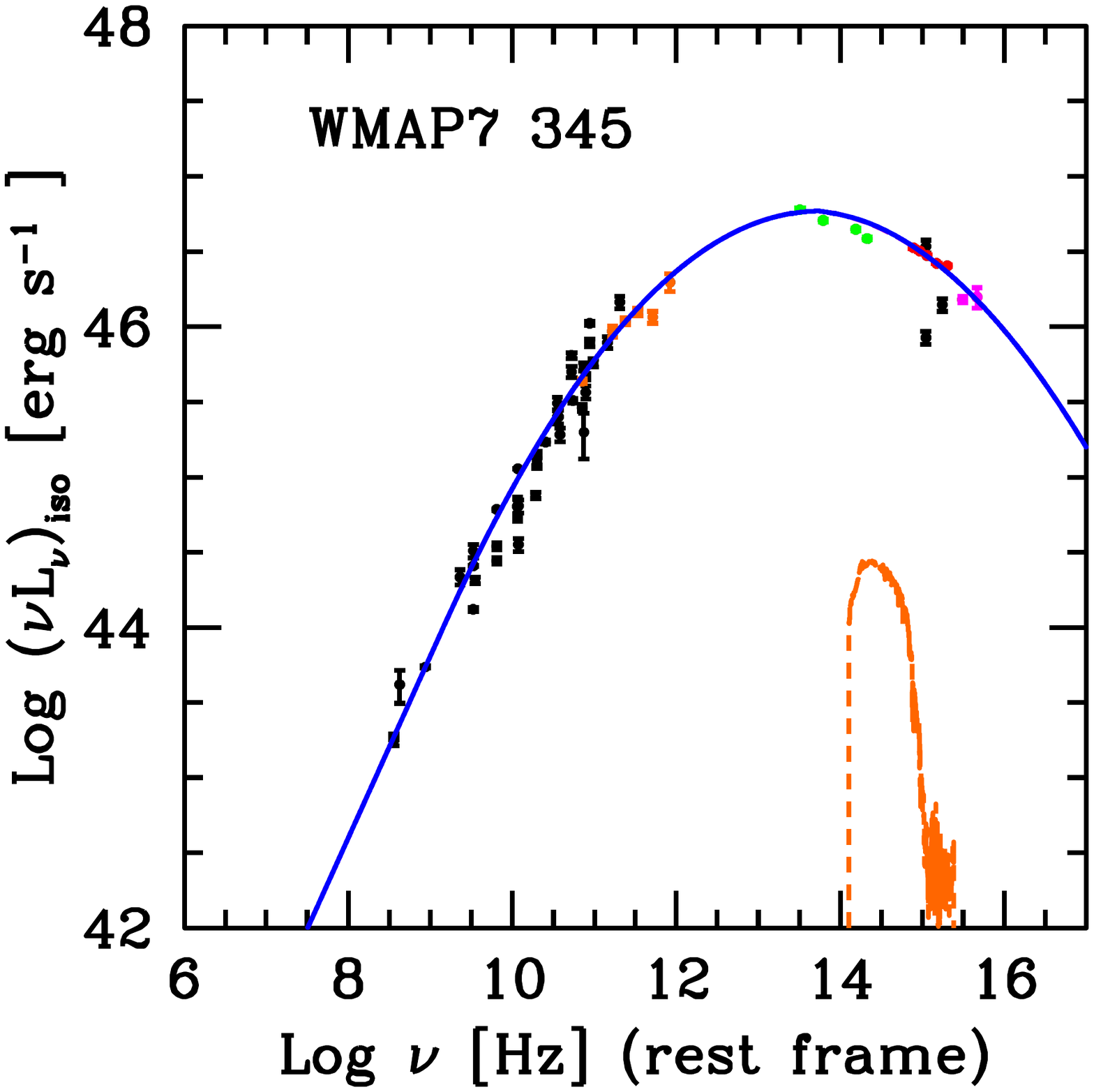}}\qquad
\caption{Continued.}
\end{figure*}

\begin{figure*} \centering
\ContinuedFloat
\subfloat{\includegraphics[width=0.32\textwidth,natwidth=610,natheight=642]{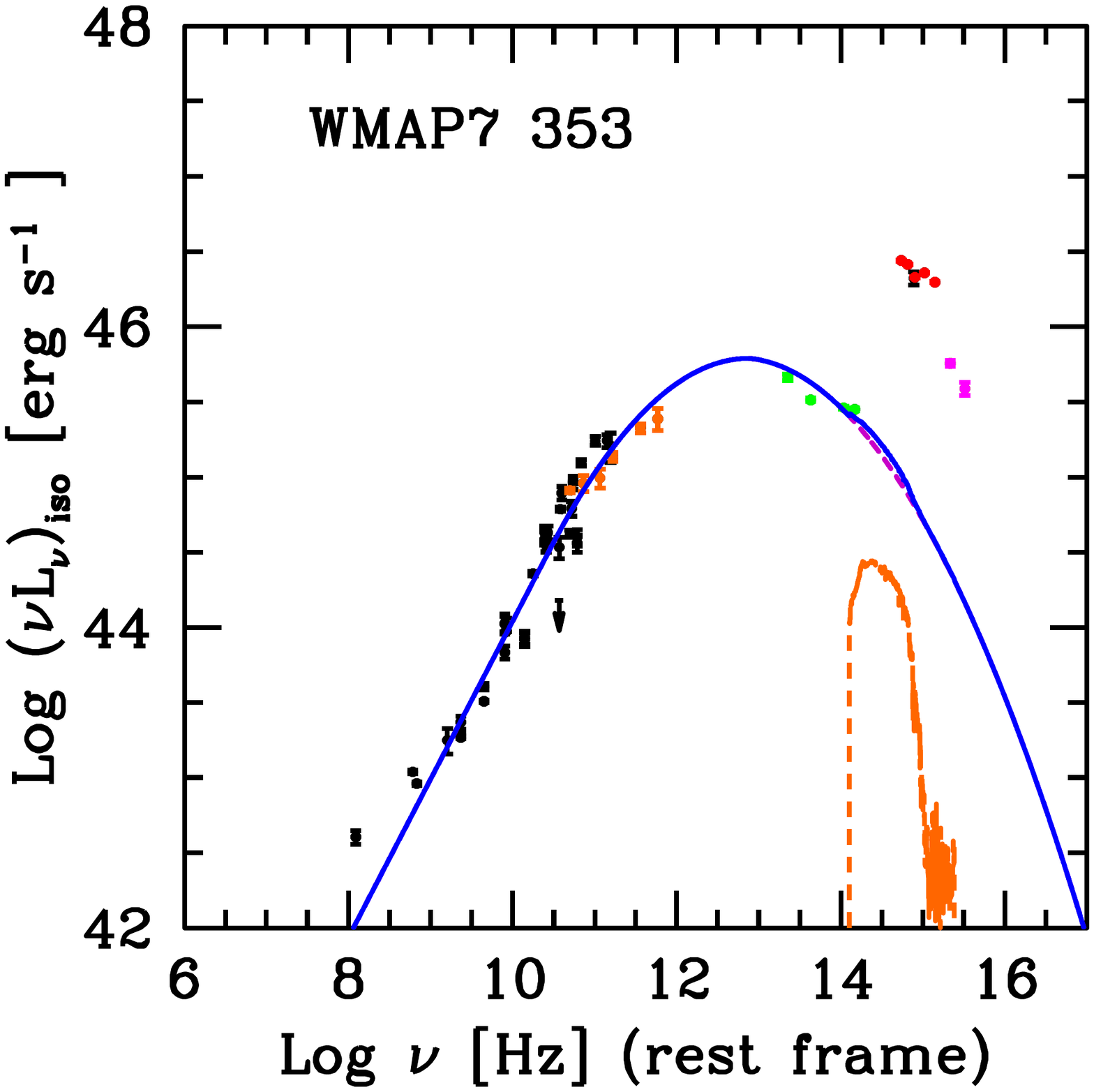}}
\subfloat{\includegraphics[width=0.32\textwidth,natwidth=610,natheight=642]{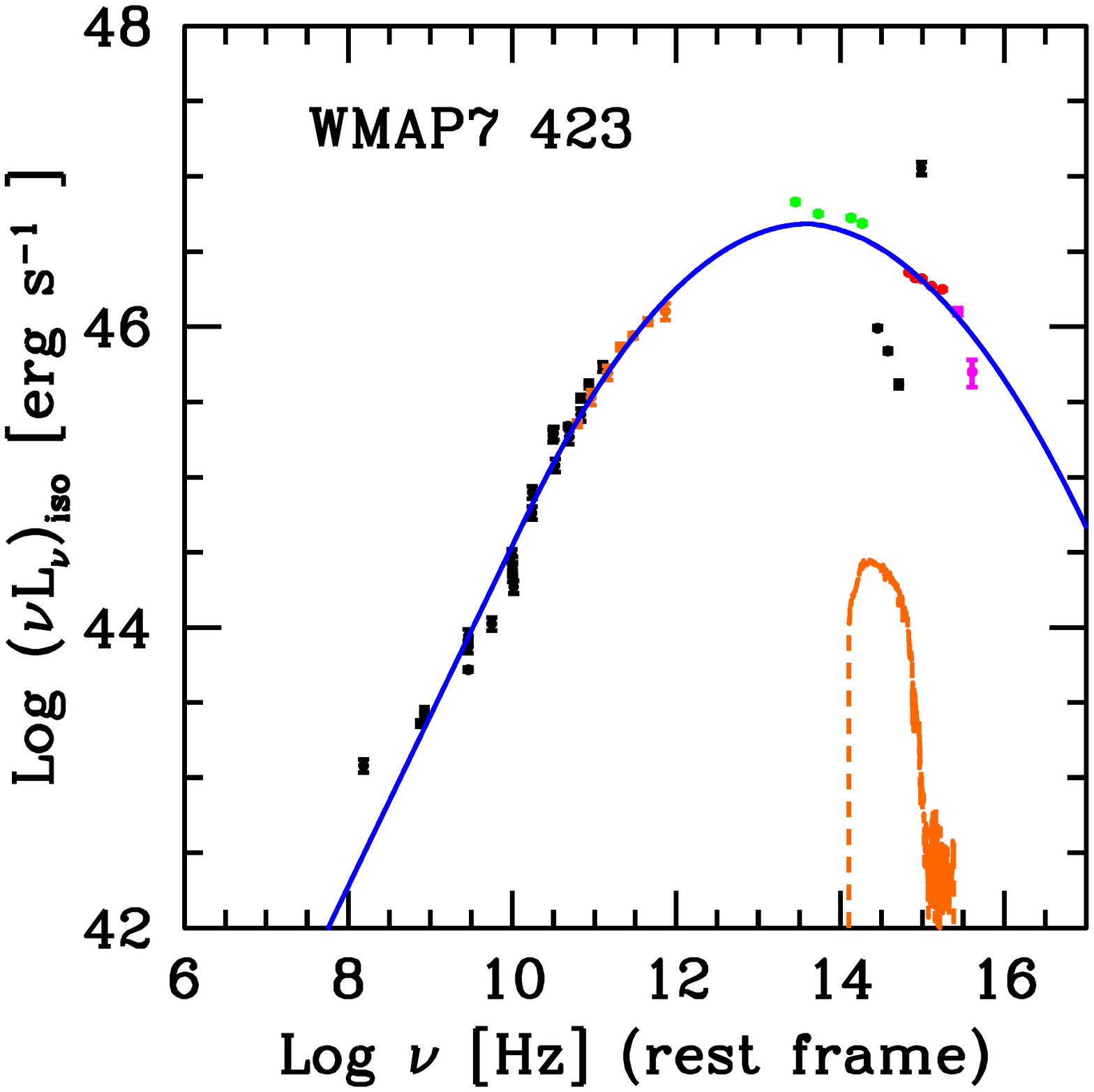}}
\subfloat{\includegraphics[width=0.32\textwidth,natwidth=610,natheight=642]{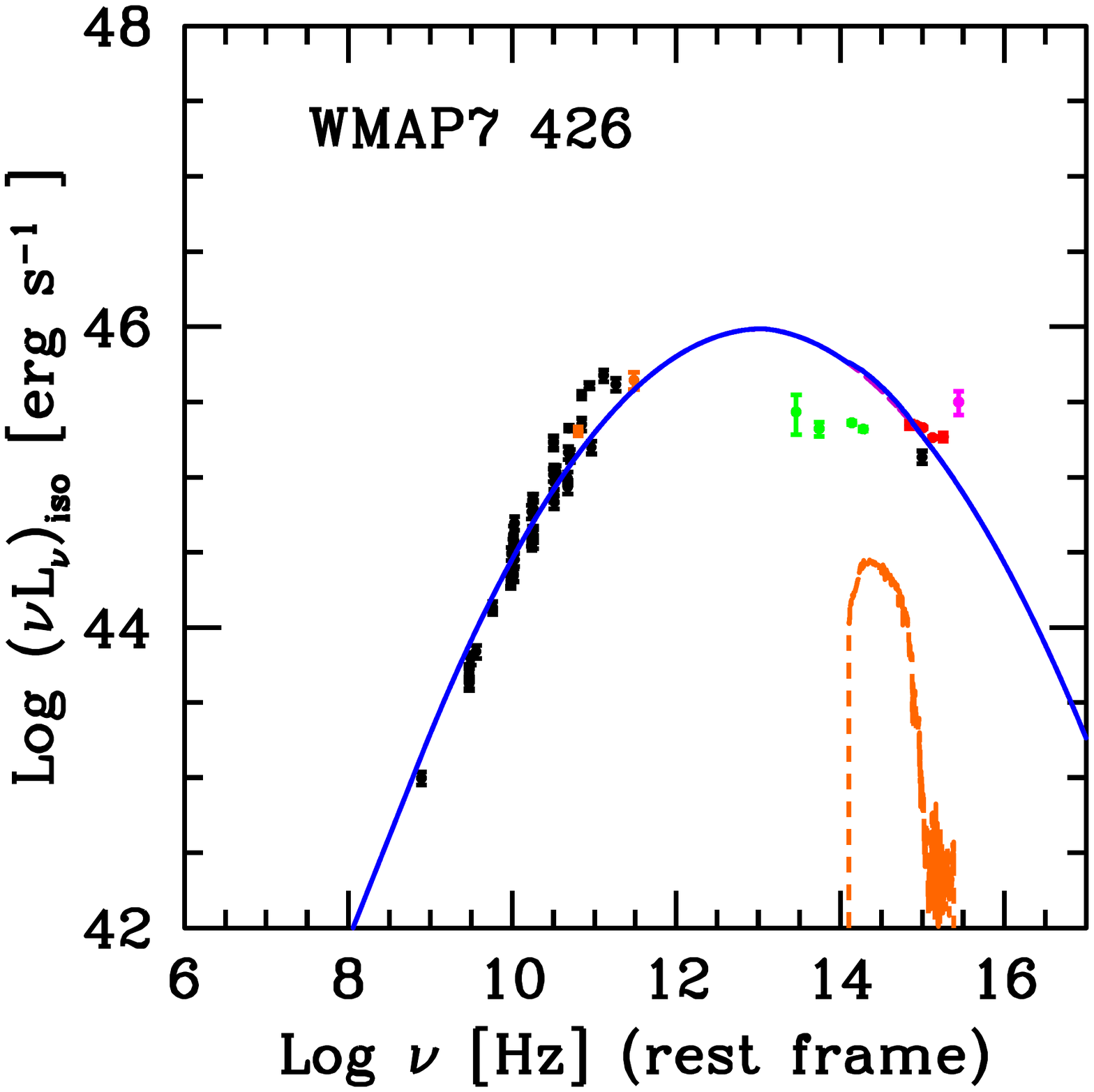}}\qquad
\subfloat{\includegraphics[width=0.32\textwidth,natwidth=610,natheight=642]{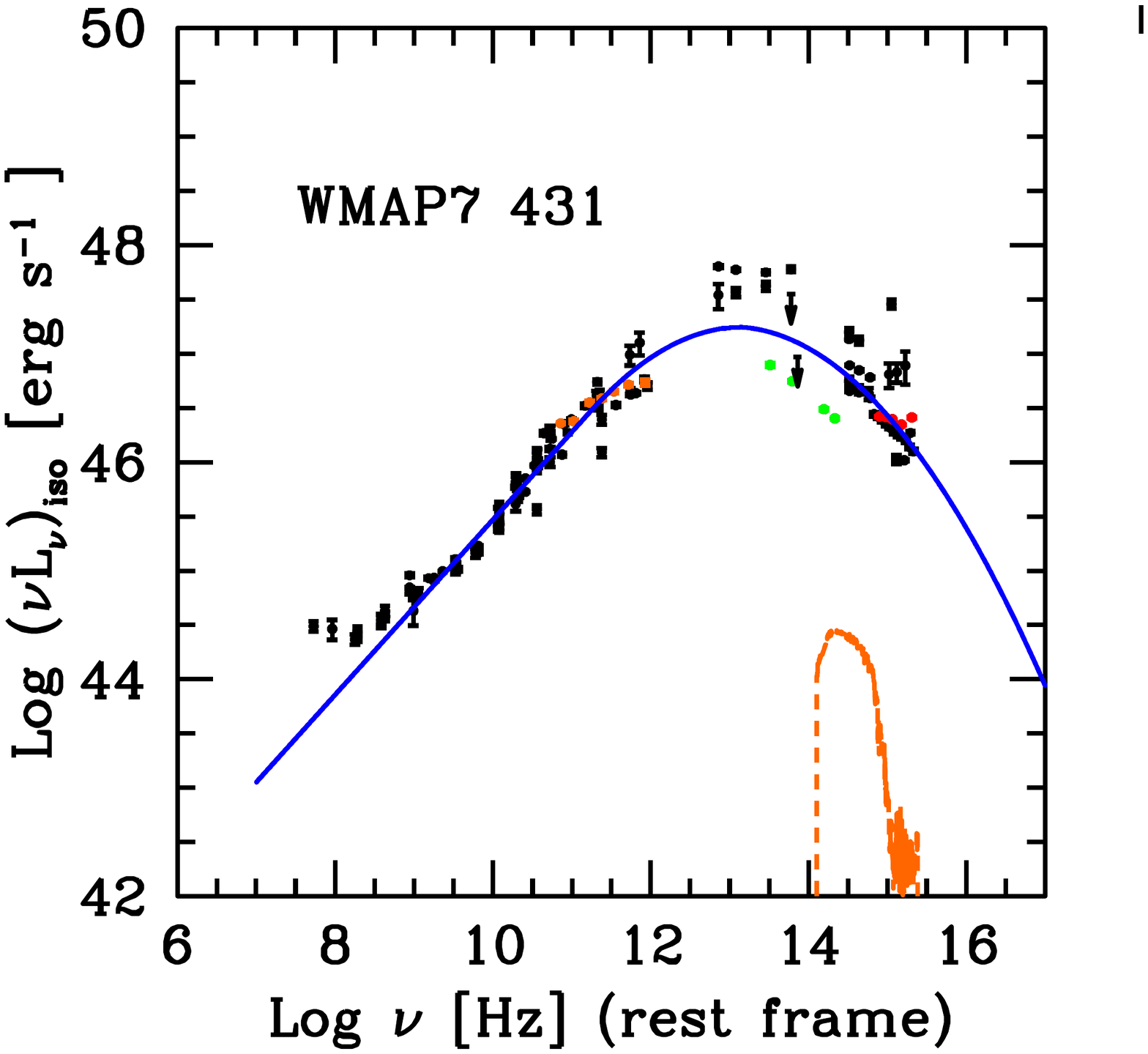}}
\subfloat{\includegraphics[width=0.32\textwidth,natwidth=610,natheight=642]{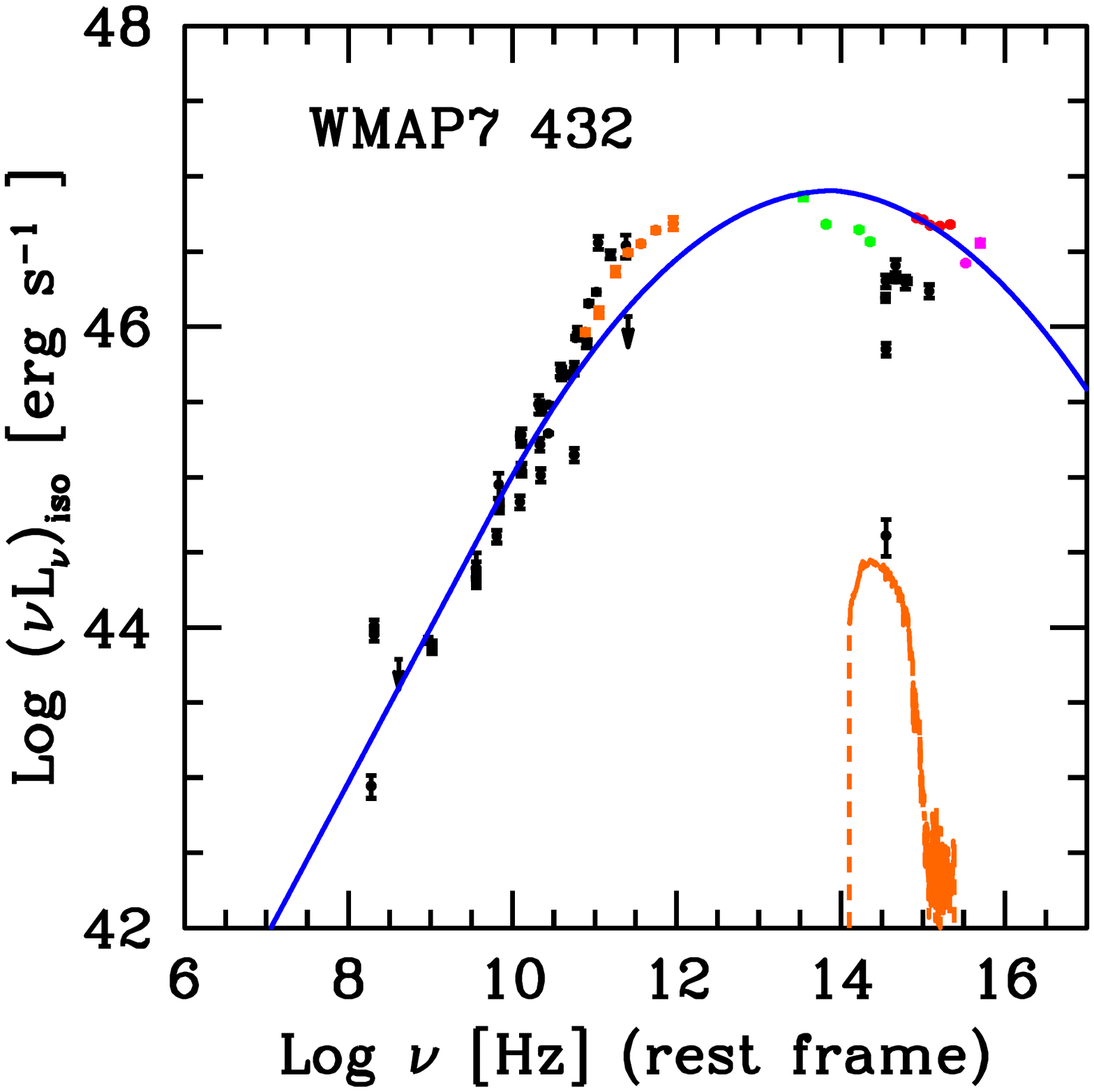}}
\subfloat{\includegraphics[width=0.32\textwidth,natwidth=610,natheight=642]{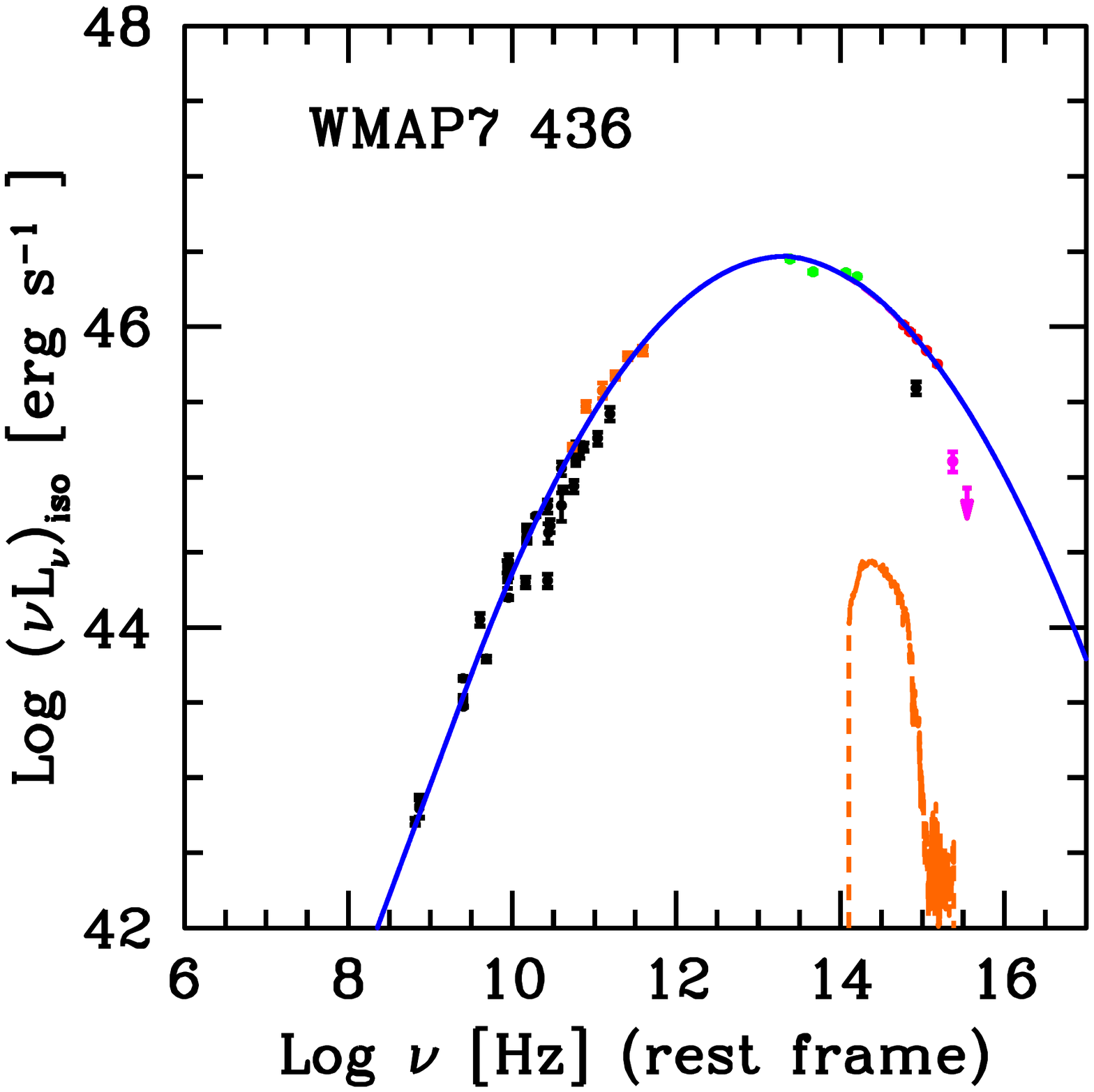}}\qquad
\subfloat{\includegraphics[width=0.32\textwidth,natwidth=610,natheight=642]{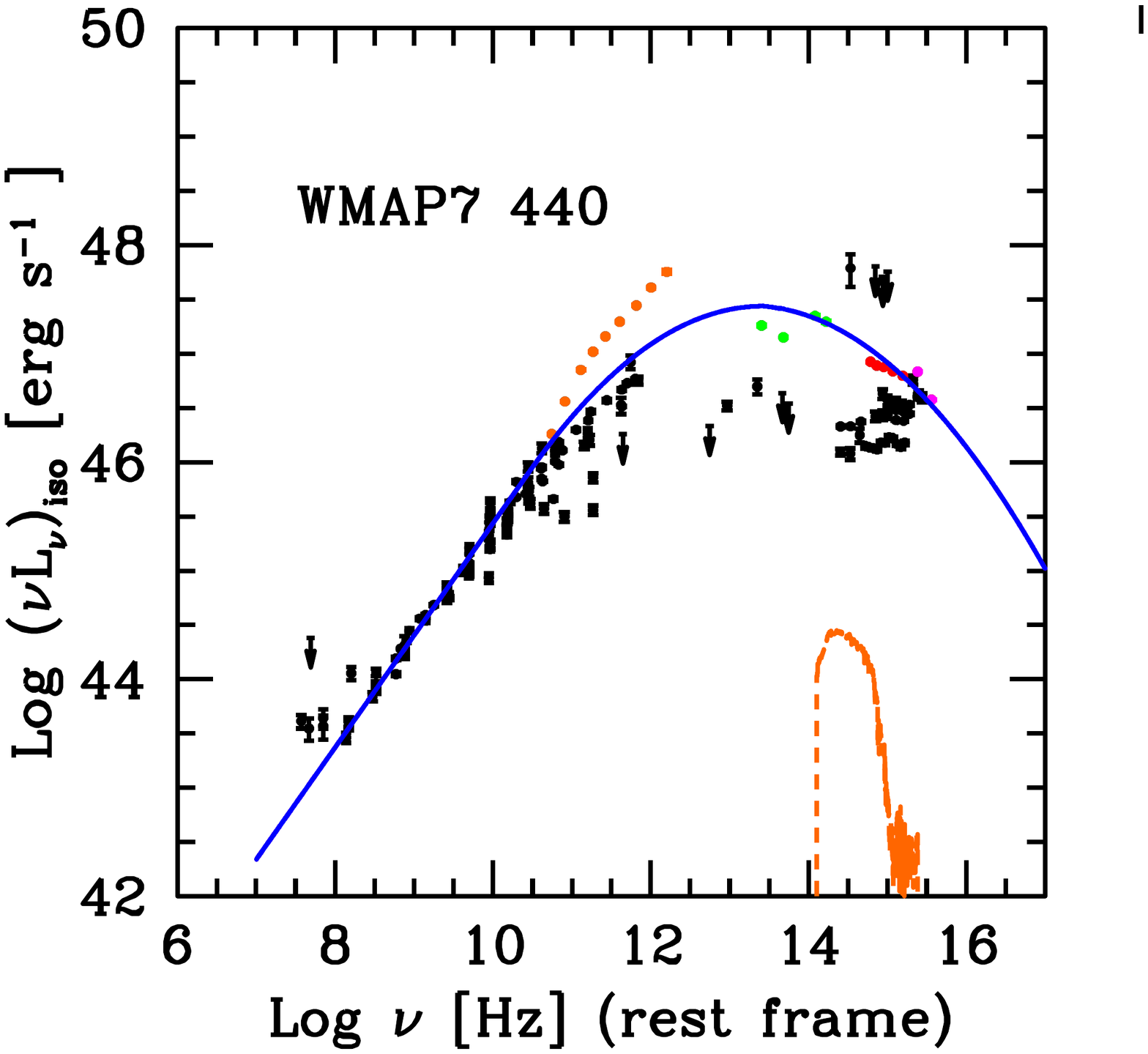}}
\caption{Continued.}
\end{figure*}

\end{appendix}
\end{document}